\documentclass[
 reprint,
 amsmath,amssymb,
 aps,
pra,
]{revtex4-2}
\usepackage{graphicx}
\usepackage{float}
\usepackage{dcolumn}
\usepackage{bm}
\usepackage{bbm}
\usepackage{braket}
\usepackage{xcolor}
\usepackage[T1]{fontenc}
\usepackage{enumerate,enumitem}
\usepackage{booktabs}
\usepackage{physics}
\usepackage{amsmath}
\usepackage{appendix}

\begin{document}

\title{Protocols for inter-module two-qubit gates mediated by time-bin encoded photons}

\author{Z.~M.~McIntyre }
\email{zoe.mcintyre@mail.mcgill.ca}
\author{W.~A.~Coish}%
 \email{william.coish@mcgill.ca}
\affiliation{%
 Department of Physics, McGill University, 3600 rue University, Montreal, QC, H3A 2T8, Canada
}%

\date{\today}

\begin{abstract} 
  As quantum devices scale to larger numbers of qubits, entangling gates between distant stationary qubits will help provide flexible, long-range connectivity in modular architectures. In this work, we present protocols for implementing long-range two-qubit gates mediated by either Fock-state or time-bin qubits---photonic encodings that are both compatible with the coplanar waveguide resonators commonly used in circuit quantum electrodynamics (QED). These protocols become deterministic in the limit of vanishing photon loss. Additionally, photon loss can be heralded, signaling a failed two-qubit gate attempt. We model the loss of a time-bin qubit to a dielectric environment consisting of an ensemble of two-level systems (TLSs), which are believed to be the dominant mechanism for dielectric loss in circuit QED architectures. The backaction (on the stationary qubits) associated with the loss of the time-bin qubit is strongly suppressed in a non-Markovian regime where the temporal separation of the time bins is short compared to the dielectric environment's correlation time. This result suggests strategies based on a combination of materials-fabrication and time-bin-qubit optimization for ensuring that the loss of a time-bin qubit is not only heralded, but also approximately backaction-free.
\end{abstract}

\maketitle

\section{Introduction}

A promising approach to large-scale quantum information processing involves coupling stationary qubits to microwave photons. The stationary qubits could be superconducting qubits~\cite{blais2021circuit} or spins housed in semiconductor heterostructures~\cite{burkard2023semiconductor}. As these systems scale up, the complexities associated with wiring and control will likely motivate the adoption of modular architectures where distributed quantum processing units (QPUs) are combined into a single device via classical processing or quantum interconnects~\cite{bravyi2022future}. In the near term, techniques like circuit cutting~\cite{bravyi2016trading,peng2020simulating} and entanglement forging~\cite{eddins2022doubling} can be used to simulate the output of larger quantum circuits by classically combining the outputs of smaller circuits run sequentially on a single QPU or in parallel across separate QPUs. Along similar lines, the statistics of long-range entanglement can be replicated using virtual gates~\cite{mitarai2021constructing,yamamoto2023error,singh2024experimental}, which simulate two-qubit gates from a quasiprobability decomposition of single-qubit rotations and projective measurements. The costs of simulating non-local virtual gates can be reduced by allowing for the exchange of classical information between QPUs~\cite{piveteau2023circuit}, as was recently demonstrated experimentally~\cite{carrera2024combining}. 

Strategies like circuit cutting and entanglement forging are best suited to computations where the structure of the problem motivates a partition into weakly interacting subsystems, each of which can be simulated individually before correlating the outcomes on a classical computer. Such a structure can be exploited to simplify many relevant problems in quantum computational chemistry~\cite{mcardle2020quantum}, such as finding the ground states of molecular Hamiltonians~\cite{eddins2022doubling}. However, entanglement between QPUs would enable a larger class of problems to be tackled in the long term~\cite{bravyi2022future}. 

In a circuit-QED or spin-circuit QED architecture, entanglement between modules housing distinct QPUs can be established via photonic interconnects capable of transmitting quantum information over distances far exceeding the centimeter-scale wavelength of microwave radiation.  Cryogenic meter-scale interconnects, combined with a protocol for applying gates across the interconnects, would form the architectural requirements for so-called $l$-type modularity~\cite{bravyi2022future}. Significant effort has gone towards developing strategies for quantum-state transfer and Bell-state generation between qubits connected by such interconnects~\cite{kurpiers2018deterministic,axline2018demand,campagne2018deterministic,zhong2019violating,kurpiers2019quantum, zhong2021deterministic,burkhart2021error,storz2023loophole,grebel2024bidirectional,mollenhauer2024high}. Bell pairs generated using these strategies could be stored and consumed on-demand as part of a gate-teleportation protocol~\cite{gottesman1999demonstrating,wan2019quantum}, enabling two-qubit gates between distant qubits. 
However, a more qubit-efficient approach may be to apply long-range gates directly to the code qubits themselves, bypassing the need for ancillary Bell pairs, Bell-pair quantum memory, possible entanglement-purification steps, and the additional gates required by the teleportation protocol itself. Long-range gates could also be used for performing parity checks of distant qubits~\cite{myers2007stabilizer,mcintyre2024flying}, implementing transversal encoded two-qubit gates~\cite{eastin2009restrictions}, reducing the impact of cosmic-ray events on encoded information~\cite{xu2022distributed}, or as a means of establishing the non-planar connectivity required for certain quantum error correcting codes (quantum low-density parity check codes)~\cite{mackay2004sparse,kovalev2013quantum,panteleev2021degenerate,bravyi2024high}.

Two-qubit gates between atom-qubits in separate optical cavities have been proposed~\cite{duan2005robust} and realized experimentally~\cite{daiss2021quantum}. These schemes leverage selection rules to entangle a qubit encoded in a multilevel system with a photonic polarization qubit~\cite{reiserer2014quantum,reiserer2015cavity}, which is then used to mediate the two-qubit gate. Such selection rules exist in many systems featuring optical transitions, including atoms~\cite{reiserer2014quantum} and excitons~\cite{tomm2024realization}. Polarization-dependent atomic transitions have also been considered as a way of implementing two-qubit gates between photonic qubits themselves~\cite{duan2004scalable,hacker2016photon}.  Although polarization qubits can readily be transmitted through either free space or optical fibers, their integration into circuit QED architectures is not commonly considered due to the fixed polarization of the electromagnetic field housed in the coplanar waveguide resonators commonly used to manipulate, measure, and couple qubits~\cite{marquardt2007efficient,blais2021circuit}. 

Here, we describe fixed-polarization protocols for performing two-qubit gates across quantum-photonic interconnects.  In contrast to entangling gates like the cross-resonance~\cite{rigetti2010fully,song2024realization} and resonator-induced-phase gates~\cite{cross2015optimized,deng2025long}, which require common coupling of the two qubits to a single spectrally isolated standing-wave mode, the approaches presented here couple the stationary qubits to propagating quasimodes of the interconnect---``flying'' qubits. The photonic degree-of-freedom required for flying qubits is provided by either a Fock-state encoding (whose basis states are defined by the presence or absence of a photon in the interconnect) or a time-bin encoding (whose basis states are the single-photon Fock states of a pair of orthogonal spatiotemporal modes of the interconnect). Time-bin encodings have been considered for linear-optical quantum computing~\cite{humphreys2013linear}, quantum key distribution~\cite{bouchard2022quantum}, Bell-state generation~\cite{kurpiers2019quantum,xie2021quantum,zheng2022entanglement}, and quantum-state transfer~\cite{kurpiers2019quantum}. The two-qubit gate protocols presented here become deterministic (they succeed with unit probability) in the limit of vanishing photon loss. The loss of a photon is a heralded error; hence, rather than contribute to a gate infidelity, errors due to photon loss are instead converted into a reduced probability of successfully executing the gate. 

In the most common (Markovian) models of photon loss, absorption of a photon by a dielectric medium occurs at a single instant in time. This absorption time is an indelible trace left in the environment that distinguishes between the two computational basis states of a photonic time-bin qubit immediately before it is absorbed. For the time-bin-qubit mediated gate described in this work, absorption of the intermediary photon in a way that distinguishes between the time-bin states leads to an error (backaction) on the stationary qubits. However, in a more general (non-Markovian) model of dielectric loss, there is room for a photon to be absorbed without introducing backaction. Standard models of photon loss do not account for details of the photon's spatiotemporal mode, but we fully account for this in the present work. Moreover, we model the loss of a time-bin qubit to a bath of two-level systems (TLSs), accounting for a structured spectral density of the TLS environment in order to determine the backaction of photon loss on the stationary qubits involved in the gate. Such TLSs are believed to be responsible for the majority of dielectric loss in circuit QED platforms~\cite{martinis2005decoherence, wang2015surface, de2018suppression,mcrae2020materials, lisenfeld2023enhancing, crowley2023disentangling,chen2024phonon,liu2024observation,zanuz2024mitigating}. We calculate the distinguishability of the TLS bath states resulting from the absorption of a photon from either the early or the late time bin. This distinguishability controls the rate of dephasing errors on stationary qubits resulting from the loss of the time-bin qubit. We show that the distinguishability of the states of the TLS ensemble (conditioned on the time-bin state absorbed) can be tuned by modifying either the TLS spectral density or the time-bin separation. This result suggests strategies for improving the performance of the gate protocols presented here in the context of a quantum error-correcting code, where the rate of errors on stationary code qubits should be kept as low as possible.

The structure of the rest of this article is as follows: In Sec.~\ref{sec:pitching}, we review how cavity-assisted Raman transitions can be used to deterministically emit and absorb single photons. In Sec.~\ref{sec:both-gates}, we present two protocols for applying a controlled-Z (CZ) gate to two distantly separated qubits: The first is appropriate for qubits separated by a distance exceeding the size of the photonic wavepacket, while the second alleviates this restriction at the cost of introducing an ancilla qudit used for measurement of the photonic qubit. In Sec.~\ref{sec:gate-simulation}, we present the results of a numerical simulation of the gate infidelity accounting for decoherence of the stationary qubits. We discuss the backaction associated with the loss of a time-bin qubit in Sec.~\ref{sec:loss-backaction}. We offer concluding remarks in Sec.~\ref{sec:conclusion}.

\section{Pitching photons}\label{sec:pitching}

The emission and subsequent reabsorption of single photons has long been recognized as a useful resource for quantum state transfer and entanglement distribution in quantum networks~\cite{cirac1997quantum}. In this section, we review how a cavity-assisted Raman transition can be used to controllably emit a single photon into a designated spatiotemporal quasimode of a transmission line coupled to the cavity.

As is well known from the study of atomic ensembles with $\Lambda$-type level structures~\cite{gerry1990dynamics,gorshkov2007photon,morin2019deterministic}, one way to emit and absorb single photons is via cavity-assisted Raman transitions. This process can be described by a Hamiltonian of the form
\begin{equation}\label{engineered-rabi}
    H[\Omega(t)]=i\Omega(t)\ketbra{g,1}{f,0}+\mathrm{h.c.},
\end{equation}
where $\ket{0}$ and $\ket{1}$ are zero- and one-photon Fock states of the cavity. The paradigmatic setup for realizing an interaction of this form [Eq.~\eqref{engineered-rabi}] is one where a cavity mode is coupled off-resonantly with detuning $\Delta$ and strength $g_0$ to the $\ket{g}\leftrightarrow\ket{e}$ transition of a $\Lambda$ system having states $\ket{g}$, $\ket{e}$, and $\ket{f}$, while the atom's $\ket{f}\leftrightarrow \ket{e}$ transition is driven off-resonantly by a classical drive with envelope $\lambda_{\mathrm{d}}(t)$. Adiabatically eliminating the $\ket{e}$ state then yields an interaction of the form $H[\Omega(t)]$ with $\Omega(t)=g_0\lambda_{\mathrm{d}}(t)/\Delta$~\cite{gerry1990dynamics}, where we have here assumed degeneracy of the $\ket{f}$ and $\ket{g}$ levels. In the dispersive regime of cavity QED (where the difference between the qubit and cavity frequencies far exceeds the strength of the transverse qubit-cavity coupling $g_0$), an interaction of the form of Eq.~\eqref{engineered-rabi} can be realized for superconducting transmon qubits by coherently driving the cavity~\cite{ilves2020demand} or qubit~\cite{pechal2014microwave,zeytinouglu2015microwave,kurpiers2018deterministic,kurpiers2019quantum,storz2023loophole} at the frequency of the dressed $\ket{f,0}\leftrightarrow \ket{g,1}$ transition. In this case, $\Omega(t)=g_0\lambda_{\mathrm{d}}(t)\alpha/(\sqrt{2}\Delta(\Delta+\alpha))$, where $\lambda_{\mathrm{d}}(t)$ again denotes the drive envelope and $\alpha$ is the transmon anharmonicity~\cite{zeytinouglu2015microwave}.  

\begin{figure}
    \centering
    \includegraphics[width=\linewidth]{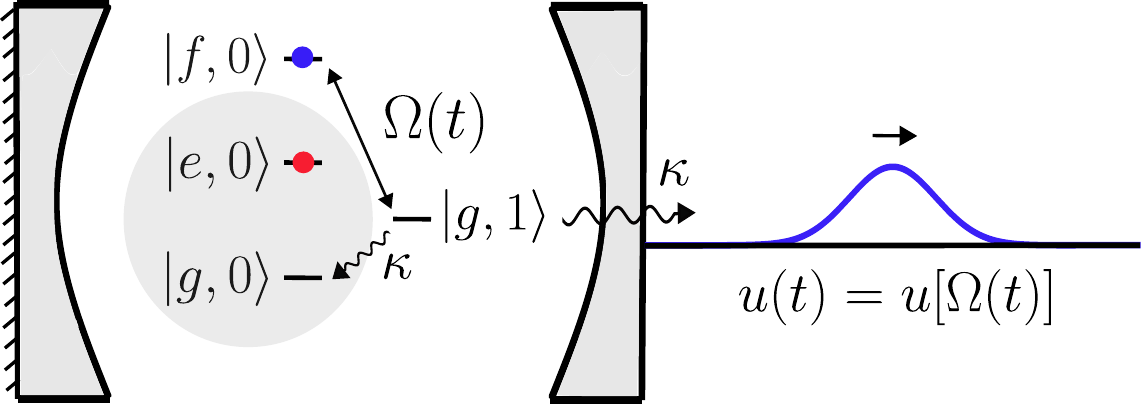}
    \caption{A cavity-assisted Raman process with tunable strength $\Omega(t)$ can be used to emit a photon into a chosen quasimode $u$ of a transmission-line resonator coupled to the cavity. }
    \label{fig:driving-setup}
\end{figure}

The interaction $H[\Omega(t)]$ can be used to emit a photon, conditioned on the atom being in state $\ket{f}$, into some quasimode $u$ [with a waveform $u(t)$ normalized according to $\int dt\:\lvert u(t)\rvert^2=1$] of a transmission-line resonator coupled to the cavity (Fig.~\ref{fig:driving-setup}). The resulting single-photon state can be expressed as $\ket{1_u}=r_u^\dagger \ket{\mathrm{vac}}$, where $\ket{\mathrm{vac}}$ is the vacuum state of the transmission line and
\begin{equation}\label{creation-operator-quasimode}
    r_{u}^\dagger=\int dt\: u(t)r_{\mathrm{out}}^\dagger(t).
\end{equation}
Here, $r_{\mathrm{out}}(t)=(2\pi)^{-1}\int d\omega e^{-i\omega t}r_\omega$ is an output-field operator written in terms of annihilation operators $r_\omega$ acting on frequency modes of the transmission line and satisfying $[r_\omega, r_{\omega'}^\dagger]=\delta(\omega-\omega')$~\cite{gardiner1985input}. Photon emission conditioned on the atomic state $\ket{f}$ can be realized in superposition to map $(\alpha\ket{e}+\beta\ket{f})\otimes\ket{\mathrm{vac}}\rightarrow \alpha\ket{e,\mathrm{vac}}+\beta\ket{g,1_u}$.

To relate the populated quasimode $u$ to the envelope $\Omega(t)=\Omega_{\mathrm{e}}(t)$ required for emitting a photon into mode $u$, we consider evolution described by the master equation
\begin{equation}\label{pitch-master-eq}
    \dot{\rho}=-i[H,\rho]+\kappa\mathcal{D}[a]\rho,
\end{equation}
where $H=H[\Omega_{\mathrm{e}}(t)]$, $\kappa$ is the rate of cavity decay into the transmission line, and $a$ is an annihilation operator that removes one photon from the cavity mode: $a\ket{1}=\ket{0}$. The damping superoperator $\mathcal{D}[a]$ acts according to $\mathcal{D}[a]\rho=a\rho a^\dagger-\{a^\dagger a,\rho\}/2$. 

For $\kappa\gg |\Omega(t)|$, the waveform $u(t)$ of the emitted photon can be related to the output field via the input-output relation  (see Ref.~\onlinecite{ranjan2020pulsed} and the supplement of Ref.~\onlinecite{mcintyre2022non} for details): $\langle r_{\mathrm{out}}\rangle_t =\sqrt{\kappa}\langle a\rangle_t=u(t)\langle \tau_{gf}\rangle_0$, where $\tau_{gf}=\ketbra{g}{f}$, and where the average is defined as $\langle \mathcal{O}\rangle_t=\mathrm{Tr}\{\mathcal{O}\rho(t)\}$ for operator $\mathcal{O}$. With the density matrix $\rho(t)$ written as $\rho(t)=\sum_{\alpha,\beta}\rho_{\alpha,\beta}(t)\ketbra{\alpha}{\beta}$ for $\ket{\alpha}\in\{\ket{g,0},\ket{g,1},\ket{f,0}\}$, we therefore have
\begin{equation}\label{waveform}
    u(t)=\sqrt{\kappa}\frac{\langle a \rangle_t}{\langle \tau_{gf}\rangle_0}=\sqrt{\kappa}\frac{\rho_{g1,g0}(t)}{\rho_{f0,g0}(0)}.
\end{equation}

To determine the dynamics of the relevant matrix element $\rho_{g1,g0}(t)$, we consider the following set of coupled differential equations obtained from Eq.~\eqref{pitch-master-eq}:
\begin{align}\label{diff-eqs}
    \begin{aligned}
        &\dot{\rho}_{g1,g0}(t)=\Omega_{\mathrm{e}}(t)\rho_{f0,g0}(t)-\frac{\kappa}{2}\rho_{g1,g0}(t),\\
        &\dot{\rho}_{f0,g0}(t)=-\Omega_{\mathrm{e}}^*(t) \rho_{g1,g0}(t).
    \end{aligned}
\end{align}
A simple expression for $\rho_{g1,g0}(t)$ can be derived in the regime where $\Omega_{\mathrm{e}}(t)$ varies slowly on a timescale $\kappa^{-1}$ and $\kappa$ far exceeds the strength of the driving, $\kappa\gg \lvert \Omega_{\mathrm{e}}(t)\rvert$. In this regime, the dynamics of the cavity field $\langle a\rangle_t=\rho_{g1,g0}(t)$ follows that of the instantaneous (equal time) atomic coherence $\langle \tau_{gf}\rangle_t=\rho_{f0,g0}(t)$, and we can adiabatically eliminate the cavity field by approximating $\dot{\rho}_{g1,g0}\simeq 0$. Solving the above system of equations then gives [cf.~Eq.~\eqref{waveform}]
\begin{equation}\label{system-sol}
    u(t)\simeq \frac{2\Omega_{\mathrm{e}}(t)}{\sqrt{\kappa}}e^{-\frac{2}{\kappa}\int_{-\infty}^tds\:\lvert \Omega_{\mathrm{e}}(s)\rvert^2}.
\end{equation}
Under the assumption that $\Omega_{\mathrm{e}}(t)$ varies slowly on a timescale $\kappa^{-1}$, this expression [Eq.~\eqref{system-sol}] indicates that $u(t)$ varies on a timescale $\sim\kappa/\lvert \Omega_{\mathrm{e}}\rvert^2$, which self-consistently exceeds the timescale $\kappa^{-1}$ provided $\kappa\gg\lvert\Omega_{\mathrm{e}}(t)\rvert$ [as initially assumed when solving the system of coupled differential equations in Eq.~\eqref{diff-eqs}].

In order to express $\Omega_{\mathrm{e}}(t)$ in terms of $u(t)$, we note that $ \lvert u(t)\rvert^2\simeq -\frac{d}{dt}e^{-\frac{4}{\kappa}\int_{-\infty}^t ds\:\lvert \Omega_{\mathrm{e}}(s)\rvert^2}$ [cf.~Eq.~\eqref{system-sol}], which can be integrated to obtain the relation
\begin{equation}
    e^{-\frac{4}{\kappa}\int_{-\infty}^t ds\:\lvert \Omega_{\mathrm{e}}(s)\rvert^2}\simeq 1-\int_{-\infty}^t ds\:\lvert u(s)\rvert^2.\label{condition-from-normalization}
\end{equation}
Using Eq.~\eqref{condition-from-normalization} in conjunction with Eq.~\eqref{system-sol} allows us to solve for $\Omega_{\mathrm{e}}(t)$ in terms of $u(t)$, giving~\cite{gorshkov2007photon, morin2019deterministic}
\begin{equation}\label{pulse-relation-2}
     \Omega_{\mathrm{e}}(t)\simeq \frac{\sqrt{\kappa}}{2}\frac{ u(t)}{\sqrt{\int_t^\infty ds\:\lvert u(s)\rvert^2}},\quad \lvert \Omega_{\mathrm{e}}(t)\rvert\ll \kappa.
\end{equation}
Here, we have made use of the normalization condition $\int dt\:\lvert u(t)\rvert^2=1$ to rewrite the range of the integral in the denominator. \textit{Absorption} of an incident photon in mode $u(t)$ can be realized by shaping the envelope for the absorption pulse $\Omega(t)=\Omega_{\mathrm{a}}(t)$ so that its time reverse $\Omega_{\mathrm{a}}^*(-t)$ would lead to the emission of a photon into the time-reversed quasimode having waveform $u^*(-t)$~\cite{gorshkov2007photon}. The pulse required for absorption of a photon with waveform $u(t)$ into a cavity having linewidth $\kappa$ can then be found from Eq.~\eqref{pulse-relation-2} and is given by
\begin{equation}\label{absorption}
    \Omega_{\mathrm{a}}(t)\simeq \frac{\sqrt{\kappa}}{2}\frac{u(t)}{\sqrt{\int_{-\infty}^tds\:\lvert u(s)\rvert^2}},\quad \lvert \Omega_{\mathrm{a}}(t)\rvert \ll \kappa.
\end{equation}
The combination of complex conjugation and mapping of $t\rightarrow {-}t$ in the waveform $u^*(-t)$ of the time-reversed quasimode can be understood in a plane-wave basis by noting that $r_\omega^\dagger$ [cf.~Eq.~\eqref{creation-operator-quasimode}] populates the plane-wave mode $\langle x \vert r_\omega^\dagger \vert\mathrm{vac}\rangle\propto e^{-i\omega (t-x/v)}$ (assuming a linear dispersion and propagation at speed $v$). Reversing the propagation direction of this plane wave can be accomplished through complex conjugating and a mapping of $t\rightarrow{-}t$, giving $e^{-i\omega(t+x/v)}$. This can be generalized from plane waves to quasimodes: For a photon in the quasimode $u$, $\langle x\vert r_u^\dagger\vert\mathrm{vac}\rangle\propto \int d\omega\:u(\omega)e^{-i\omega(t-x/v)}$, where $u(\omega)=\int dt\:e^{i\omega t}u(t)$ is the Fourier transform of the waveform $u(t)$. In this case, complex conjugation and a mapping of $t\rightarrow{-}t$ leads to an expression involving $u^*(\omega)$, corresponding to the Fourier transform of $u^*(-t)$---the waveform of the time-reversed mode.

While the model assumed here [Eq.~\eqref{engineered-rabi}] leads to the emission of a photon with unit efficiency, a finite lifetime of the $\ket{f}$ state will generally reduce the emission efficiency due to quantum jumps $\ket{f,0}\rightarrow\ket{e,0}$ during the emission process. This reduction in efficiency can be estimated as the ratio of the duration $\sim \pi/\Omega_{\mathrm{max}}$ of the photon-emission process relative to the $T_1^{f\rightarrow e}$ time of the $\ket{f}$ state, where here, $\Omega_{\mathrm{max}}$ is the maximal value of $\Omega(t)$. For transmon qubits, achievable values of $\Omega_{\mathrm{max}}$ are typically limited by the transmon anharmonicitiy $\alpha$~\cite{zeytinouglu2015microwave}, which tends to be on the order of a few hundred MHz. A value of $\Omega_{\mathrm{max}}/2\pi\approx 10$ MHz has been achieved with $\alpha/2\pi=-421$ MHz, corresponding to photon emission on a timescale $\sim\pi/\Omega_{\mathrm{max}}\approx 50$ ns~\cite{pechal2014microwave}. Although the $\ket{f}$ state in Ref.~\cite{pechal2014microwave} had a lifetime of only 550 ns, $T_1^{f\rightarrow e}$ times in excess of 40 $\mu$s have been measured~\cite{peterer2015coherence}, and emission of photons on faster timescales could likely be achieved by replacing the transmon with a more anharmonic qubit like the fluxonium~\cite{pechal2014microwave}. An additional source of inefficiency affecting the eventual re-absorption of the photon is the possibility of unintentional waveform distortion $u(t)\rightarrow u'(t)$ due to, e.g., dispersion. In this scenario, driving with the pulse $\Omega_{\mathrm{a}}(t)$ [Eq.~\eqref{absorption}] tailored for re-absorption of a photon with waveform $u(t)$  will lead to an imperfect recapture. Since the photon-absorption process is the time-reverse of the emission process~\cite{gorshkov2007photon}, the probability of absorbing a photon in quasimode $u'(t)$ with driving designed to absorb a photon in quasimode $u(t)$ is simply given by the probability of emitting a photon into $u'(t)$ with a pulse $\Omega_{\mathrm{e}}(t)$ designed to emit a photon into $u(t)$. This probability is itself given by the quasimode overlap $\int dt \: u^*(t) u'(t)$. 

In the next section, we describe two protocols for performing CZ gates between distantly separated stationary qubits. Both protocols require the emission and subsequent re-absorption of a photonic qubit, encoded in either a Fock basis or in a time-bin basis. In the notation established above, a photon encoding a time-bin qubit has a waveform $u_{\mathrm{E}}(t)$ if the photon is emitted at the early time, or $u_{\mathrm{L}}(t)=u_{\mathrm{E}}(t-\tau_{\mathrm{sep}})$ if it is emitted at the late time. Here, $\tau_{\mathrm{sep}}$ is a temporal offset chosen to be sufficiently large for the two time-bin states to be treated as orthogonal [$\int dt\: u_{\mathrm{E}}^*(t)u_{\mathrm{L}}(t)=0$]. Since these CZ gates are heralded, imperfect photon emission and reabsorption as described in the preceding paragraph will contribute to the probability of gate failure due to photon loss, rather than to the infidelity of the gate conditioned on a successful heralding outcome.

\section{CZ gates mediated by Fock-state or time-bin qubits}\label{sec:both-gates}

We consider a setup where two stationary qubits (described by two-level subspaces of three-level qudits $\mathrm{Q}_1$ and $\mathrm{Q}_2$) are each coupled to their own cavity. The cavities are themselves coupled by a transmission line, allowing the radiation emitted from one cavity to be channeled into the other. In cases where there is no ambiguity, we also denote by $\mathrm{Q}_i$ the qubit encoded in the three-level system $\mathrm{Q}_i$. 

A summary of this section is as follows: In Sec.~\ref{entangle-time-bin}, we first give steps for realizing an entangling operation between a stationary, cavity-coupled qubit and a photonic qubit encoded in either a Fock-state or time-bin basis. This operation is used as a primitive in constructing the long-range gates presented in Secs.~\ref{meas-free-gate} and \ref{gate-steps}. The two CZ-gate protocols differ mainly in whether or not they require a measurement of the flying qubit, which, as we later explain, could be realized by mapping the photonic-qubit basis states onto the states of an ancilla.  The ``measurement-free'' protocol, described in Sec.~\ref{meas-free-gate}, does not require measurement of the flying qubit but does require a minimum spatial separation of $\mathrm{Q}_1$ and $\mathrm{Q}_2$, set by the size of the photonic wavepacket. The ``ancilla-assisted'' protocol, described in Sec.~\ref{gate-steps}, alleviates this requirement by introducing a measurement of the flying qubit. Both protocols take advantage of the multilevel structure of $\Lambda$-type emitters to herald the loss of the flying qubit. The ancilla-assisted CZ-gate protocol could be mediated by a Fock-state encoded photon in principle, but we choose to present the protocol with a time-bin encoding for the single photon. This is due to the possibility---explored in Sec.~\ref{sec:loss-backaction}---of combining this gate protocol with engineering of the transmission line's dielectric environment to achieve a situation where, despite the orthogonality of the time-bin basis states, the loss of the time-bin qubit is not only heralded but also approximately backaction-free.

\subsection{A CZ gate between flying and stationary qubits}\label{entangle-time-bin}

\begin{figure}
    \centering
    \includegraphics[width=\linewidth]{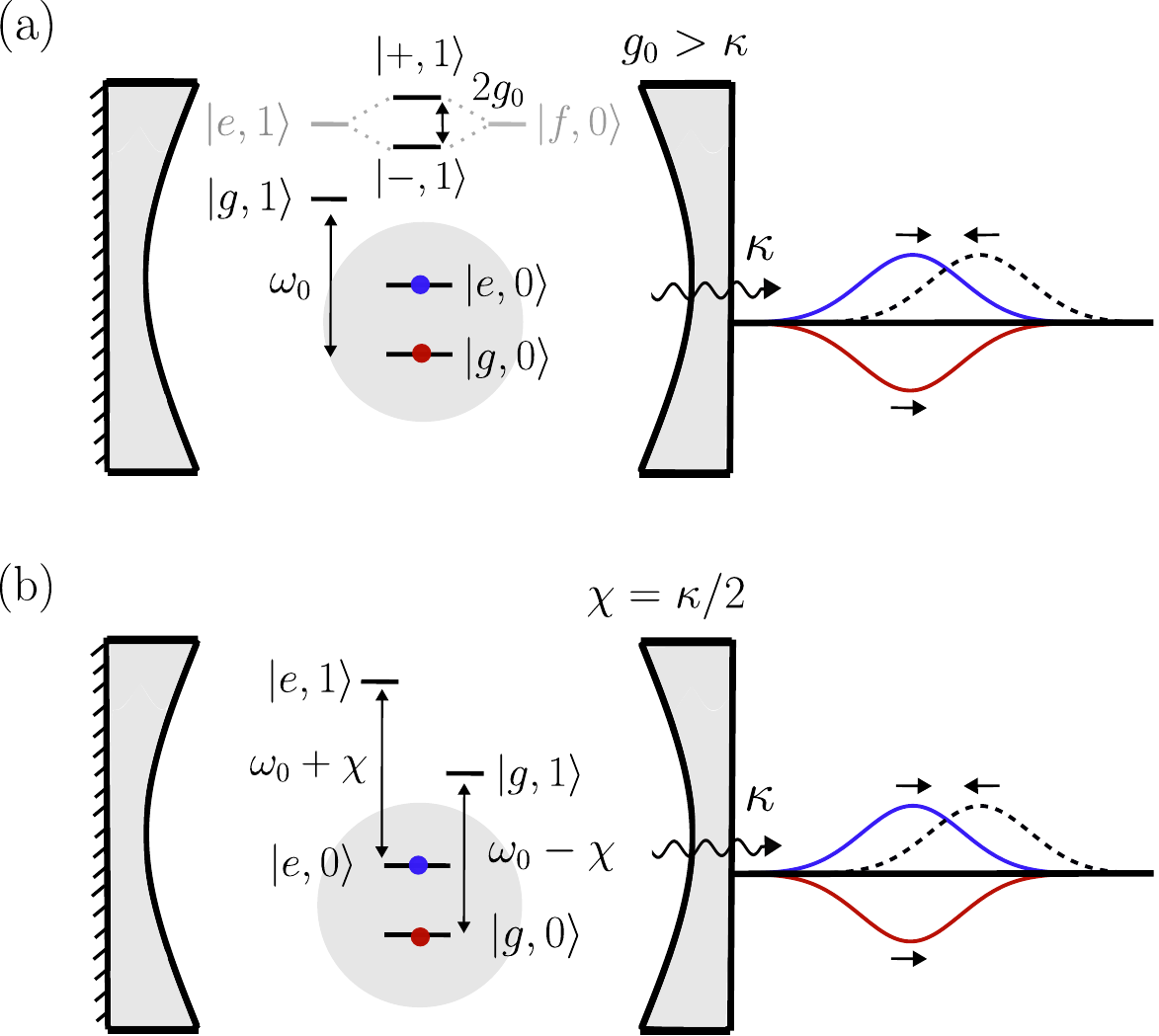}
    \caption{Two ways of realizing qubit-state-conditioned phase shifts on the photonic waveform: (a) Strong resonant coupling of the $\ket{e}$ and $\ket{f}$ states suppresses the cavity density-of-states at the bare cavity frequency $\omega_0$ for a qubit in state $\ket{e}$, while for a qubit in state $\ket{g}$, it remains peaked at $\omega_0$. For an incoming photon (dashed wavepacket) that is resonant with $\omega_0$ and has a narrow bandwidth compared to $\kappa$, the result is a $\pi$ phase shift incurred upon reflection, conditioned on state $\ket{g}$ (red and blue wavepackets). This mechanism has been demonstrated experimentally in Refs.~\cite{besse2018single,hacker2019deterministic,besse2020parity}. (b) Dispersive coupling with strength $\lvert \chi\rvert=\kappa/2$ can be used to realize a $\pm \pi/2$ phase shift conditioned on the qubit state, where the sign depends on the sign of $\chi$. This operation is locally equivalent to the $\pi$ phase shift described in (a) and has been demonstrated experimentally in Refs.~\cite{kono2018quantum,wang2022flying}.}
    \label{fig:phase_shifts}
\end{figure}

A crucial ingredient for realizing the long-range gate protocol of Sec.~\ref{meas-free-gate} is an entangling operation of the form 
\begin{equation}\label{local-cz-2}
    U=e^{i\pi\ketbra{1_u,g}}.
\end{equation}
This CZ gate acts on a stationary qubit having basis states $\ket{e}$ and $\ket{g}$ and a Fock-state encoded photonic qubit having basis states $\ket{1_u}$ and $\ket{\mathrm{vac}}$. The gate protocol presented in Sec.~\ref{gate-steps}, below, relies on the same operation for a time-bin encoded photon:
\begin{align}\label{local-cz}
\begin{aligned}
    &U=e^{i\pi \ketbra{\mathrm{L},g}}.
\end{aligned}
\end{align}  
In this case, the photon encodes a time-bin qubit having basis states $\ket{\mathrm{E}}$ and $\ket{\mathrm{L}}$, where $\ket{\mathrm{E}}$ ($\ket{\mathrm{L}}$) is a transmission-line state containing one photon in the spatiotemporal mode with waveform $u_{\mathrm{E}}(t)$  [$u_{\mathrm{L}}(t)$]. There are several ways these CZ gates could be realized using phase shifts of the photonic waveform conditioned on the state of a stationary qubit. 

Given a three-level system having states $\ket{g},\ket{e},\ket{f}$, a qubit-conditioned phase shift can be realized by resonantly coupling the system's $\ket{e}\leftrightarrow\ket{f}$ transition to a single-sided cavity [Fig.~\ref{fig:phase_shifts}(a)]. This cavity-qubit interaction leads to hybridization of the bare qubit and cavity states into states of the form
$\ket{\pm,1}\propto \ket{f,0}\pm\ket{e,1}$. These states
have eigenenergies $\omega_0\pm g_0$, where here $\omega_0$ is the bare cavity frequency and $g_0$ is the strength of the cavity-qubit coupling. For a qubit in state $\ket{g}$, an incoming photon resonant with the bare cavity frequency will undergo a $\pi$ phase shift provided the photon bandwidth is narrow relative to the cavity linewidth $\kappa$~\cite{besse2018single,hacker2019deterministic,besse2020parity}. (This bandwidth condition can be achieved if the photon wavepacket duration $\tau$ satisfies $\tau^{-1}\ll\kappa$.) However, for a qubit in state $\ket{e}$, the cavity density of states at $\omega_0$ is significantly suppressed by the cavity-qubit hybridization provided $g_0\gg \kappa$. An incoming photon will therefore be reflected \textit{without} a $\pi$ phase shift for a qubit in state $\ket{e}$ [Fig.~\ref{fig:phase_shifts}(a)]. For a Fock-state encoded photonic qubit, this $\pi$ phase shift implemented conditioned on the stationary qubit being in $\ket{g}$ provides a direct realization of the CZ gate given in Eq.~\eqref{local-cz-2}. Similarly, provided only a photon in state $\ket{\mathrm{L}}$ is allowed to interact with the stationary qubit, the same mechanism could be used to realize a CZ gate between a stationary qubit and a time-bin qubit [Eq.~\eqref{local-cz}].  The time-bin conditioned interaction with the stationary qubit could be realized using a fast switch~\cite{pechal2016superconducting,li2024quantum} or tunable coupling to the transmission line~\cite{grebel2024bidirectional}. 

We give an alternate mechanism for realizing the operation $U$ with a time-bin qubit [Eq.~\eqref{local-cz}]: For a stationary qubit which is instead \textit{dispersively} coupled to a cavity, an entangling operation locally equivalent to $U$ could be realized by dynamically switching the cavity susceptibility so that it differs for a photon in state $\ket{\mathrm{E}}$ versus $\ket{\mathrm{L}}$ [Fig.~\ref{fig:phase_shifts}(b)]. In this scenario, the cavity-qubit coupling can be modeled by a time-dependent dispersive interaction of the form
\begin{equation}
    H(t)=\chi(t) \sigma_z a^\dagger a,
\end{equation}
where
\begin{equation}\label{dispersive-coupling-time-dependent}
    \chi(t)=\frac{\kappa}{2}\Theta\left(t-T/2\right).
\end{equation}
Here, $\sigma_z$ is a Pauli-Z operator acting on the qubit, $a$ is an annihilation operator that removes one photon from the cavity, and $\Theta$ is a Heaviside step function with $T$ chosen so that $\Theta(t-T/2)=0$ [$\Theta(t-T/2)=1$] for an early (late) photon. With this choice of $\chi(t)$, and for a photon  waveform $u_\mathrm{E,L}(t)$ having a bandwidth narrow compared to $\kappa$ ($\tau^{-1}\ll \kappa$), a photon in state $\ket{\mathrm{L}}$ will ideally undergo a $+\pi/2$ ($-\pi/2$) phase shift when the stationary qubit is in state $\ket{g}$ ($\ket{e}$), while a photon in state $\ket{\mathrm{E}}$ will undergo a $\pi $ phase shift independent of the state of the stationary qubit. In terms of elementary gates and up to a global phase of $\pi$, this operation can be described as the combined action of (i) the required CZ gate $U$ [Eq.~\eqref{local-cz}] between the time-bin qubit and the stationary qubit and (ii) an $S^\dagger$ gate on the time-bin qubit, where in the basis $\{\ket{E},\ket{L}\}$, $S$ has matrix elements 1 and $i$ along the diagonal.

Since the quasimode waveforms $u_{\mathrm{E,L}}(t)$ are temporally well separated (to ensure orthogonality), the stepwise behavior of $\chi(t)$ [Eq.~\eqref{dispersive-coupling-time-dependent}] could be smoothed out in practice provided the timescale for turning $\chi$ on or off is short compared to the time-bin separation, and consequently, the operation of this CZ gate could be made largely insensitive to the temporal details of the $\chi(t)$ ramp-up provided it takes place well before the arrival of a photon in state $\ket{\mathrm{L}}$. For circuit QED systems, tuning of the dispersive shift from zero to non-zero values can be achieved on nanosecond timescales using SQUID-based tunable couplers~\cite{noh2023strong,gard2024fast}.

\subsection{Measurement-free gate}\label{meas-free-gate}

\begin{figure}
    \centering
    \includegraphics[width=\linewidth]{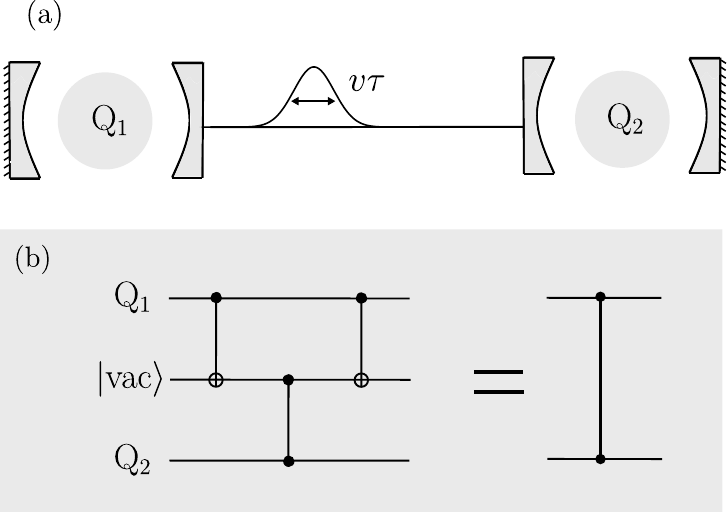}
    \caption{(a) Two qubits $\mathrm{Q}_1$ and $\mathrm{Q}_2$, each coupled to their own cavity mode, are connected by a long transmission line. A CZ gate between $\mathrm{Q}_1$ and $\mathrm{Q}_2$ can be realized by sending a Fock-state qubit, encoded in the presence or absence of a photon having a waveform of duration $\tau$, from $\mathrm{Q}_1$ to $\mathrm{Q}_2$ and back. The spatial extent of the photonic waveform is set by $v\tau$, where $v$ is the speed of light in the transmission line. (b) Circuit for applying a CZ gate to $\mathrm{Q}_1$ and $\mathrm{Q}_2$ using controlled-NOT (CNOT) operations with an auxiliary degree of freedom. These CNOT gates schematically represent conditional photon emission and absorption [cf.~Eq.~\eqref{emit}]. }
    \label{fig:meas-free}
\end{figure}

The first long-range gate we present does not require measurement of the photonic qubit, but does require that the size of the photonic wavepacket [set by the spatial extent of its waveform $u(t)$] be less than the distance separating the two qubits $\mathrm{Q}_1$ and $\mathrm{Q}_2$ [Fig.~\ref{fig:meas-free}(a)]. We begin by summarizing the steps involved in performing the gate [Fig.~\ref{fig:meas-free}(b)], before giving details of how each can be realized in practice. We assume an initial state of the two stationary qubits and transmission line of the form
\begin{equation}\label{initial-state}
    \ket{\Psi_0}\ket{\mathrm{vac}}=(\alpha\ket{gg}+\beta\ket{ge}+\gamma\ket{eg}+\delta\ket{ee})\ket{\mathrm{vac}},
\end{equation}
where $\ket{s_1s_2}=\ket{s_1}\otimes\ket{s_2}$ denotes $\mathrm{Q}_i$ in state $\ket{s_i}$, $\ket{\mathrm{vac}}$ is the vacuum state of the transmission line, and where $\alpha,\beta,\gamma,\delta$ are coefficients satisfying the normalization condition $[\lvert\alpha\rvert^2+\lvert\beta\rvert^2+\lvert \gamma\rvert^2+\lvert\delta\rvert^2]^{1/2}=1$. The gate begins by entangling $\mathrm{Q}_1$ with a Fock-state encoded photonic qubit (having basis states $\ket{1_u}$ and $\ket{\mathrm{vac}}$) via the mapping
\begin{align}\label{emit}
\begin{aligned}
     &\ket{e}\ket{\mathrm{vac}}\rightarrow\ket{e}\ket{\mathrm{vac}},\\
     &\ket{g}\ket{\mathrm{vac}}\rightarrow\ket{g}\ket{1_u}.
\end{aligned}
\end{align}
The photon is then sent to $\mathrm{Q}_2$, where the photonic qubit and $\mathrm{Q}_2$ undergo a CZ gate described by $U$ [Eq.\eqref{local-cz-2}]. The photon is then sent back to and re-absorbed by $\mathrm{Q}_1$, reversing the mapping of Eq.~\eqref{emit}. As indicated via the circuit identity shown in Fig.~\ref{fig:meas-free}(b), these steps realize a CZ gate between $\mathrm{Q}_1$ and $\mathrm{Q}_2$.

Having given this overview, we now provide a more detailed sequence of steps for realizing the measurement-free gate. Starting from Eq.~\eqref{initial-state}, the mapping of Eq.~\eqref{emit} can be realized by first applying to $\mathrm{Q}_1$ a $\pi_{fe}$ pulse followed by a $\pi_{eg}$ pulse followed by another $\pi_{fe}$ pulse. Here, $\pi_{ab}$ denotes a $\pi$ pulse that exchanges amplitude between levels $\ket{a}$ and $\ket{b}$. These $\pi$ pulses applied to $\mathrm{Q}_1$ produce the state
\begin{equation}\label{first-pitch}
    (\alpha\ket{fg}+\beta\ket{fe}+\gamma\ket{eg}+\delta\ket{ee})\ket{\mathrm{vac}}.
\end{equation}
At this point, a photon can be emitted into the quasimode $u$ (thereby populating the state $\ket{1_u}$) via a cavity-assisted Raman transition, conditioned on $\mathrm{Q}_1$ having a nonzero amplitude for the state $\ket{f}$. This completes the mapping of Eq.~\eqref{emit}. The photon is sent to $\mathrm{Q}_2$, after which the photonic qubit and $\mathrm{Q}_2$ undergo a CZ gate [Eq.~\eqref{local-cz}]. Following this CZ gate, the state of $\mathrm{Q}_1$, $\mathrm{Q}_2$, and the transmission line is given by
\begin{equation}\label{intermediate}
    (-\alpha\ket{gg}+\beta\ket{ge})\ket{1_u}+(\gamma\ket{eg}+\delta\ket{ee})\ket{\mathrm{vac}}.
\end{equation}
Next, the photon is sent back to and re-absorbed by $\mathrm{Q}_1$, sending $\ket{g}\ket{1_u}\rightarrow \ket{f}\ket{\mathrm{vac}}$ and disentangling the state of $\mathrm{Q}_1$ and $\mathrm{Q}_2$ from the state of the transmission line. Applying a $\pi_{eg}$ pulse, followed by a $\pi_{fe}$ pulse, followed by another $\pi_{eg}$ pulse to $\mathrm{Q}_1$ will then yield the final state
\begin{equation}\label{final-state}
    (-\alpha\ket{gg}+\beta\ket{ge}+\gamma\ket{eg}+\delta\ket{ee})\ket{\mathrm{vac}},
\end{equation}
consistent with the application of a CZ gate $e^{i\pi \ketbra{gg}}$ to the initial state of $\mathrm{Q}_1$ and $\mathrm{Q}_2$. If the photon was lost in transit, then the drive used to reabsorb the photon will act trivially [since the Hamiltonian in Eq.~\eqref{engineered-rabi} does not have a finite matrix element linking $\ket{g,0}$ and $\ket{f,0}$]. The final three $\pi$ pulses used to produce Eq.~\eqref{final-state} will then lead to a finite amplitude of having $\mathrm{Q}_1$ in state $\ket{f}$ conditioned on a photon having been emitted. This enables heralding of photon loss through a binary measurement described by the POVM $\{\ketbra{f},1-\ketbra{f}\}$~\cite{jerger2016realization}. 

Since the photonic qubit is emitted and ultimately re-absorbed by the same qubit ($\mathrm{Q}_1$), the drives used to emit and absorb the photon cannot overlap in time. The spatial extent of the quasimode $u$ must therefore be comparable to or shorter than the distance separating $\mathrm{Q}_1$ and $\mathrm{Q}_2$. As an example, this requirement sets a minimum distance of approximately $v\tau \simeq 0.1$ m for a $\tau \simeq 1$-ns photonic pulse traveling at a speed of $v\simeq 0.5\,c$, where $c$ is the speed of light in vacuum. A more precise estimate for $v$ is \cite{grebel2024bidirectional} $v=c/\sqrt{(1+\epsilon_r)/2}\approx 0.43c$, assuming a dielectric constant $\epsilon_r=10$ (for sapphire). The minimum distance $v\tau$ must in turn be balanced against the requirement that $\tau>\kappa^{-1}$, which, as explained in Sec.~\ref{entangle-time-bin}, is required for realizing the operation $U$ [Eq.~\eqref{local-cz-2}]. For $\kappa/2\pi=50$ MHz~\cite{grebel2024bidirectional}, for instance,  a pulse duration $\tau> 3.2$ ns would be required, corresponding to a pulse length $>0.41$ m (assuming the value for $v$ given above).

In the next section, we present a second protocol for applying a CZ gate to $\mathrm{Q}_1$ and $\mathrm{Q}_2$. This second protocol makes use of an ancilla that can be used to reabsorb the photon and therefore does not require the photon emission and absorption to be temporally separated.  The protocol is thus compatible with qubits that are separated by a distance smaller than the spatial extent of the photon waveform.

\subsection{Ancilla-assisted gate}\label{gate-steps}

\begin{figure*}
    \centering
    \includegraphics[width=0.8\linewidth]{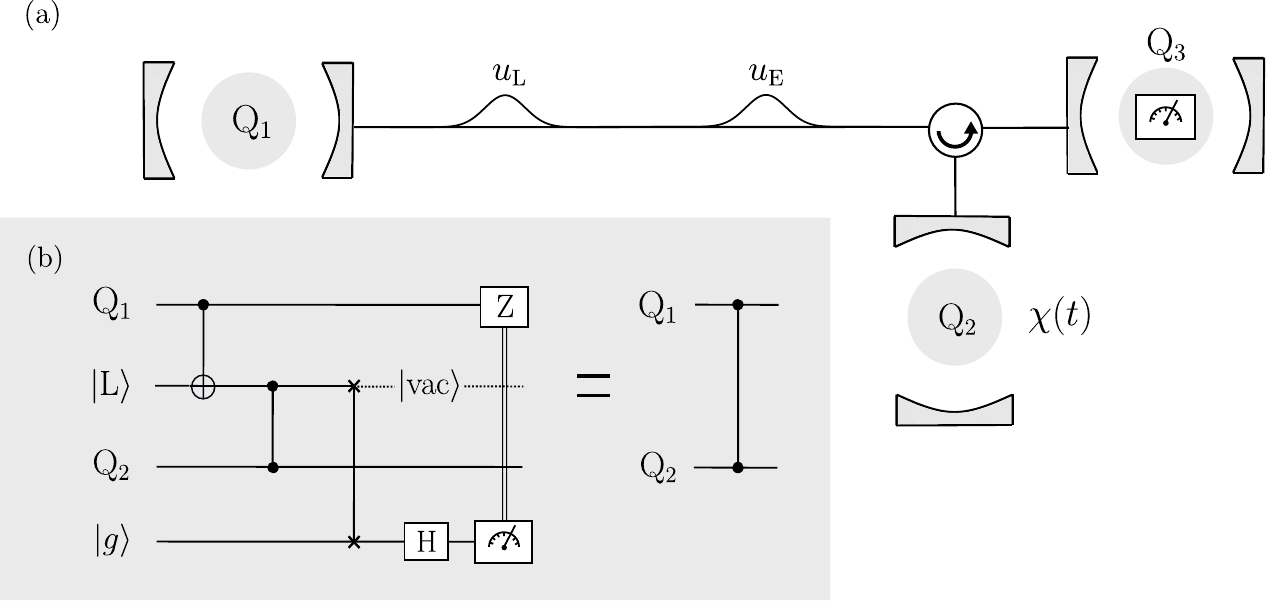}
    \caption{(a) Schematic of the imagined setup for the ancilla-assisted CZ gate: A two-qubit gate can be applied to cavity-coupled qubits $\mathrm{Q}_1$ and $\mathrm{Q}_2$ by emitting and re-absorbing a time-bin qubit. Qubit $\mathrm{Q}_1$ plays the role of the quantum emitter that pitches the time-bin qubit, which is subsequently re-absorbed by the ancilla $\mathrm{Q}_3$. While in transit, the photon encoding the time-bin qubit imparts a time-bin conditioned phase to $\mathrm{Q}_2$, enabled by a dispersive shift $\chi(t)$ that can be toggled on during the interaction with either the early ($u_\mathrm{E}$) or late ($u_{L}$) time-bin waveforms [Eq.~\eqref{dispersive-coupling-time-dependent}]. (b)  Quantum-circuit description of the suggested protocol for applying a CZ gate to $\mathrm{Q}_1$ and $\mathrm{Q}_2$ using a time-bin intermediary, with the ancilla $\mathrm{Q}_3$ initially prepared in its ground state $\ket{g}$. The entangling operation given in Eq.~\eqref{entangle} is represented schematically as an initial preparation of the time-bin qubit in state $\ket{\mathrm{L}}$, followed by a CNOT gate acting on $\mathrm{Q}_1$ and the time-bin qubit. }
    \label{fig:gate-circuit}
\end{figure*}

In this subsection, we present a second protocol for applying a CZ gate to $\mathrm{Q}_1$ and $\mathrm{Q}_2$ (Fig.~\ref{fig:gate-circuit}). In contrast to the measurement-free protocol presented in Sec.~\ref{meas-free-gate}, the photonic qubit mediating the gate presented here only travels from $\mathrm{Q}_1$ to $\mathrm{Q}_2$, rather than there and back. After interacting with $\mathrm{Q}_2$, the photonic qubit is measured with the help of an ancilla, $\mathrm{Q}_3$. This measurement projects $\mathrm{Q}_1$ and $\mathrm{Q}_2$ into a state consistent with the application of a CZ gate, $e^{i\pi \ketbra{gg}}$, up to a potential single-qubit rotation (conditioned on the measurement outcome). Although this gate could be mediated by a Fock-state encoded photonic qubit in principle, we consider a time-bin encoding as it offers the additional possibility of ``erasing'' the backaction associated with photon loss---a possibility explored in Sec.~\ref{sec:loss-backaction}. The time-bin qubit mediating the CZ gate presented in this section plays a role analogous to that of the photonic polarization qubit in the proposal of Ref.~\cite{duan2005robust}, where a similar protocol for performing a long-range CZ gate was suggested. In the protocol of Ref.~\cite{duan2005robust}, a photon encoding a polarization qubit is successively reflected off two distant cavities containing qubits, becoming entangled with both qubits in the process. A final measurement of the photon in an appropriate polarization basis then completes the gate. This protocol has been experimentally implemented between trapped-atom stationary qubits with a polarization-qubit intermediary \cite{daiss2021quantum}.

Much like the measurement-free gate of Sec.~\ref{meas-free-gate}, the ancilla-assisted gate is realized by entangling $\mathrm{Q}_1$ with a photonic qubit, here taken to be a time-bin qubit:
\begin{align}
\begin{aligned}\label{entangle}
    &\ket{e}\ket{\mathrm{vac}}\rightarrow\ket{e}\ket{\mathrm{E}},\\
    &\ket{g}\ket{\mathrm{vac}}\rightarrow\ket{g}\ket{\mathrm{L}}.
\end{aligned}
\end{align}
The time-bin qubit can then be entangled with $\mathrm{Q}_2$ using the unitary operation $U$ [cf.~Eq.~\eqref{local-cz}]. The gate is completed by measuring the state of the time-bin qubit in the $X$ basis $\ket{\mathrm{E}}\pm \ket{\mathrm{L}}$. In an optical implementation, the required time-bin Hadamard could be realized using delays and switches~\cite{soudagar2007cluster,lo2020quantum}, or by mapping the time-bin qubit to a polarization degree-of-freedom and applying the Hadamard to the polarization qubit~\cite{bussieres2010testing}. In the event that a time-bin Hadamard is unavailable or impractical, as would likely be the case in a circuit QED setup, the required $X$-basis measurement could be performed by mapping the state of the time-bin qubit onto the state of an ancillary three-level system $\mathrm{Q}_3$ initialized in its ground state [Fig.~\ref{fig:gate-circuit}(a)] using the mapping~\cite{kurpiers2019quantum}  
\begin{align}\label{catch-mapping}
\begin{aligned}
\ket{g}\ket{\mathrm{E}}&\rightarrow \ket{e}\ket{\mathrm{vac}},\\
\ket{g}\ket{\mathrm{L}}&\rightarrow \ket{g}\ket{\mathrm{vac}},\\
\ket{g}\ket{\mathrm{vac}}&\rightarrow\ket{f}\ket{\mathrm{vac}}.
\end{aligned}
\end{align}
The mappings of Eqs.~\eqref{entangle} and \eqref{catch-mapping} have been demonstrated experimentally with superconducting transmon qubits for the purpose of Bell-state generation~\cite{kurpiers2019quantum} and can be realized using $\pi$ pulses together with cavity-assisted Raman transitions. The $X$-basis measurement of the time-bin qubit can then be realized by applying a Hadamard to $\mathrm{Q}_3$ (mapping $\ket{e}\rightarrow \ket{+}$ and $\ket{g}\rightarrow \ket{-}$) and measuring $\mathrm{Q}_3$ in the basis $\{\ket{g}, \ket{e},\ket{f}\}$. In the absence of errors, a measurement of $\ket{g}$ or $\ket{e}$ heralds success: If the measurement outcome is $\ket{e}$, then the post-measurement state of $\mathrm{Q}_1$ and $\mathrm{Q}_2$ corresponds to the state obtained by applying the CZ gate $U_{\mathrm{CZ}}=e^{i\pi \ketbra{gg}}$ to the initial state of $\mathrm{Q}_1$ and $\mathrm{Q}_2$ [Eq.~\eqref{initial-state}]. If the outcome is $\ket{g}$, then the applied operation can be transformed to $U_{\mathrm{CZ}}$ by applying a $Z$ gate to $\mathrm{Q}_1$ [Fig.~\ref{fig:gate-circuit}(b)]. We quantify the backaction resulting from photon loss, heralded by a measurement of $\mathrm{Q}_3$ in state $\ket{f}$, in Sec.~\ref{sec:loss-backaction}.

The method considered here for coherently mapping the state of a stationary qubit onto a time-bin degree-of-freedom [Eq.~\eqref{entangle}] requires the availability of an auxiliary level or degree-of-freedom beyond the two states $\ket{e}$ and $\ket{g}$ spanning the computational subspace. Although we here assume the availability of an auxiliary level $\ket{f}$ (Fig.~\ref{fig:driving-setup}), the required auxiliary system could alternatively consist of, e.g., a long-lived nuclear spin as described in Ref.~\cite{tissot2024efficient}.

As with the measurement-free gate, we now give a more detailed sequence of steps for realizing the ancilla-assisted gate protocol: Starting again from the initial state given in Eq.~\eqref{initial-state}, the amplitude for the state $\ket{g}$ ($\ket{e}$) of $\mathrm{Q}_1$ is first transferred to $\ket{e}$ ($\ket{f}$) using a $\pi_{fe}$ pulse followed by a $\pi_{eg}$ pulse. Using a cavity-assisted Raman transition, a photon can be transferred into the cavity coupled to $\mathrm{Q}_1$, conditioned on $\mathrm{Q}_1$ starting in the state $\ket{f}$. This photon will leak into the transmission line and populate the state $\ket{\mathrm{E}}$, leaving the cavity in its vacuum state independent of the state of $\mathrm{Q}_1$. The state of $\mathrm{Q}_1$, $\mathrm{Q}_2$, $\mathrm{Q}_3$ (initialized in $\ket{g}$) and the transmission line is then
\begin{equation}\label{in-between-pitches}
    (\alpha\ket{eg}+\beta\ket{ee})\ket{g}\ket{\mathrm{vac}}+(\gamma\ket{gg}+\delta\ket{ge})\ket{g}\ket{\mathrm{E}}.
\end{equation}
The basis ordering for the stationary qubits in Eq.~\eqref{in-between-pitches} is $\ket{\mathrm{Q}_1,\mathrm{Q}_2}\ket{\mathrm{Q}_3}$. We assume a switch~\cite{pechal2016superconducting,li2024quantum}, a tunable coupler~\cite{grebel2024bidirectional}, or a time-dependent dispersive shift $\chi(t)$ [Fig.~\ref{fig:gate-circuit}(b)] is used to ensure that a photon in state $\ket{\mathrm{E}}$ does not interact with $\mathrm{Q}_2$. This action results in an $\ket{\mathrm{E}}$-conditioned identity operation on $\mathrm{Q}_2$ [cf.~Eq.~\eqref{local-cz}], up to a phase in the case of $\chi(t)$ (see Sec.~\ref{entangle-time-bin}). A photon in state $\ket{\mathrm{E}}$ is then re-absorbed by $\mathrm{Q}_3$, giving
\begin{equation}
    (\alpha\ket{eg}+\beta\ket{ee})\ket{g}\ket{\mathrm{vac}}+(\gamma\ket{gg}+\delta\ket{ge})\ket{f}\ket{\mathrm{vac}}.
\end{equation}
Next, a $\pi_{fe}$ pulse is applied to $\mathrm{Q}_1$, followed by a $\pi_{eg}$ pulse. A $\pi_{fe}$ pulse is concurrently applied to $\mathrm{Q}_3$, so that the state of $\mathrm{Q}_3$ is $\ket{e}$ conditioned on the photon having been in state $\ket{\mathrm{E}}$ [cf.~Eq.~\eqref{catch-mapping}]. Driving the $\ket{f,0}\leftrightarrow\ket{g,1}$ transition of $\mathrm{Q}_1$ [Eq.~\eqref{engineered-rabi}] will again transfer a photon into the transmission line, conditioned on $\mathrm{Q}_1$ being in state $\ket{f}$. However, a photon emitted at this stage will instead populate the state $\ket{\mathrm{L}}$, leading to the state
\begin{equation}\label{intermediate}
    (\alpha\ket{gg}+\beta\ket{ge})\ket{g}\ket{\mathrm{L}}+(\gamma\ket{eg}+\delta\ket{ee})\ket{e}\ket{\mathrm{vac}}.
\end{equation}
During the time that $\ket{\mathrm{L}}$ is populated, the switch, tunable coupler, or dispersive shift $\chi(t)$ is instead configured so that a photon in state $\ket{\mathrm{L}}$ \textit{does} interact with $\mathrm{Q}_2$, picking up a phase shift of $\pi$ conditioned on $\mathrm{Q}_2$ being in state $\ket{g}$ [cf.~Eq.~\eqref{local-cz}]. This interaction sends $\alpha\rightarrow-\alpha$ in Eq.~\eqref{intermediate}, above. 

Finally, a photon in state $\ket{\mathrm{L}}$ is re-absorbed by $\mathrm{Q}_3$, at which point the transmission line (in state $\ket{\mathrm{vac}}$) becomes disentangled from $\mathrm{Q}_1$, $\mathrm{Q}_2$, and $\mathrm{Q}_3$, whose state is now given by
\begin{equation}
    (-\alpha\ket{gg}+\beta\ket{ge})\ket{f}+(\gamma\ket{eg}+\delta\ket{ee})\ket{e}.
\end{equation}
A $\pi$-pulse sequence consisting of back-to-back $\pi_{eg}$, $\pi_{fe}$, and $\pi_{eg}$ pulses applied to $\mathrm{Q}_3$ can be used to transfer population in state $\ket{f}$ of $\mathrm{Q}_3$ to $\ket{g}$. The final state of $\mathrm{Q}_3$ is therefore $\ket{g}$ ($\ket{e}$) conditioned on the time-bin qubit having been in state $\ket{\mathrm{L}}$ ($\ket{\mathrm{E}}$) [Eq.~\eqref{catch-mapping}]:
\begin{equation}
    (-\alpha\ket{gg}+\beta\ket{ge})\ket{g}+(\gamma\ket{eg}+\delta\ket{ee})\ket{e}.
\end{equation}
If the photon was lost in transit, then the drives [leading to $\Omega(t)$] applied to $\mathrm{Q}_3$ as part of the time-bin-qubit reabsorption both act trivially [since there is no matrix element linking $\ket{g,0}$ and $\ket{f,0}$ in Eq.~\eqref{engineered-rabi}]. The ancilla $\mathrm{Q}_3$ therefore remains in state $\ket{g}$ until the final three $\pi$ pulses, which ultimately leave $\mathrm{Q}_3$ in state $\ket{f}$ [Eq.~\eqref{catch-mapping}]. 

As explained above, an $X$-basis measurement of the time-bin qubit can then be realized by applying a Hadamard to $\mathrm{Q}_3$ and measuring $\mathrm{Q}_3$ in the computational basis [Fig.~\ref{fig:gate-circuit}(b)]. Photon loss is heralded by a measurement of $\mathrm{Q}_3$ in state $\ket{f}$. 

\section{Gate infidelity}\label{sec:gate-simulation}

In this section, we report estimates for the gate infidelity 
\begin{equation}
    \varepsilon=1-F
\end{equation}
of the ancilla-assisted gate, post-selected on not having lost the photon encoding the time-bin qubit (a.k.a.~post-selected on a measurement of $\mathrm{Q}_3$ in $\ket{g}$ or $\ket{e}$). Here,
\begin{equation}\label{gate-fidelity}
   F= \int d\psi \:\langle\psi \vert U_{\mathrm{CZ}}^\dagger \mathcal{M} (\ketbra{\psi}) U_{\mathrm{CZ}} \vert \psi\rangle
\end{equation}
is the post-selected gate fidelity, where $d\psi$ is the two-qubit Haar measure, $U_{\mathrm{CZ}}=e^{i\pi\ketbra{gg}}$ is a unitary describing the action of an ideal CZ gate, and $\mathcal{M}(\ketbra{\psi})$ is the two-qubit state obtained in the presence of errors in the CZ-gate protocol, modeled in the manner we now describe. We focus on the ancilla-assisted gate for this analysis due to its compatibility with photon wavepackets longer than the distance separating the qubits. This compatibility yields better flexibility for optimizing the duration $\tau$ of the photonic waveform given fixed values of the qubit and cavity parameters.

Given some initial state $\ket{\psi}\otimes \ket{g}$ of $\mathrm{Q}_1$, $\mathrm{Q}_2$, and $\mathrm{Q}_3$, we numerically evaluate $\mathcal{M}(\ketbra{\psi})$ by solving a master equation 
\begin{equation}\label{master-eq-simul}    
\dot{\rho}=-i[H_{\mathrm{tot}}(t), \rho]+\sum_{j=0}\mathcal{D}[L_j]\rho\equiv \mathcal{L}_{\mathrm{tot}}\rho
\end{equation}
for the joint state $\rho(T)$ of $\mathrm{Q}_i$ ($i=1,2,3$) prior to the final $X$-basis measurement of $\mathrm{Q}_3$ at time $T$. Here, $\mathcal{D}[L]\rho=L\rho L^\dagger-\frac{1}{2}\{L^\dagger L, \rho\}$ is the usual damping superoperator. The state $\mathcal{M}(\ketbra{\psi})$ of $\mathrm{Q}_1$ and $\mathrm{Q}_2$ entering Eq.~\eqref{gate-fidelity} corresponds to the state obtained following the measurement of $\mathrm{Q}_3$, including the $Z$ gate applied to $\mathrm{Q}_1$ conditioned on a measurement of $\mathrm{Q}_3$ in state $\ket{g}$ [Fig.~\ref{fig:gate-circuit}(b)]. The Hamiltonian $H_{\mathrm{tot}}(t)$ includes $H[\Omega(t)]$ [cf.~Eq.~\eqref{engineered-rabi}] and $\pi$ pulses (here treated as ideal and instantaneous) applied to $\mathrm{Q}_1$ and $\mathrm{Q}_3$ as part of the time-bin-qubit emission and absorption, as well as a time-dependent dispersive coupling $\chi(t)$ to $\mathrm{Q}_2$ [Eq.~\eqref{dispersive-coupling-time-dependent}]. The unidirectional propagation of radiation from one cavity to the next is modeled via the SLH formalism~\cite{combes2017slh}. The specific terms entering $H_{\mathrm{tot}}(t)$ are defined in the Appendix.   

We denote by $\tau_{\alpha\beta,i}$ the operator $\tau_{\alpha\beta}=\ketbra{\alpha}{\beta}$ acting on $\mathrm{Q}_i$. To account for qubit decoherence, we include damping processes generated by $L_j\in \{\sqrt{\Gamma}\tau_{eg,i},\sqrt{\Gamma/2}(\tau_{ee,i}-\tau_{gg,i})\}_{i=1,2,3}\cup \{\sqrt{\Gamma}\tau_{fe,i},\sqrt{\Gamma/2}(\tau_{ff,i}-\tau_{ee,i})\}_{i=1,3}$. 
These generators describe relaxation of the $\ket{f}$ states of $\mathrm{Q}_1$ and $\mathrm{Q}_3$ into the qubits' respective $\ket{e}$ states, as well as relaxation $\ket{e}\rightarrow \ket{g}$ for each of $\mathrm{Q}_i$, $i=1,2,3$. (The $\ket{f}$ level of $\mathrm{Q}_2$ is not populated at any point.) Qubit dephasing is assumed to be relaxation-limited, leading to a factor-of-$2$ difference between the rates associated with relaxation and dephasing. For simplicity, we take all relaxation rates to be equal. The linewidths $\kappa_i$ of the cavities coupled to $\mathrm{Q}_i$ ($i=1,2,3$) are taken to be equal as well: $\kappa_i=\kappa$.

\begin{figure}
    \centering
    \includegraphics[width=\linewidth]{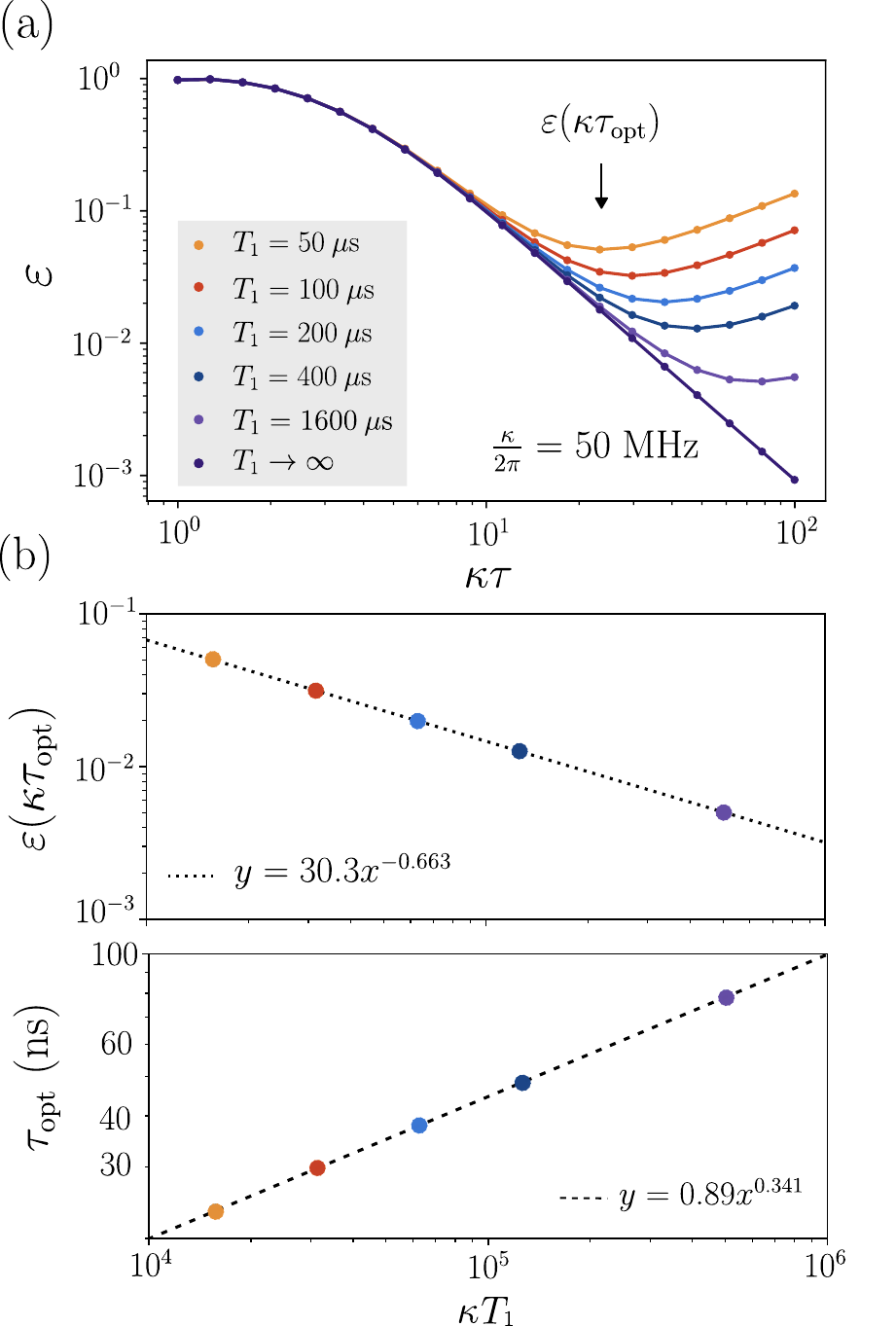}
    \caption{(a) The gate infidelity $\varepsilon=1-F$ evaluated as a function of $\tau$ for a fixed value of $\kappa/2\pi=50$ MHz and a total gate duration $T=16\tau$. In the absence of qubit decoherence ($T_1\rightarrow\infty$), the gate infidelity is proportional to $(\kappa\tau)^{-2}$ for $\kappa\tau\gtrsim 10$, while for a finite $T_1$, competition between the errors due to the finite pulse bandwidth and the errors due to decoherence leads to a minimum in $\varepsilon(\kappa\tau)$, and consequently to an optimal time-bin duration $\tau_{\mathrm{opt}}$. Points indicate values of $\kappa\tau$ used in simulation, while solid lines are intended to provide a guide to the eye. (b) Top panel: The minimum error $\varepsilon(\kappa\tau_{\mathrm{opt}})$ as a function of $T_1$ for the same values of $T_1$ as considered in (a). The dotted line is a least-squares fit whose slope $m=-0.663$ gives a numerical estimate of the scaling exponent $\zeta$ in Eq.~\eqref{scaling-exponent}. Bottom panel: The optimal time-bin duration $\tau_{\mathrm{opt}}$, also as a function of $T_1$ for the same values of $T_1$ as shown in (a). The slope $m=0.341$ of the line of best fit gives a numerical estimate of the scaling exponent $\xi$.}
    \label{fig:gate-fidelity}
\end{figure}

The measurement of $\mathrm{Q}_3$ is treated as ideal in simulation. We provide an estimate of the gate infidelity due to measurement errors at the end of this section. Evaluating the gate infidelity $\varepsilon=1-F$ [cf.~Eq.~\eqref{gate-fidelity}] in the manner described above yields the results shown in Fig.~\ref{fig:gate-fidelity} for $\kappa/2\pi=50$ MHz. The quality of the ``slowly varying'' approximation made in deriving the drive envelopes $\Omega_{\mathrm{e,a}}(t)$ [Eqs.~\eqref{pulse-relation-2}-\eqref{absorption}] is expected to increase with $\kappa\tau$ as the dynamics of the cavity mode coupled to $\mathrm{Q}_1$  ($\mathrm{Q}_3$) become more instantaneously related to those of $\mathrm{Q}_1$ ($\mathrm{Q_3}$). As described in Sec.~\ref{entangle-time-bin}, the quality of the CZ gate between $\mathrm{Q}_2$ and the time-bin qubit will also increase with increasing $\kappa\tau$ since realizing a well-defined phase $\phi$ on a photonic waveform (sending, e.g., $\ket{\mathrm{E}}\rightarrow e^{i\phi}\ket{\mathrm{E}}$) requires that the bandwidth $\sim \tau^{-1}$ of the waveform be narrow compared to the cavity linewidth $\kappa$. For a Gaussian $u(t)$ with standard deviation $\tau$, and in the absence of other sources of error, we expect these finite-bandwidth effects to control the gate infidelity ($\varepsilon=\varepsilon_{\kappa\tau}$) according to~\cite{mcintyre2024flying}
\begin{equation}\label{error-kappatau}
    \varepsilon_{\kappa\tau} = \frac{A}{(\kappa\tau)^2},\quad\kappa\tau\gg 1,\quad \Gamma=T_1^{-1}\rightarrow 0,
\end{equation}
where $A$ is a numerical prefactor related to the slope of the $T_1\rightarrow \infty$ line in Fig.~\ref{fig:gate-fidelity}(a). For a finite $T_1$, however, increasing $\tau$ results in a longer gate duration and consequently an increased contribution $\varepsilon_{T_1}$ to the gate infidelity due to qubit decoherence, which we expect to scale as
\begin{equation}\label{error-gamma}
    \varepsilon_{T_1}= B\frac{\tau}{T_1},\quad \tau<T_1.
\end{equation}
Here, $B$ is again a numerical prefactor. The competition between $\varepsilon_{\kappa\tau}$ and $\varepsilon_{T_1}$ leads to a minimum in $\varepsilon(\kappa\tau)$ (Fig.~\ref{fig:gate-fidelity}), implying the existence of an optimal $\tau=\tau_{\mathrm{opt}}$. This minimum occurs at the value of $\tau$ for which $d(\varepsilon_{\kappa\tau}+\varepsilon_{T_1})/d\tau=0$, leading to a predicted scaling for $\tau_{\mathrm{opt}}$ given by
\begin{equation}\label{tau-opt-scaling}
    \tau_{\mathrm{opt}}=\frac{K}{\kappa}(\kappa T_1)^{\xi},\quad \xi=1/3,
\end{equation}
where $K=(2A/B)^{1/3}$.
The minimum $\varepsilon(\kappa\tau_{\mathrm{opt}})=\mathrm{inf}_\tau \varepsilon(\kappa\tau)=(\varepsilon_{\kappa\tau}+\varepsilon_{T_1})\vert_{\tau_{\mathrm{opt}}}$ is therefore predicted to scale according to
\begin{align}\label{scaling-exponent}
    \varepsilon(\kappa\tau_{\mathrm{opt}})& =D (\kappa T_1)^{\zeta},\quad \zeta=-2/3,
\end{align}
where $D=3(AB^2/2)^{1/3}$. The predicted scaling exponents $\xi$ and $\zeta$ [Eqs.~\eqref{tau-opt-scaling} and \eqref{scaling-exponent}]  resulting from this simple analysis are relatively consistent with the numerical results shown in Fig.~\ref{fig:gate-fidelity}(b), where the slopes of the lines of best fit (corresponding to the scaling exponents) are 
\begin{align}
\begin{aligned}
    &\xi_{\mathrm{est}}=0.341\pm 0.004,\\
    &\zeta_{\mathrm{est}}=-0.6628\pm 0.0008,
\end{aligned}
\end{align}
and where the prefactors $K$ and $D$ extracted from the same fits are $K=0.89\pm 0.04$ and $D=30.3\pm 0.3$. The deviation from the predicted scaling exponents of $\xi=1/3$ and $\zeta=-2/3$ can most likely be attributed to a more sophisticated interplay between finite-bandwidth effects and qubit decoherence than what was assumed in deriving Eqs.~\eqref{tau-opt-scaling} and \eqref{scaling-exponent}.

The simulation results displayed in Fig.~\ref{fig:gate-fidelity} were obtained under the assumption that the measurement of $\mathrm{Q}_3$ was ideal. Measurement errors occurring with probability $p_{\mathrm{m}}$ will instead lead to a fidelity
\begin{equation}
    F(p_{\mathrm{m}})=(1-p_{\mathrm{m}})F(0)+p_{\mathrm{m}}F',
\end{equation}
where an expression for $F=F(0)$ was given in Eq.~\eqref{gate-fidelity}, and where $F'$ is the gate fidelity obtained in the presence of a measurement error. Such a measurement error simply leads to the wrong correction operator being applied: a $Z$ gate is applied to $\mathrm{Q}_1$ when the measurement outcome is $\ket{e}$, or not applied when the outcome is $\ket{g}$. In the absence of other sources of error (e.g.~qubit decoherence), we therefore have the following simple expression for the gate fidelity conditioned on a measurement error:
\begin{equation}
    F'=\int d\psi\langle \psi\lvert U_{\mathrm{CZ}}^\dagger Z_1 U_{\mathrm{CZ}}\vert \psi\rangle=0.
\end{equation}
In the presence of measurement errors, the gate fidelity is therefore given by
\begin{equation}
    F(p_{\mathrm{m}})= (1-p_{\mathrm{m}})F(0).
\end{equation}
A finite $p_{\mathrm{m}}$ can therefore be accounted for through a simple rescaling of the $\varepsilon$ axis of Fig.~\ref{fig:gate-fidelity} by a factor of $1-p_{\mathrm{m}}$.

\section{Time-bin qubit erasure}\label{sec:loss-backaction}

Dielectric loss due to two-level systems (TLSs) has been identified as a key limitation in current solid-state quantum-computing platforms~\cite{martinis2005decoherence,wang2015surface,de2018suppression,mcrae2020materials,lisenfeld2023enhancing,crowley2023disentangling,chen2024phonon,liu2024observation,zanuz2024mitigating}. These TLSs may interact with electric fields via a charge dipole and thus limit the lifetimes of superconducting qubits and microwave resonators. In this section, we estimate the size of the errors introduced due to the loss of a time-bin encoded photonic qubit during the ancilla-assisted gate. Photon loss can be heralded by measuring $\mathrm{Q}_3$ as described in Sec.~\ref{gate-steps}. We identify a regime where the loss of the photon encoding the time-bin qubit is approximately backaction-free, leading to a reduction in the occurrence of photon-loss-related errors on stationary qubits. The avenues given here for reducing backaction-induced errors are only relevant to the ancilla-assisted gate: Since the photonic qubit mediating the measurement-free gate of Sec.~\ref{meas-free-gate} travels from $\mathrm{Q}_1$ to $\mathrm{Q}_2$ \textit{and back}, simply heralding the loss of the photon would not provide any information about whether the photon was lost before or after interacting with $\mathrm{Q}_2$. By contrast, in the case of the ancilla-assisted gate, we can reasonably assume that a photon lost in transit is lost before reaching $\mathrm{Q}_2$.

In order to quantify the backaction caused by photon loss during the ancilla-assisted gate, we model the interaction of the time-bin encoded photon with an ensemble of TLSs located along the transmission line. We take the initial state of $\mathrm{Q}_1$, $\mathrm{Q}_2$, the time-bin qubit, and the TLS bath to be of the form $\ket{\Psi}\ket{\mathcal{E}_0}$, where $\ket{\mathcal{E}_0}$ is the initial state of the TLS bath, and where $\ket{\Psi}$ is the state resulting from the mapping of Eq.~\eqref{entangle} applied to an initial state $\ket{\Psi_0}\ket{\mathrm{vac}}$ [Eq.~\eqref{initial-state}]. We can express $\ket{\Psi}$ as a state of the form
\begin{equation}
    \ket{\Psi}=\sum_{\lambda=\mathrm{E,L}}\sqrt{p_\lambda}\ket{\Psi_{\lambda}}\ket{\lambda},
\end{equation}
where $p_{\mathrm{E}}=\lvert \gamma\rvert^2+\lvert \delta\rvert^2$ and $p_{\mathrm{L}}=\lvert \alpha\rvert^2+\lvert \beta\rvert^2$ are the probabilities of emitting the photon in the early (E) and late (L) time bins, respectively, and where $\ket{\Psi_\lambda}=\langle\lambda\vert \Psi\rangle/\sqrt{p_\lambda}$.

For both the early and late time bins ($\lambda=\mathrm{E},\mathrm{L}$), we assume that the photon is lost with probability $q$ while travelling between $\mathrm{Q}_1$ and $\mathrm{Q}_2$: 
\begin{align}
    \ket{\lambda,\mathcal{E}_0}&\rightarrow \sqrt{1-q}\ket{\lambda,\mathcal{E}_0}+\sqrt{q}\ket{\mathrm{vac},\mathcal{E}_{\lambda}}.\label{early-loss}
\end{align}
In Eq.~\eqref{early-loss}, $\ket{\mathcal{E}_\lambda}$ is the state of the environment conditioned on having lost a photon from time-bin $\lambda$.  

As explained in Sec.~\ref{gate-steps}, the loss of the time-bin qubit is heralded by a measurement of $\mathrm{Q}_3$ in state $\ket{f}$. The post-measurement state $\rho_f$ of $\mathrm{Q}_1$ and $\mathrm{Q}_2$, conditioned on such a measurement outcome $\ket{f}$, is obtained by tracing over the state of the environment and is given by
\begin{equation}\label{post-meas-loss}
\rho_{f}=\sum_{\lambda}p_\lambda\ketbra{\Psi_{\lambda}}+\sqrt{p_{\mathrm{E}}p_{\mathrm{L}}}\left(C\ketbra{\Psi_{\mathrm{L}}}{\Psi_{\mathrm{E}}}+\mathrm{h.c.}\right),
\end{equation}
where the coherence factor $C$ depends on the overlap
\begin{equation}\label{overlap-environment}
    C=\langle \mathcal{E}_{\mathrm{E}}\vert\mathcal{E}_{\mathrm{L}}\rangle=\lvert C\rvert e^{i\varphi},\quad \varphi=\mathrm{arg}\:C.
\end{equation}
In the ideal case, $C=1$ and the original two-qubit state $\ketbra{\Psi_0}=\rho_f$ remains pure despite the loss of the photon: $\mathrm{Tr}\:\rho_f^2=2\lvert C\rvert^2-1=1$. In general, however, it may be the case that $C\neq 1$. For $C\neq 1$, the distinguishability of the environmental states $\ket{\mathcal{E}_\lambda}$ [cf.~Eq.~\eqref{overlap-environment}] leads to a loss of coherence in an effective two-dimensional subspace spanned by $\ket{\Psi_\lambda}$. Since $\ket{\Psi_\lambda}$ are product states with $\mathrm{Q}_1$ in state $\ket{e}$ ($\ket{g}$) for $\ket{\Psi_{\mathrm{E}}}$ ($\ket{\Psi_{\mathrm{L}}}$), the state $\rho_f$ with $C\neq 1$ can be expressed as the output of a single-qubit dephasing channel applied to the original state $\ket{\Psi_0}$ [Eq.~\eqref{initial-state}]:
\begin{equation}\label{dephasing-channel}
    e^{-i\frac{\varphi}{2}Z_1}\rho_f e^{i \frac{\varphi}{2}Z_1}=(1-\eta)\ketbra{\Psi_0}+\eta Z_1\ketbra{\Psi_0}Z_1,
\end{equation}
where $\eta=(1-\lvert C\rvert)/2$. In the event that $C$ has a phase ($\varphi\neq 0$) which is known, it can be compensated deterministically through a rotation of $\mathrm{Q}_1$ as shown above. The backaction of photon loss therefore amounts to a phase flip on $\mathrm{Q}_1$ occurring with probability $\eta$. Such a phase flip could be detected and corrected via an error-correcting code prior to re-attempting the failed gate.

To estimate the size of $C$, which sets the strength of the dephasing channel [Eq.~\eqref{dephasing-channel}], we model the absorption of the photon by a dielectric medium. We assume the dielectric is composed of many two-level systems (TLSs) located at positions $x_j$ along the transmission line housing the time-bin qubit, and that these TLSs couple to photons in the waveguide with a linear interaction. We therefore take the total lab-frame Hamiltonian $H$ of the transmission line and TLS environment to be of the form 
\begin{equation}
    H=H_0+H_{\mathrm{int}},
\end{equation}
where
\begin{equation}
    H_0=\sum_k\:\omega_k r_k^\dagger r_k+\sum_j \omega_j \ketbra{\uparrow_j}
\end{equation}
generates the decoupled evolution of the transmission-line modes $r_k$ (having frequency $\omega_k=v\lvert k\rvert$) and TLS environment. Here, the annihilation operators $r_k$ satisfy  $[r_k, r_{k'}^\dagger]=\delta_{k,k'}$, and $\ketbra{\uparrow_j}$ is a projector onto the excited state of TLS $j$ with frequency splitting $\omega_j$. The Hamiltonian $H_{\mathrm{int}}$ describing the interaction between the transmission-line modes and TLS environment is of the form
\begin{equation}\label{interaction}
    H_{\mathrm{int}}=\sqrt{v\tau}\sum_j g_j\left[\psi(x_j)+\psi^\dagger(x_j)\right]\sigma_j^x,
\end{equation}
where $\sigma_j^x=\sigma_j^++\sigma_j^-$ is a Pauli-X operator for TLS $j$ with $\sigma_j^-=\ketbra{\downarrow_j}{\uparrow_j}=(\sigma_j^+)^\dagger$. The operator $\psi(x)$ in Eq.~\eqref{interaction} is a field operator for the transmission line, given by $\psi(x)=L^{-1/2}\sum_k e^{ik x}r_k$ for a transmission line of length $L$, while $\sqrt{v\tau}$ is a prefactor with dimensions of $[\mathrm{length}]^{1/2}$ required by dimensional considerations. For this choice, $g_j$ is the single-photon coupling strength between TLS $j$ and a photon in quasimode $u$ having an effective mode volume $\propto v\tau$. The only essential restrictions on Eq.~\eqref{interaction} are that the coupling is linear in the (electric or magnetic) field of the transmission line and that the interaction between the field and TLS is short-ranged and local at each site $x_j$. The form given in Eq.~\eqref{interaction} is consistent, e.g., with electric-dipole coupling, in which case $g_j$ would be proportional to the transition dipole of TLS $j$, but this form could equally well describe coupling between the magnetic field in the transmission line and the magnetic dipoles of environmental spins. In Eq.~\eqref{interaction}, we have neglected longitudinal coupling terms $\sigma_j^z$, which are also generically present but average out in a rotating-wave approximation~\cite{mcintyre2024photonic}.                                                  

In an interaction picture defined with respect to $H_0$, $\tilde{H}(t)=U_0H U_0^\dagger-iU_0^\dagger \dot{U}_0=\tilde{H}_{\mathrm{int}}(t)+\mathrm{counter {\text -}rot.}$ with
\begin{equation}\label{interaction-rwa}
    \tilde{H}_{\mathrm{int}}(t)=\sqrt{\frac{v\tau}{L}}\sum_{j,k}g_j\left[e^{i(kx_j-\omega_k t)}e^{i\omega_j t}r_k\sigma_j^++\mathrm{h.c.}\right],
\end{equation}
where we have neglected counter-rotating terms in a rotating-wave approximation valid for $\lvert g_j\rvert\ll \lvert \omega_j\rvert$.  For the purpose of the calculation presented in this section, it will be helpful to introduce an orthonormal set $\{u'\}$ of quasimodes associated with bosonic operators $r_{u'}$ satisfying the usual commutation relation $[r_{u'},r_{u''}^\dagger]=\delta_{u',u''}$. For $u'=u$, the operator $r_{u}$ annihilates the quasimode $u$ introduced in Eq.~\eqref{creation-operator-quasimode}. We can write $\psi(x)$ in terms of these operators as $\psi(x)=\sum_{u'} \phi_{u'}(x)r_{u'}$, where $\phi_{u'}(x)=L^{-1/2}\sum_k e^{ikx}\phi_{u'k}$ for $\phi_{u'k}=\langle k \vert u'\rangle$. Using the relation
\begin{equation}
    r_k=\sum_{u'}\phi_{u'k}r_{u'},
\end{equation}
we can then rewrite Eq.~\eqref{interaction-rwa} as 
\begin{equation}\label{interaction-rwa-2}
    \tilde{H}_{\mathrm{int}}(t)=\sqrt{v\tau}\sum_{j,u'}g_j\left[\phi_{u'}(x_j-v t)e^{i\omega_jt}r_{u'}\sigma_j^++\mathrm{h.c.}\right],
\end{equation}
where we have used the fact that $\phi_{u'}(x_j-vt)=L^{-1/2}\sum_k e^{i(kx_j-\omega_k t)}\phi_{u'k}$ with $\omega_k=v |k|$.

To calculate the coherence factor $C$, we consider the evolution of the initial state $\ket{1_u,\mathcal{E}_0}$, where $\ket{1_u}=r_u^\dagger\ket{\mathrm{vac}}$ [cf.~Eq.~\eqref{creation-operator-quasimode}] and $\ket{\mathcal{E}_0}=\ket{\downarrow \downarrow\cdots\downarrow}$. When the evolution is dominated by a single action of $\tilde{H}_\mathrm{int}(t)$ (Born approximation) the state will approximately take the form
\begin{equation}\label{wavefunctionansatz}
    \ket{1_u(t)}\simeq \alpha_0(t)\ket{1_u,\mathcal{E}_0}+\sum_j \alpha_j(t)\ket{\mathrm{vac},j},
\end{equation}
where $\alpha_j(t)$ are coefficients and the state $\ket{j}$ is given by $\ket{j}=\sigma_j^+\ket{\mathcal{E}_0}=\ket{\downarrow\cdots\uparrow_j\cdots\downarrow}$. Knowing how a photon decays out of a specific quasimode $u$ will allow us to calculate the post-measurement states $\ket{\mathcal{E}_\lambda}$ of the TLS bath conditioned on a measurement of $\mathrm{Q}_3$ in state $\ket{f}$. Recall that the state $\ket{\mathcal{E}_\lambda}$ is obtained due to decay out of the quasimode $u=u_\lambda$ defining the time-bin qubit [cf.~Eq.~\eqref{early-loss}]. 

We calculate the leading-order transition amplitude $\alpha_j(t)$ associated with the loss of a photon initially in quasimode $u$ due to absorption by TLS $j$, $\alpha_j(t) = \bra{\mathrm{vac},j}\tilde{U}(t)\ket{1_u,\mathcal{E}_0}$, where
\begin{equation}
    \tilde{U}(t)=\mathcal{T}e^{-i\int_{-\infty}^t dt' \tilde{H}_\mathrm{int}(t')}\simeq 1-i\int_{-\infty}^t dt' \tilde{H}_\mathrm{int}(t').
\end{equation}
This gives 
\begin{equation}\label{alphaj}
    \alpha_j(t)\simeq -i \sqrt{v\tau}g_j\int_{-\infty}^t dt'\:\phi_u(x_j-vt')e^{i\omega_j t'}.
\end{equation}
Since the wavepacket $\phi_u(x-vt)$ propagates without dispersion, the quasimode $u$ can equivalently be described by a function of time at a single point in space, given by the waveform $u(t)=\sqrt{v}\phi_u(-vt)e^{i\omega_0 t}$, where the factor $\sqrt{v}$ is required to satisfy the normalization condition $\int dx\:\lvert\phi_u(x)\rvert^2=\int dt\:\lvert u(t)\rvert^2=1$, and where the phase factor $e^{i\omega_0 t}$ results from the relationship between the lab-frame wavepacket $\phi_u(x-vt)$ and an interaction-picture waveform $u(t)$. Inserting this relationship into Eq.~\eqref{alphaj} gives
\begin{equation}\label{alphaj-coefficients}
    \alpha_j(t)\simeq -ig_j\sqrt{\tau}\int_{-\infty}^t dt'\:u(t'-x_j/v)e^{i\delta\omega_j t'},
\end{equation}
where we have introduced the detuning $\delta\omega_j$ between TLS $j$ and the central frequency $\omega_0$ of the photon:
\begin{equation}
    \delta\omega_j=\omega_j-\omega_0.
\end{equation}
For times satisfying $t-x_j/v\gg \tau$, the integral in Eq.~\eqref{alphaj-coefficients} no longer picks up any weight since the propagating single-photon pulse is no longer interacting with the TLS at position $x_j$. We can therefore extend the upper integration limit in Eq.~\eqref{alphaj-coefficients} to $+\infty$, yielding the steady-state amplitudes
\begin{align}\label{tls-saturation}
    \alpha_j(t)\simeq \bar{\alpha}_j = -i g_j\sqrt{\tau}\:e^{i\delta\omega_j x_j/v}u(\delta\omega_j).
\end{align}
Note that $u(\omega)$ has dimensions of $[\mathrm{time}]^{1/2}$, following from the normalization condition $(2\pi)^{-1}\int d\omega\:\lvert u(\omega)\rvert^2=1$. The quantity $\lvert \bar{\alpha}_j\rvert^2$ gives the excited-state population of TLS $j$ resulting from its interaction with a photon in quasimode $u$. Rather intuitively, this population is proportional to the spectral weight $\propto |u(\delta\omega_j)|^2$ of the photon at the detuning $\delta\omega_j$ of TLS $j$ from the central frequency of the photon.

To relate this result back to the map of Eq.~\eqref{early-loss}, we substitute Eq.~\eqref{tls-saturation} into  Eq.~\eqref{wavefunctionansatz}  for times $t$ satisfying $t-x_j/v\gg \tau$, giving 
\begin{align}\label{mapping}
    \ket{1_{u_\lambda}(t)}\simeq \sqrt{1-q}\ket{\lambda,\mathcal{E}_0}+\sqrt{q}\ket{\mathrm{vac}}\sum_j \frac{\bar{\alpha}_{j,\lambda}}{\sqrt{q}}\ket{j},
\end{align}
where $\bar{\alpha}_{j,\lambda}$ is given by Eq.~\eqref{tls-saturation} with $u(\omega)=u_\lambda(\omega)$. Under the assumption that $u_{\mathrm{L}}(t-\tau_{\mathrm{sep}})=u_{\mathrm{E}}(t)\equiv u(t)$, we have the simple relation $u_{\mathrm{L}}(\omega)=e^{-i\omega \tau_{\mathrm{sep}}}u(\omega)$ between the frequency-domain waveform of a photon occupying $\ket{\mathrm{E}}$ or $\ket{\mathrm{L}}$. In terms of quantities characterizing the quasimode and TLS's, the probability $q$ of losing the photon is given by
\begin{align}
    q=\sum_j \lvert \bar{\alpha}_j\rvert^2= \tau\int \frac{d\omega}{2\pi} J(\omega)\lvert u(\omega)\rvert^2,
\end{align}
where $J(\omega)=2\pi\sum_j\lvert g_j\rvert^2\delta(\omega-\delta\omega_j)$ is the spectral density of the TLS bath. 
In terms of the amplitudes $\bar{\alpha}_{j,\lambda}$, we can therefore write the environmental states $\ket{\mathcal{E}_\lambda}$ conditioned on having lost a photon from time bin $\lambda=\mathrm{E},\mathrm{L}$  as
\begin{equation}\label{bin-state}
    \ket{\mathcal{E}_{\lambda}}=\frac{1}{\sqrt{q}}\sum_j \bar{\alpha}_{j,\lambda}\ket{j}.
\end{equation}
Using Eq.~\eqref{bin-state}, we can then re-express the coherence factor $C$ as 
\begin{equation}\label{coherence}
    C=\frac{\int d\omega\: J(\omega)\lvert u(\omega)\rvert^2 e^{-i\omega \tau_{\mathrm{sep}}}}{\int d\omega \:J(\omega)\lvert u(\omega)\rvert^2}.
\end{equation} 
As expected, the loss of qubit coherence ($C\neq 1$) results from the temporal separation $\tau_{\mathrm{sep}}$ of the quasimodes defining the early and late time-bin states.  However, it is interesting to note that the orthogonality of the time bin states $\ket{\mathrm{E}}$ and $\ket{\mathrm{L}}$ does not imply orthogonality of the environmental states $\ket{\mathcal{E}_\mathrm{E}},\ket{\mathcal{E}_\mathrm{L}}$. This in turn implies that there are conditions under which photon loss can have zero backaction on the stationary qubits. In other words, for an appropriate environment spectral density, time-bin waveform, and time-bin separation, the time-bin information can be erased with the loss of the time-bin photon. 

One regime where photon loss \textit{does} lead to significant backaction is the regime where the spectral density $J(\omega)$ is approximately flat over the support of a Gaussian $u(\omega)$ having standard deviation $\tau$, for which
\begin{equation}\label{eq:u_omega_gaussian}
    \lvert u(\omega)\rvert^2=2\sqrt{\pi}\tau e^{-\tau^2\omega^2}.
\end{equation}
In that case, setting $J(\omega)=J_0$ in Eq.~\eqref{coherence} gives 
\begin{equation}\label{coherence-wide-bandwidth}
    C=e^{-\left(\frac{\tau_{\mathrm{sep}}}{2\tau}\right)^2},\quad J(\omega)=J_0,
\end{equation}
which in turn gives $\eta\approx 1/2$ [cf.~Eq.~\eqref{dephasing-channel}] since $\tau_{\mathrm{sep}}>\tau$ is needed to ensure orthogonality of the time-bin states. The loss of a photon into an environment having a bandwidth that is wide compared to $\tau_{\mathrm{sep}}^{-1}$ therefore gives rise to significant dephasing of the stationary qubits [Eq.~\eqref{dephasing-channel}]. 

The form of $C$ [Eq.~\eqref{coherence}] suggests that if $J(\omega)$ is instead narrow in frequency compared to the photon bandwidth, then it --- rather than $u(\omega)$ --- will determine the qualitative features of $C$. For simplicity, we consider a Gaussian spectral density, 
\begin{equation}\label{gaussian-spectral-density}
    J(\omega)= \sqrt{2\pi}\frac{g^2}{\Lambda}e^{-\frac{(\omega-\delta\bar{\omega})^2}{2\Lambda^2}},
\end{equation}
where $\Lambda$ sets the bandwidth of the environment, $\delta\bar{\omega}$ gives the average of $\delta\omega_j=\omega_j-\omega_0$, and where $g^2=\sum_j\lvert g_j\rvert^2$. For the same Gaussian $u(\omega)$ as in Eq.~\eqref{eq:u_omega_gaussian}, this choice of $J(\omega)$ [Eq.~\eqref{gaussian-spectral-density}] gives
\begin{align}
    &q=2\sqrt{\pi}(g\tau)^2\frac{e^{-\frac{(\delta\bar{\omega}\tau)^2}{1+2\Lambda^2\tau^2}}}{\sqrt{1+2\Lambda^2\tau^2}},\\
    &C=e^{-i\delta\bar{\omega}\tau_{\mathrm{sep}}}e^{-\frac{(\Lambda\tau_{\mathrm{sep}})^2}{2+4\Lambda^2\tau^2}},
\end{align}
where the phase $\varphi=\mathrm{arg}\:C$ depending on $\delta\bar{\omega}$ could be corrected via a single-qubit rotation provided $\delta\bar{\omega}$ is known [cf.~Eq.~\eqref{dephasing-channel}]. As expected, the photon-loss probability $q$ decreases as the average $\bar{\omega}$ of the TLS frequency distribution moves further off resonance from the central frequency $\omega_0$ of the photon. Notably, however, there is no corresponding decrease in $C$ with $\delta\bar{\omega}$ since $C$ is post-selected on having already lost the photon.  In the wide-bandwidth regime $\Lambda\gg \tau_{\mathrm{sep}}^{-1}$, we recover the exponential suppression of $C$ in the ratio of timescales $(\tau_{\mathrm{sep}}/\tau)^2\gtrsim 1$ already given in Eq.~\eqref{coherence-wide-bandwidth}.  In the opposite, narrow-bandwidth regime $\Lambda\ll\tau_{\mathrm{sep}}^{-1}$, we instead find that 
\begin{equation}
    \lvert C\rvert=1-\frac{1}{2}(\Lambda\tau_{\mathrm{sep}})^2+O((\Lambda\tau_{\mathrm{sep}})^4),
\end{equation}
yielding dephasing errors occurring with probability $\eta\approx (\Lambda\tau_{\mathrm{sep}})^2/4\ll 1$. In this regime, the backaction due to photon loss, as quantified by the rate of dephasing errors, is strongly suppressed relative to the wide-bandwidth (Markovian) case. This is a consequence of the TLS bath having a correlation time $\tau_{\mathrm{c}}\sim \Lambda^{-1}$ that is long compared to the time-bin separation $\tau_{\mathrm{sep}}$.

For a time-bin separation of $\tau_{\mathrm{sep}}\approx 100$ ns, consistent with the optimal values of $\tau=\tau_{\mathrm{opt}}$ shown in Fig.~\ref{fig:gate-fidelity}(c), we have $\Lambda\tau_{\mathrm{sep}}=1$ for $\Lambda/(2\pi)\approx 1.6$ MHz, corresponding to a bath correlation time of $\tau_{\mathrm{c}}\approx 0.1$ $\mu$s. For this choice of $\tau_{\mathrm{sep}}$, a correlation time $\gg0.1$ $\mu$s would therefore be required to ensure that the loss of the time-bin qubit is approximately backaction-free. Realistic values for the correlation time will depend on the nature of the coupling $g_j$ between the quasimode field $\psi(x)$ and the environmental TLSs, the amount of inhomogeneous broadening experienced by the TLSs, and the lifetime of the TLSs. Charge-like two-level systems with lifetimes on the order of milliseconds have been measured~\cite{liu2024observation,chen2024phonon}. Although the spectral characteristics of charge-like TLSs would likely be difficult to control, naturally occurring narrow-band mechanisms associated with, e.g., well-defined transitions between hyperfine-split spin sublevels~\cite{de2018suppression} in surface hydrogen may eventually become the limiting source of photon loss as materials-fabrication techniques continue to yield improvements in the densities of charge-like TLSs in materials commonly used for circuit QED. Given a certain dielectric material, a simple measurement exists to determine whether it enables backaction-free loss of time-bin encoded photons: After entangling a time-bin qubit with a stationary qubit according to Eq.~\eqref{entangle}, the photon is sent directly to a single-photon detector. If no click is registered (signaling the loss of the photon), then the qubit is measured in the $X$ or $Y$ basis to yield information about the real and imaginary parts of $C=\langle \mathcal{E}_{\mathrm{E}}\vert\mathcal{E}_{\mathrm{L}}\rangle$.

\section{Conclusion}\label{sec:conclusion}
In this work, we have presented two possible strategies for realizing entangling gates between qubits connected by meter-scale photonic interconnects. Such interconnects, equipped with a two-qubit entangling gate protocol, would provide the quantum links required for so-called $l$-type modular architectures~\cite{bravyi2022future}. Both CZ gate protocols presented here are compatible with a transmission-line free spectral range of zero and could be used to implement gates over arbitrarily long distances in principle. The gates are deterministic with heralded photon loss, and notably, they do not rely on the use of a polarization degree-of-freedom for the photonic qubit encoding. This ensures compatibility with the coplanar waveguide resonators commonly used in circuit QED architectures. The measurement-free gate requires that the distance between the two qubits exceed the size of the photonic wavepacket and does not require measurement of the photonic qubit. The ancilla-assisted gate lifts this requirement on the minimum qubit separation at the cost of introducing a measurement of the photonic (time-bin) qubit, which could be realized straightforwardly with the help of an ancilla.

We have also analyzed the dephasing channel associated with the loss of a time-bin qubit, which cannot be treated using existing loss models since the details of the spatiotemporal modes defining the time-bin qubit must be taken into account. We show that in spite of the orthogonality of the time-bin states, there are conditions under which the loss of the photon causes only minimal dephasing of the stationary qubits. Whether or not these conditions are met depends on the bandwidth (and consequently, correlation time) of the dielectric environment that absorbs the photon, relative to the temporal separation of the time-bin states. This result highlights the role that materials fabrication could potentially play in reducing errors on stationary qubits connected by photonic time-bin qubits. An investigation of whether current dielectric materials could be made to satisfy the requirements for backaction-free loss of time-bin encoded photons would provide an interesting line of future inquiry, beyond the scope of the present work.

\begin{acknowledgments}
We acknowledge funding from the Natural Sciences and Engineering Research Council of Canada (NSERC) and from the Fonds de Recherche du Qu\'ebec--Nature et technologies (FRQNT). 
\end{acknowledgments}

\appendix*
\section{Master-equation simulation}
In this appendix, we provide details concerning the master-equation simulation of the gate infidelity $\varepsilon$ [Eq.~\eqref{gate-fidelity}] presented in Sec.~\ref{sec:gate-simulation}. For simplicity, we implement the simulation using Gaussian time-bin waveforms having a width set by $\tau$:
\begin{equation}\label{gaussian-waveform}
    u(t,t_{\mathrm{peak}})=\frac{e^{-\frac{(t-t_{\mathrm{peak}})^2}{2\tau^2}}}{\pi^{1/4}\sqrt{\tau}}.
\end{equation}
We also take the total duration of the gate to be fixed at $T=16\tau$.

Given some initial state $\ket{\psi}$ of $\mathrm{Q}_1$ and $\mathrm{Q}_2$, we evaluate the joint final state $\rho_\psi(T)$ of $\mathrm{Q}_1$, $\mathrm{Q}_2$, and $\mathrm{Q}_3$ (prepared in the ground state) by solving the master equation given in Eq.~\eqref{master-eq-simul}. The formal solution of Eq.~\eqref{master-eq-simul} can be written as
\begin{equation} \label{evolved-state}
    \rho_\psi(T)=\underbrace{\mathcal{T}e^{-i\int_0^T dt\:\mathcal{L}_{\mathrm{tot}}(t)}}_{\equiv \Lambda_{\mathrm{tot}}(T)}\rho_{\psi}(0),
\end{equation}
where $\rho_\psi(0)=\pi_{fg,1}\ketbra{\psi}\pi_{fg,1}\otimes \ketbra{g}$, $\mathcal{T}$ is a time-ordering operator, and $\pi_{ab,i}$ is a $\pi$ pulse between levels $\ket{a}$ and $\ket{b}$ of qubit $\mathrm{Q}_i$. Concretely, the simulation proceeds by evaluating $\Lambda_{\mathrm{tot}}(T)$ as
\begin{equation}
\Lambda_{\mathrm{tot}}(T)=\Lambda(T,T/2)L_\pi\Lambda(T/2,0).
\end{equation}
Here, $L_\pi$ describes the action of (ideal) $\pi_{fe,1}$, $\pi_{eg,1}$, and $\pi_{fe,3}$ pulses applied to $\mathrm{Q}_1$ and $\mathrm{Q}_3$ as part of the time-bin photon emission and absorption processes. The superoperator $\Lambda$ describing evolution before and after the $\pi$ pulses is given by
\begin{equation}
    \Lambda(t_f,t_i)=\mathcal{T}e^{-i\int_{t_i}^{t_f}dt\:\mathcal{L}(t)},
\end{equation}
where $\mathcal{L}(t)$ generates evolution according to 
\begin{equation}
    \dot{\rho}=-i[H_{\mathrm{tot}}(t), \rho]+\sum_{j=0}\mathcal{D}[L_j]\rho\equiv -i\mathcal{L}\rho.
\end{equation}
Here, the damping superoperator $\mathcal{D}[L]$ acts according to $\mathcal{D}[L]\rho=L\rho L^\dagger-\frac{1}{2}\{L^\dagger L, \rho\}$, and the sum over $L_j$ describes the contribution of various dissipative processes (explained in the main text). We model the directed routing of photons via the SLH formalism~\cite{combes2017slh}, which eliminates the need to explicitly simulate the field propagating between cavities. This simplification can be justified provided the propagation time is short compared to the timescale set by the cavity linewidth $\kappa_i$~\cite{combes2017slh}. Following the SLH approach, we therefore include the dissipator
\begin{align}
L_0=\sum_{j=1}^3\sqrt{\kappa_j}a_j
\end{align}  
in the sum over $\mathcal{D}[L_j]$, together with the following bilinear terms $\propto \sqrt{\kappa_i\kappa_j}$ in the Hamiltonian $H(t)$ governing the coherent evolution of the qubits and cavities:
\begin{equation}
    H_{\mathrm{tot}}(t)=\sum_{j=1}^3 H_{j}(t)+ \frac{i}{2}\sum_{\substack{i=1\\j>i}}(\sqrt{\kappa_i\kappa_j}a_i^\dagger a_j-\mathrm{h.c.}).\label{H-early}
\end{equation}
The particular combination of coherent terms $\propto \sqrt{\kappa_i\kappa_j}$ and dissipator $\mathcal{D}[L_0]$ ensures a master-equation description that is consistent with the unidirectional propagation of photons~\cite{combes2017slh}. We have also assumed in writing Eq.~\eqref{H-early} that all cavities are resonant.

In addition to the terms $\propto \sqrt{\kappa_i\kappa_j}$, the Hamiltonian $H(t)$ also includes terms $H_i(t)$ describing cavity-qubit coupling for qubits $\mathrm{Q}_i$ ($i=1,2,3$). The Hamiltonians $H_1(t)$ and $H_3(t)$ describe the emission (by $\mathrm{Q}_1$) and subsequent re-absorption (by $\mathrm{Q}_3$) of a photon with waveform $u(t,t_{\mathrm{peak}})$, conditioned on $\mathrm{Q}_1$ being in the $\ket{f}$ state: 
\begin{equation}
    H_{j}(t)=i\Omega_{j}(t)\left(\tau_{fg,j}a_j^\dagger-\mathrm{h.c.}\right),\quad j=1,3.
\end{equation}
Here, $\tau_{ab,j}$ denotes the pseudospin operator $\tau_{ab}=\ketbra{a}{b}$ for qubit $\mathrm{Q}_{j}$, and $a_j$ is a bosonic annihilation operator that removes one photon from the cavity mode coupled to $\mathrm{Q}_j$. The drives $\Omega_j$ are defined piecewise according to
\begin{equation}
    \Omega_j(t)=\sum_{\alpha=\mathrm{E,L}}\Omega_{\mathrm{\alpha},j}(t)\Pi_{\mathrm{\alpha}}(t),
\end{equation}
where $\Pi_{\mathrm{E}}(t)=\Theta(t)\Theta(T/2-t)$ is equal to 1 for $t\in (0,T/2)$ and 0 otherwise. Similarly, $\Pi_{\mathrm{L}}(t)=\Theta(t-T/2)\Theta(T-t)$ is equal to 1 for $t\in(T/2,T)$ and 0 otherwise. The drive $\Omega_{\alpha,1}(t)$ acting on $\mathrm{Q}_1$ is given by Eq.~\eqref{pulse-relation-2} with $u(t)=u_{\alpha}(t)$, while the drive $\Omega_{\alpha,3}(t)$ acting on $\mathrm{Q}_3$ is given by Eq.~\eqref{absorption} with $u(t)=u_{\alpha}(t)$. In both cases, $u_{\mathrm{E}}(t)$ and $u_{\mathrm{L}}(t)$ are given by Eq.~\eqref{gaussian-waveform} with $t_{\mathrm{peak}}=T/4$ and $t_{\mathrm{peak}}=3T/4$, respectively. Finally, the Hamiltonian $H(t)$ includes a term describing the dispersive coupling of $\mathrm{Q}_2$ to its cavity, given by
\begin{equation}\label{disp}
    H_{2}(t)= \chi(t) \left(\tau_{ee,2}-\tau_{gg,2}\right)a_2^\dagger a_2,
\end{equation}
where $\chi(t)$ is given by Eq.~\eqref{dispersive-coupling-time-dependent}. 

Given $\rho_\psi(T)$, we apply an $S$ gate to $\mathrm{Q}_1$ and measure $\mathrm{Q}_3$ in the $X$ basis (Fig.~\ref{fig:gate-circuit}b). The $S$ gate ensures that in the absence of errors, the final state of $\mathrm{Q}_1$ and $\mathrm{Q}_2$ is given by $U_{\mathrm{CZ}}\ket{\psi}$ with $U_{\mathrm{CZ}}=e^{i\pi\ketbra{gg}}$. The need for such an $S$ gate is a consequence of the mechanism [Eq.~\eqref{disp}] used to realize an entangling gate between the time-bin qubit and $\mathrm{Q}_2$. As discussed in Sec.~\ref{entangle-time-bin}, the mechanism of Eq.~\eqref{disp} produces an operation equal to the product of $U=e^{i\pi\ketbra{\mathrm{L},g}}$ [Eq.~\eqref{local-cz}] and an $S^\dagger$ gate on the time-bin qubit. Since the state of $\mathrm{Q}_1$ is correlated with that of the time-bin qubit via the mapping of Eq.~\eqref{entangle}, the $S^\dagger$ gate on the time-bin qubit can be canceled by applying an $S$ gate to $\mathrm{Q}_1$. Such a gate would not be required in practice (it can be tracked in software) but is included in Eq.~\eqref{post-measurement-state}, below, to allow for comparison to an ideal operation of the form $U_{\mathrm{CZ}}=e^{i\pi\ketbra{gg}}$.

The post-measurement state $\rho_{\psi,\pm}(T)$ of $\mathrm{Q}_1$ and $\mathrm{Q}_2$, conditioned on a measurement of $\mathrm{Q}_3$ in state $\ket{\pm}\propto\ket{e}\pm\ket{g}$, is then given by 
\begin{equation}\label{post-measurement-state}
    \rho_{\psi,\pm} = \frac{\mathrm{Tr}_3\{\ketbra{\pm}S \rho_\psi(T) S^\dagger\}}{P_\pm},
\end{equation}
where $\mathrm{Tr}_3$ denotes a partial trace over the state of $\mathrm{Q}_3$ and $P_\pm=\mathrm{Tr}\{\ketbra{\pm} \rho_\psi(T)\}$ gives the probability of obtaining a measurement outcome $\ket{\pm}$. Conditioned on a measurement of $\mathrm{Q}_3$ in state $\ket{-}$, a $Z$ gate is applied to $\mathrm{Q}_1$ [Fig.~\ref{fig:gate-circuit}(b)]. Both the measurement and the $Z$ gate are treated as ideal. This procedure yields the state $\mathcal{M}(\ketbra{\psi})$ entering the expression for $F$ [Eq.~\eqref{gate-fidelity}]:
\begin{equation}\label{channel}
    \mathcal{M}(\ketbra{\psi})=\rho_{\psi,+}=Z_1\rho_{\psi,-}Z_1.
\end{equation}

\providecommand{\noopsort}[1]{}\providecommand{\singleletter}[1]{#1}%


\begin{thebibliography}{84}%
\makeatletter
\providecommand \@ifxundefined [1]{%
 \@ifx{#1\undefined}
}%
\providecommand \@ifnum [1]{%
 \ifnum #1\expandafter \@firstoftwo
 \else \expandafter \@secondoftwo
 \fi
}%
\providecommand \@ifx [1]{%
 \ifx #1\expandafter \@firstoftwo
 \else \expandafter \@secondoftwo
 \fi
}%
\providecommand \natexlab [1]{#1}%
\providecommand \enquote  [1]{``#1''}%
\providecommand \bibnamefont  [1]{#1}%
\providecommand \bibfnamefont [1]{#1}%
\providecommand \citenamefont [1]{#1}%
\providecommand \href@noop [0]{\@secondoftwo}%
\providecommand \href [0]{\begingroup \@sanitize@url \@href}%
\providecommand \@href[1]{\@@startlink{#1}\@@href}%
\providecommand \@@href[1]{\endgroup#1\@@endlink}%
\providecommand \@sanitize@url [0]{\catcode `\\12\catcode `\$12\catcode
  `\&12\catcode `\#12\catcode `\^12\catcode `\_12\catcode `\%12\relax}%
\providecommand \@@startlink[1]{}%
\providecommand \@@endlink[0]{}%
\providecommand \url  [0]{\begingroup\@sanitize@url \@url }%
\providecommand \@url [1]{\endgroup\@href {#1}{\urlprefix }}%
\providecommand \urlprefix  [0]{URL }%
\providecommand \Eprint [0]{\href }%
\providecommand \doibase [0]{https://doi.org/}%
\providecommand \selectlanguage [0]{\@gobble}%
\providecommand \bibinfo  [0]{\@secondoftwo}%
\providecommand \bibfield  [0]{\@secondoftwo}%
\providecommand \translation [1]{[#1]}%
\providecommand \BibitemOpen [0]{}%
\providecommand \bibitemStop [0]{}%
\providecommand \bibitemNoStop [0]{.\EOS\space}%
\providecommand \EOS [0]{\spacefactor3000\relax}%
\providecommand \BibitemShut  [1]{\csname bibitem#1\endcsname}%
\let\auto@bib@innerbib\@empty
\bibitem [{\citenamefont {Blais}\ \emph {et~al.}(2021)\citenamefont {Blais},
  \citenamefont {Grimsmo}, \citenamefont {Girvin},\ and\ \citenamefont
  {Wallraff}}]{blais2021circuit}%
  \BibitemOpen
  \bibfield  {author} {\bibinfo {author} {\bibfnamefont {A.}~\bibnamefont
  {Blais}}, \bibinfo {author} {\bibfnamefont {A.~L.}\ \bibnamefont {Grimsmo}},
  \bibinfo {author} {\bibfnamefont {S.~M.}\ \bibnamefont {Girvin}},\ and\
  \bibinfo {author} {\bibfnamefont {A.}~\bibnamefont {Wallraff}},\ }\bibfield
  {title} {\bibinfo {title} {Circuit quantum electrodynamics},\ }\href@noop {}
  {\bibfield  {journal} {\bibinfo  {journal} {Rev.~Mod.~Phys.}\ }\textbf
  {\bibinfo {volume} {93}},\ \bibinfo {pages} {025005} (\bibinfo {year}
  {2021})}\BibitemShut {NoStop}%
\bibitem [{\citenamefont {Burkard}\ \emph {et~al.}(2023)\citenamefont
  {Burkard}, \citenamefont {Ladd}, \citenamefont {Pan}, \citenamefont
  {Nichol},\ and\ \citenamefont {Petta}}]{burkard2023semiconductor}%
  \BibitemOpen
  \bibfield  {author} {\bibinfo {author} {\bibfnamefont {G.}~\bibnamefont
  {Burkard}}, \bibinfo {author} {\bibfnamefont {T.~D.}\ \bibnamefont {Ladd}},
  \bibinfo {author} {\bibfnamefont {A.}~\bibnamefont {Pan}}, \bibinfo {author}
  {\bibfnamefont {J.~M.}\ \bibnamefont {Nichol}},\ and\ \bibinfo {author}
  {\bibfnamefont {J.~R.}\ \bibnamefont {Petta}},\ }\bibfield  {title} {\bibinfo
  {title} {Semiconductor spin qubits},\ }\href@noop {} {\bibfield  {journal}
  {\bibinfo  {journal} {Rev.~Mod.~Phys.}\ }\textbf {\bibinfo {volume} {95}},\
  \bibinfo {pages} {025003} (\bibinfo {year} {2023})}\BibitemShut {NoStop}%
\bibitem [{\citenamefont {Bravyi}\ \emph {et~al.}(2022)\citenamefont {Bravyi},
  \citenamefont {Dial}, \citenamefont {Gambetta}, \citenamefont {Gil},\ and\
  \citenamefont {Nazario}}]{bravyi2022future}%
  \BibitemOpen
  \bibfield  {author} {\bibinfo {author} {\bibfnamefont {S.}~\bibnamefont
  {Bravyi}}, \bibinfo {author} {\bibfnamefont {O.}~\bibnamefont {Dial}},
  \bibinfo {author} {\bibfnamefont {J.~M.}\ \bibnamefont {Gambetta}}, \bibinfo
  {author} {\bibfnamefont {D.}~\bibnamefont {Gil}},\ and\ \bibinfo {author}
  {\bibfnamefont {Z.}~\bibnamefont {Nazario}},\ }\bibfield  {title} {\bibinfo
  {title} {The future of quantum computing with superconducting qubits},\
  }\href@noop {} {\bibfield  {journal} {\bibinfo  {journal} {J.~Appl.~Phys.}\
  }\textbf {\bibinfo {volume} {132}} (\bibinfo {year} {2022})}\BibitemShut
  {NoStop}%
\bibitem [{\citenamefont {Bravyi}\ \emph {et~al.}(2016)\citenamefont {Bravyi},
  \citenamefont {Smith},\ and\ \citenamefont {Smolin}}]{bravyi2016trading}%
  \BibitemOpen
  \bibfield  {author} {\bibinfo {author} {\bibfnamefont {S.}~\bibnamefont
  {Bravyi}}, \bibinfo {author} {\bibfnamefont {G.}~\bibnamefont {Smith}},\ and\
  \bibinfo {author} {\bibfnamefont {J.~A.}\ \bibnamefont {Smolin}},\ }\bibfield
   {title} {\bibinfo {title} {Trading classical and quantum computational
  resources},\ }\href@noop {} {\bibfield  {journal} {\bibinfo  {journal}
  {Phys.~Rev.~X}\ }\textbf {\bibinfo {volume} {6}},\ \bibinfo {pages} {021043}
  (\bibinfo {year} {2016})}\BibitemShut {NoStop}%
\bibitem [{\citenamefont {Peng}\ \emph {et~al.}(2020)\citenamefont {Peng},
  \citenamefont {Harrow}, \citenamefont {Ozols},\ and\ \citenamefont
  {Wu}}]{peng2020simulating}%
  \BibitemOpen
  \bibfield  {author} {\bibinfo {author} {\bibfnamefont {T.}~\bibnamefont
  {Peng}}, \bibinfo {author} {\bibfnamefont {A.~W.}\ \bibnamefont {Harrow}},
  \bibinfo {author} {\bibfnamefont {M.}~\bibnamefont {Ozols}},\ and\ \bibinfo
  {author} {\bibfnamefont {X.}~\bibnamefont {Wu}},\ }\bibfield  {title}
  {\bibinfo {title} {Simulating large quantum circuits on a small quantum
  computer},\ }\href@noop {} {\bibfield  {journal} {\bibinfo  {journal}
  {Phys.~Rev.~Lett.}\ }\textbf {\bibinfo {volume} {125}},\ \bibinfo {pages}
  {150504} (\bibinfo {year} {2020})}\BibitemShut {NoStop}%
\bibitem [{\citenamefont {Eddins}\ \emph {et~al.}(2022)\citenamefont {Eddins},
  \citenamefont {Motta}, \citenamefont {Gujarati}, \citenamefont {Bravyi},
  \citenamefont {Mezzacapo}, \citenamefont {Hadfield},\ and\ \citenamefont
  {Sheldon}}]{eddins2022doubling}%
  \BibitemOpen
  \bibfield  {author} {\bibinfo {author} {\bibfnamefont {A.}~\bibnamefont
  {Eddins}}, \bibinfo {author} {\bibfnamefont {M.}~\bibnamefont {Motta}},
  \bibinfo {author} {\bibfnamefont {T.~P.}\ \bibnamefont {Gujarati}}, \bibinfo
  {author} {\bibfnamefont {S.}~\bibnamefont {Bravyi}}, \bibinfo {author}
  {\bibfnamefont {A.}~\bibnamefont {Mezzacapo}}, \bibinfo {author}
  {\bibfnamefont {C.}~\bibnamefont {Hadfield}},\ and\ \bibinfo {author}
  {\bibfnamefont {S.}~\bibnamefont {Sheldon}},\ }\bibfield  {title} {\bibinfo
  {title} {Doubling the size of quantum simulators by entanglement forging},\
  }\href@noop {} {\bibfield  {journal} {\bibinfo  {journal} {PRX Quantum}\
  }\textbf {\bibinfo {volume} {3}},\ \bibinfo {pages} {010309} (\bibinfo {year}
  {2022})}\BibitemShut {NoStop}%
\bibitem [{\citenamefont {Mitarai}\ and\ \citenamefont
  {Fujii}(2021)}]{mitarai2021constructing}%
  \BibitemOpen
  \bibfield  {author} {\bibinfo {author} {\bibfnamefont {K.}~\bibnamefont
  {Mitarai}}\ and\ \bibinfo {author} {\bibfnamefont {K.}~\bibnamefont
  {Fujii}},\ }\bibfield  {title} {\bibinfo {title} {Constructing a virtual
  two-qubit gate by sampling single-qubit operations},\ }\href@noop {}
  {\bibfield  {journal} {\bibinfo  {journal} {New J.~Phys.}\ }\textbf {\bibinfo
  {volume} {23}},\ \bibinfo {pages} {023021} (\bibinfo {year}
  {2021})}\BibitemShut {NoStop}%
\bibitem [{\citenamefont {Yamamoto}\ and\ \citenamefont
  {Ohira}(2023)}]{yamamoto2023error}%
  \BibitemOpen
  \bibfield  {author} {\bibinfo {author} {\bibfnamefont {T.}~\bibnamefont
  {Yamamoto}}\ and\ \bibinfo {author} {\bibfnamefont {R.}~\bibnamefont
  {Ohira}},\ }\bibfield  {title} {\bibinfo {title} {Error suppression by a
  virtual two-qubit gate},\ }\href@noop {} {\bibfield  {journal} {\bibinfo
  {journal} {J.~Appl.~Phys.}\ }\textbf {\bibinfo {volume} {133}} (\bibinfo
  {year} {2023})}\BibitemShut {NoStop}%
\bibitem [{\citenamefont {Singh}\ \emph {et~al.}(2024)\citenamefont {Singh},
  \citenamefont {Mitarai}, \citenamefont {Suzuki}, \citenamefont {Heya},
  \citenamefont {Tabuchi}, \citenamefont {Fujii},\ and\ \citenamefont
  {Nakamura}}]{singh2024experimental}%
  \BibitemOpen
  \bibfield  {author} {\bibinfo {author} {\bibfnamefont {A.~P.}\ \bibnamefont
  {Singh}}, \bibinfo {author} {\bibfnamefont {K.}~\bibnamefont {Mitarai}},
  \bibinfo {author} {\bibfnamefont {Y.}~\bibnamefont {Suzuki}}, \bibinfo
  {author} {\bibfnamefont {K.}~\bibnamefont {Heya}}, \bibinfo {author}
  {\bibfnamefont {Y.}~\bibnamefont {Tabuchi}}, \bibinfo {author} {\bibfnamefont
  {K.}~\bibnamefont {Fujii}},\ and\ \bibinfo {author} {\bibfnamefont
  {Y.}~\bibnamefont {Nakamura}},\ }\bibfield  {title} {\bibinfo {title}
  {Experimental demonstration of a high-fidelity virtual two-qubit gate},\
  }\href@noop {} {\bibfield  {journal} {\bibinfo  {journal} {Phys.~Rev.~Res.}\
  }\textbf {\bibinfo {volume} {6}},\ \bibinfo {pages} {013235} (\bibinfo {year}
  {2024})}\BibitemShut {NoStop}%
\bibitem [{\citenamefont {Piveteau}\ and\ \citenamefont
  {Sutter}(2023)}]{piveteau2023circuit}%
  \BibitemOpen
  \bibfield  {author} {\bibinfo {author} {\bibfnamefont {C.}~\bibnamefont
  {Piveteau}}\ and\ \bibinfo {author} {\bibfnamefont {D.}~\bibnamefont
  {Sutter}},\ }\bibfield  {title} {\bibinfo {title} {Circuit knitting with
  classical communication},\ }\href@noop {} {\bibfield  {journal} {\bibinfo
  {journal} {IEEE Transactions on Information Theory}\ } (\bibinfo {year}
  {2023})}\BibitemShut {NoStop}%
\bibitem [{\citenamefont {Carrera~Vazquez}\ \emph {et~al.}(2024)\citenamefont
  {Carrera~Vazquez}, \citenamefont {Tornow}, \citenamefont {Rist{\`e}},
  \citenamefont {Woerner}, \citenamefont {Takita},\ and\ \citenamefont
  {Egger}}]{carrera2024combining}%
  \BibitemOpen
  \bibfield  {author} {\bibinfo {author} {\bibfnamefont {A.}~\bibnamefont
  {Carrera~Vazquez}}, \bibinfo {author} {\bibfnamefont {C.}~\bibnamefont
  {Tornow}}, \bibinfo {author} {\bibfnamefont {D.}~\bibnamefont {Rist{\`e}}},
  \bibinfo {author} {\bibfnamefont {S.}~\bibnamefont {Woerner}}, \bibinfo
  {author} {\bibfnamefont {M.}~\bibnamefont {Takita}},\ and\ \bibinfo {author}
  {\bibfnamefont {D.~J.}\ \bibnamefont {Egger}},\ }\bibfield  {title} {\bibinfo
  {title} {Combining quantum processors with real-time classical
  communication},\ }\href@noop {} {\bibfield  {journal} {\bibinfo  {journal}
  {Nature}\ ,\ \bibinfo {pages} {1}} (\bibinfo {year} {2024})}\BibitemShut
  {NoStop}%
\bibitem [{\citenamefont {McArdle}\ \emph {et~al.}(2020)\citenamefont
  {McArdle}, \citenamefont {Endo}, \citenamefont {Aspuru-Guzik}, \citenamefont
  {Benjamin},\ and\ \citenamefont {Yuan}}]{mcardle2020quantum}%
  \BibitemOpen
  \bibfield  {author} {\bibinfo {author} {\bibfnamefont {S.}~\bibnamefont
  {McArdle}}, \bibinfo {author} {\bibfnamefont {S.}~\bibnamefont {Endo}},
  \bibinfo {author} {\bibfnamefont {A.}~\bibnamefont {Aspuru-Guzik}}, \bibinfo
  {author} {\bibfnamefont {S.~C.}\ \bibnamefont {Benjamin}},\ and\ \bibinfo
  {author} {\bibfnamefont {X.}~\bibnamefont {Yuan}},\ }\bibfield  {title}
  {\bibinfo {title} {Quantum computational chemistry},\ }\href@noop {}
  {\bibfield  {journal} {\bibinfo  {journal} {Rev.~Mod.~Phys.}\ }\textbf
  {\bibinfo {volume} {92}},\ \bibinfo {pages} {015003} (\bibinfo {year}
  {2020})}\BibitemShut {NoStop}%
\bibitem [{\citenamefont {Kurpiers}\ \emph {et~al.}(2018)\citenamefont
  {Kurpiers}, \citenamefont {Magnard}, \citenamefont {Walter}, \citenamefont
  {Royer}, \citenamefont {Pechal}, \citenamefont {Heinsoo}, \citenamefont
  {Salath{\'e}}, \citenamefont {Akin}, \citenamefont {Storz}, \citenamefont
  {Besse} \emph {et~al.}}]{kurpiers2018deterministic}%
  \BibitemOpen
  \bibfield  {author} {\bibinfo {author} {\bibfnamefont {P.}~\bibnamefont
  {Kurpiers}}, \bibinfo {author} {\bibfnamefont {P.}~\bibnamefont {Magnard}},
  \bibinfo {author} {\bibfnamefont {T.}~\bibnamefont {Walter}}, \bibinfo
  {author} {\bibfnamefont {B.}~\bibnamefont {Royer}}, \bibinfo {author}
  {\bibfnamefont {M.}~\bibnamefont {Pechal}}, \bibinfo {author} {\bibfnamefont
  {J.}~\bibnamefont {Heinsoo}}, \bibinfo {author} {\bibfnamefont
  {Y.}~\bibnamefont {Salath{\'e}}}, \bibinfo {author} {\bibfnamefont
  {A.}~\bibnamefont {Akin}}, \bibinfo {author} {\bibfnamefont {S.}~\bibnamefont
  {Storz}}, \bibinfo {author} {\bibfnamefont {J.-C.}\ \bibnamefont {Besse}},
  \emph {et~al.},\ }\bibfield  {title} {\bibinfo {title} {Deterministic quantum
  state transfer and remote entanglement using microwave photons},\ }\href@noop
  {} {\bibfield  {journal} {\bibinfo  {journal} {Nature}\ }\textbf {\bibinfo
  {volume} {558}},\ \bibinfo {pages} {264} (\bibinfo {year}
  {2018})}\BibitemShut {NoStop}%
\bibitem [{\citenamefont {Axline}\ \emph {et~al.}(2018)\citenamefont {Axline},
  \citenamefont {Burkhart}, \citenamefont {Pfaff}, \citenamefont {Zhang},
  \citenamefont {Chou}, \citenamefont {Campagne-Ibarcq}, \citenamefont
  {Reinhold}, \citenamefont {Frunzio}, \citenamefont {Girvin}, \citenamefont
  {Jiang} \emph {et~al.}}]{axline2018demand}%
  \BibitemOpen
  \bibfield  {author} {\bibinfo {author} {\bibfnamefont {C.~J.}\ \bibnamefont
  {Axline}}, \bibinfo {author} {\bibfnamefont {L.~D.}\ \bibnamefont
  {Burkhart}}, \bibinfo {author} {\bibfnamefont {W.}~\bibnamefont {Pfaff}},
  \bibinfo {author} {\bibfnamefont {M.}~\bibnamefont {Zhang}}, \bibinfo
  {author} {\bibfnamefont {K.}~\bibnamefont {Chou}}, \bibinfo {author}
  {\bibfnamefont {P.}~\bibnamefont {Campagne-Ibarcq}}, \bibinfo {author}
  {\bibfnamefont {P.}~\bibnamefont {Reinhold}}, \bibinfo {author}
  {\bibfnamefont {L.}~\bibnamefont {Frunzio}}, \bibinfo {author} {\bibfnamefont
  {S.~M.}\ \bibnamefont {Girvin}}, \bibinfo {author} {\bibfnamefont
  {L.}~\bibnamefont {Jiang}}, \emph {et~al.},\ }\bibfield  {title} {\bibinfo
  {title} {On-demand quantum state transfer and entanglement between remote
  microwave cavity memories},\ }\href@noop {} {\bibfield  {journal} {\bibinfo
  {journal} {Nat.~Phys.}\ }\textbf {\bibinfo {volume} {14}},\ \bibinfo {pages}
  {705} (\bibinfo {year} {2018})}\BibitemShut {NoStop}%
\bibitem [{\citenamefont {Campagne-Ibarcq}\ \emph {et~al.}(2018)\citenamefont
  {Campagne-Ibarcq}, \citenamefont {Zalys-Geller}, \citenamefont {Narla},
  \citenamefont {Shankar}, \citenamefont {Reinhold}, \citenamefont {Burkhart},
  \citenamefont {Axline}, \citenamefont {Pfaff}, \citenamefont {Frunzio},
  \citenamefont {Schoelkopf} \emph {et~al.}}]{campagne2018deterministic}%
  \BibitemOpen
  \bibfield  {author} {\bibinfo {author} {\bibfnamefont {P.}~\bibnamefont
  {Campagne-Ibarcq}}, \bibinfo {author} {\bibfnamefont {E.}~\bibnamefont
  {Zalys-Geller}}, \bibinfo {author} {\bibfnamefont {A.}~\bibnamefont {Narla}},
  \bibinfo {author} {\bibfnamefont {S.}~\bibnamefont {Shankar}}, \bibinfo
  {author} {\bibfnamefont {P.}~\bibnamefont {Reinhold}}, \bibinfo {author}
  {\bibfnamefont {L.}~\bibnamefont {Burkhart}}, \bibinfo {author}
  {\bibfnamefont {C.}~\bibnamefont {Axline}}, \bibinfo {author} {\bibfnamefont
  {W.}~\bibnamefont {Pfaff}}, \bibinfo {author} {\bibfnamefont
  {L.}~\bibnamefont {Frunzio}}, \bibinfo {author} {\bibfnamefont {R.~J.}\
  \bibnamefont {Schoelkopf}}, \emph {et~al.},\ }\bibfield  {title} {\bibinfo
  {title} {Deterministic remote entanglement of superconducting circuits
  through microwave two-photon transitions},\ }\href@noop {} {\bibfield
  {journal} {\bibinfo  {journal} {Phys.~Rev.~Lett.}\ }\textbf {\bibinfo
  {volume} {120}},\ \bibinfo {pages} {200501} (\bibinfo {year}
  {2018})}\BibitemShut {NoStop}%
\bibitem [{\citenamefont {Zhong}\ \emph {et~al.}(2019)\citenamefont {Zhong},
  \citenamefont {Chang}, \citenamefont {Satzinger}, \citenamefont {Chou},
  \citenamefont {Bienfait}, \citenamefont {Conner}, \citenamefont {Dumur},
  \citenamefont {Grebel}, \citenamefont {Peairs}, \citenamefont {Povey} \emph
  {et~al.}}]{zhong2019violating}%
  \BibitemOpen
  \bibfield  {author} {\bibinfo {author} {\bibfnamefont {Y.}~\bibnamefont
  {Zhong}}, \bibinfo {author} {\bibfnamefont {H.-S.}\ \bibnamefont {Chang}},
  \bibinfo {author} {\bibfnamefont {K.}~\bibnamefont {Satzinger}}, \bibinfo
  {author} {\bibfnamefont {M.-H.}\ \bibnamefont {Chou}}, \bibinfo {author}
  {\bibfnamefont {A.}~\bibnamefont {Bienfait}}, \bibinfo {author}
  {\bibfnamefont {C.}~\bibnamefont {Conner}}, \bibinfo {author} {\bibfnamefont
  {{\'E}.}~\bibnamefont {Dumur}}, \bibinfo {author} {\bibfnamefont
  {J.}~\bibnamefont {Grebel}}, \bibinfo {author} {\bibfnamefont
  {G.}~\bibnamefont {Peairs}}, \bibinfo {author} {\bibfnamefont
  {R.}~\bibnamefont {Povey}}, \emph {et~al.},\ }\bibfield  {title} {\bibinfo
  {title} {Violating {B}ell’s inequality with remotely connected
  superconducting qubits},\ }\href@noop {} {\bibfield  {journal} {\bibinfo
  {journal} {Nat.~Phys.}\ }\textbf {\bibinfo {volume} {15}},\ \bibinfo {pages}
  {741} (\bibinfo {year} {2019})}\BibitemShut {NoStop}%
\bibitem [{\citenamefont {Kurpiers}\ \emph {et~al.}(2019)\citenamefont
  {Kurpiers}, \citenamefont {Pechal}, \citenamefont {Royer}, \citenamefont
  {Magnard}, \citenamefont {Walter}, \citenamefont {Heinsoo}, \citenamefont
  {Salath{\'e}}, \citenamefont {Akin}, \citenamefont {Storz}, \citenamefont
  {Besse} \emph {et~al.}}]{kurpiers2019quantum}%
  \BibitemOpen
  \bibfield  {author} {\bibinfo {author} {\bibfnamefont {P.}~\bibnamefont
  {Kurpiers}}, \bibinfo {author} {\bibfnamefont {M.}~\bibnamefont {Pechal}},
  \bibinfo {author} {\bibfnamefont {B.}~\bibnamefont {Royer}}, \bibinfo
  {author} {\bibfnamefont {P.}~\bibnamefont {Magnard}}, \bibinfo {author}
  {\bibfnamefont {T.}~\bibnamefont {Walter}}, \bibinfo {author} {\bibfnamefont
  {J.}~\bibnamefont {Heinsoo}}, \bibinfo {author} {\bibfnamefont
  {Y.}~\bibnamefont {Salath{\'e}}}, \bibinfo {author} {\bibfnamefont
  {A.}~\bibnamefont {Akin}}, \bibinfo {author} {\bibfnamefont {S.}~\bibnamefont
  {Storz}}, \bibinfo {author} {\bibfnamefont {J.-C.}\ \bibnamefont {Besse}},
  \emph {et~al.},\ }\bibfield  {title} {\bibinfo {title} {Quantum communication
  with time-bin encoded microwave photons},\ }\href@noop {} {\bibfield
  {journal} {\bibinfo  {journal} {Phys.~Rev.~App.}\ }\textbf {\bibinfo {volume}
  {12}},\ \bibinfo {pages} {044067} (\bibinfo {year} {2019})}\BibitemShut
  {NoStop}%
\bibitem [{\citenamefont {Zhong}\ \emph {et~al.}(2021)\citenamefont {Zhong},
  \citenamefont {Chang}, \citenamefont {Bienfait}, \citenamefont {Dumur},
  \citenamefont {Chou}, \citenamefont {Conner}, \citenamefont {Grebel},
  \citenamefont {Povey}, \citenamefont {Yan}, \citenamefont {Schuster} \emph
  {et~al.}}]{zhong2021deterministic}%
  \BibitemOpen
  \bibfield  {author} {\bibinfo {author} {\bibfnamefont {Y.}~\bibnamefont
  {Zhong}}, \bibinfo {author} {\bibfnamefont {H.-S.}\ \bibnamefont {Chang}},
  \bibinfo {author} {\bibfnamefont {A.}~\bibnamefont {Bienfait}}, \bibinfo
  {author} {\bibfnamefont {{\'E}.}~\bibnamefont {Dumur}}, \bibinfo {author}
  {\bibfnamefont {M.-H.}\ \bibnamefont {Chou}}, \bibinfo {author}
  {\bibfnamefont {C.~R.}\ \bibnamefont {Conner}}, \bibinfo {author}
  {\bibfnamefont {J.}~\bibnamefont {Grebel}}, \bibinfo {author} {\bibfnamefont
  {R.~G.}\ \bibnamefont {Povey}}, \bibinfo {author} {\bibfnamefont
  {H.}~\bibnamefont {Yan}}, \bibinfo {author} {\bibfnamefont {D.~I.}\
  \bibnamefont {Schuster}}, \emph {et~al.},\ }\bibfield  {title} {\bibinfo
  {title} {Deterministic multi-qubit entanglement in a quantum network},\
  }\href@noop {} {\bibfield  {journal} {\bibinfo  {journal} {Nature}\ }\textbf
  {\bibinfo {volume} {590}},\ \bibinfo {pages} {571} (\bibinfo {year}
  {2021})}\BibitemShut {NoStop}%
\bibitem [{\citenamefont {Burkhart}\ \emph {et~al.}(2021)\citenamefont
  {Burkhart}, \citenamefont {Teoh}, \citenamefont {Zhang}, \citenamefont
  {Axline}, \citenamefont {Frunzio}, \citenamefont {Devoret}, \citenamefont
  {Jiang}, \citenamefont {Girvin},\ and\ \citenamefont
  {Schoelkopf}}]{burkhart2021error}%
  \BibitemOpen
  \bibfield  {author} {\bibinfo {author} {\bibfnamefont {L.~D.}\ \bibnamefont
  {Burkhart}}, \bibinfo {author} {\bibfnamefont {J.~D.}\ \bibnamefont {Teoh}},
  \bibinfo {author} {\bibfnamefont {Y.}~\bibnamefont {Zhang}}, \bibinfo
  {author} {\bibfnamefont {C.~J.}\ \bibnamefont {Axline}}, \bibinfo {author}
  {\bibfnamefont {L.}~\bibnamefont {Frunzio}}, \bibinfo {author} {\bibfnamefont
  {M.~H.}\ \bibnamefont {Devoret}}, \bibinfo {author} {\bibfnamefont
  {L.}~\bibnamefont {Jiang}}, \bibinfo {author} {\bibfnamefont {S.~M.}\
  \bibnamefont {Girvin}},\ and\ \bibinfo {author} {\bibfnamefont {R.~J.}\
  \bibnamefont {Schoelkopf}},\ }\bibfield  {title} {\bibinfo {title}
  {Error-detected state transfer and entanglement in a superconducting quantum
  network},\ }\href@noop {} {\bibfield  {journal} {\bibinfo  {journal} {PRX
  Quantum}\ }\textbf {\bibinfo {volume} {2}},\ \bibinfo {pages} {030321}
  (\bibinfo {year} {2021})}\BibitemShut {NoStop}%
\bibitem [{\citenamefont {Storz}\ \emph {et~al.}(2023)\citenamefont {Storz},
  \citenamefont {Sch{\"a}r}, \citenamefont {Kulikov}, \citenamefont {Magnard},
  \citenamefont {Kurpiers}, \citenamefont {L{\"u}tolf}, \citenamefont {Walter},
  \citenamefont {Copetudo}, \citenamefont {Reuer}, \citenamefont {Akin} \emph
  {et~al.}}]{storz2023loophole}%
  \BibitemOpen
  \bibfield  {author} {\bibinfo {author} {\bibfnamefont {S.}~\bibnamefont
  {Storz}}, \bibinfo {author} {\bibfnamefont {J.}~\bibnamefont {Sch{\"a}r}},
  \bibinfo {author} {\bibfnamefont {A.}~\bibnamefont {Kulikov}}, \bibinfo
  {author} {\bibfnamefont {P.}~\bibnamefont {Magnard}}, \bibinfo {author}
  {\bibfnamefont {P.}~\bibnamefont {Kurpiers}}, \bibinfo {author}
  {\bibfnamefont {J.}~\bibnamefont {L{\"u}tolf}}, \bibinfo {author}
  {\bibfnamefont {T.}~\bibnamefont {Walter}}, \bibinfo {author} {\bibfnamefont
  {A.}~\bibnamefont {Copetudo}}, \bibinfo {author} {\bibfnamefont
  {K.}~\bibnamefont {Reuer}}, \bibinfo {author} {\bibfnamefont
  {A.}~\bibnamefont {Akin}}, \emph {et~al.},\ }\bibfield  {title} {\bibinfo
  {title} {Loophole-free bell inequality violation with superconducting
  circuits},\ }\href@noop {} {\bibfield  {journal} {\bibinfo  {journal}
  {Nature}\ }\textbf {\bibinfo {volume} {617}},\ \bibinfo {pages} {265}
  (\bibinfo {year} {2023})}\BibitemShut {NoStop}%
\bibitem [{\citenamefont {Grebel}\ \emph {et~al.}(2024)\citenamefont {Grebel},
  \citenamefont {Yan}, \citenamefont {Chou}, \citenamefont {Andersson},
  \citenamefont {Conner}, \citenamefont {Joshi}, \citenamefont {Miller},
  \citenamefont {Povey}, \citenamefont {Qiao}, \citenamefont {Wu} \emph
  {et~al.}}]{grebel2024bidirectional}%
  \BibitemOpen
  \bibfield  {author} {\bibinfo {author} {\bibfnamefont {J.}~\bibnamefont
  {Grebel}}, \bibinfo {author} {\bibfnamefont {H.}~\bibnamefont {Yan}},
  \bibinfo {author} {\bibfnamefont {M.-H.}\ \bibnamefont {Chou}}, \bibinfo
  {author} {\bibfnamefont {G.}~\bibnamefont {Andersson}}, \bibinfo {author}
  {\bibfnamefont {C.~R.}\ \bibnamefont {Conner}}, \bibinfo {author}
  {\bibfnamefont {Y.~J.}\ \bibnamefont {Joshi}}, \bibinfo {author}
  {\bibfnamefont {J.~M.}\ \bibnamefont {Miller}}, \bibinfo {author}
  {\bibfnamefont {R.~G.}\ \bibnamefont {Povey}}, \bibinfo {author}
  {\bibfnamefont {H.}~\bibnamefont {Qiao}}, \bibinfo {author} {\bibfnamefont
  {X.}~\bibnamefont {Wu}}, \emph {et~al.},\ }\bibfield  {title} {\bibinfo
  {title} {Bidirectional multiphoton communication between remote
  superconducting nodes},\ }\href@noop {} {\bibfield  {journal} {\bibinfo
  {journal} {Phys.~Rev.~Lett.}\ }\textbf {\bibinfo {volume} {132}},\ \bibinfo
  {pages} {047001} (\bibinfo {year} {2024})}\BibitemShut {NoStop}%
\bibitem [{\citenamefont {Mollenhauer}\ \emph {et~al.}(2024)\citenamefont
  {Mollenhauer}, \citenamefont {Irfan}, \citenamefont {Cao}, \citenamefont
  {Mandal},\ and\ \citenamefont {Pfaff}}]{mollenhauer2024high}%
  \BibitemOpen
  \bibfield  {author} {\bibinfo {author} {\bibfnamefont {M.}~\bibnamefont
  {Mollenhauer}}, \bibinfo {author} {\bibfnamefont {A.}~\bibnamefont {Irfan}},
  \bibinfo {author} {\bibfnamefont {X.}~\bibnamefont {Cao}}, \bibinfo {author}
  {\bibfnamefont {S.}~\bibnamefont {Mandal}},\ and\ \bibinfo {author}
  {\bibfnamefont {W.}~\bibnamefont {Pfaff}},\ }\bibfield  {title} {\bibinfo
  {title} {A high-efficiency plug-and-play superconducting qubit network},\
  }\href@noop {} {\bibfield  {journal} {\bibinfo  {journal} {arXiv preprint
  arXiv:2407.16743}\ } (\bibinfo {year} {2024})}\BibitemShut {NoStop}%
\bibitem [{\citenamefont {Gottesman}\ and\ \citenamefont
  {Chuang}(1999)}]{gottesman1999demonstrating}%
  \BibitemOpen
  \bibfield  {author} {\bibinfo {author} {\bibfnamefont {D.}~\bibnamefont
  {Gottesman}}\ and\ \bibinfo {author} {\bibfnamefont {I.~L.}\ \bibnamefont
  {Chuang}},\ }\bibfield  {title} {\bibinfo {title} {Demonstrating the
  viability of universal quantum computation using teleportation and
  single-qubit operations},\ }\href@noop {} {\bibfield  {journal} {\bibinfo
  {journal} {Nature}\ }\textbf {\bibinfo {volume} {402}},\ \bibinfo {pages}
  {390} (\bibinfo {year} {1999})}\BibitemShut {NoStop}%
\bibitem [{\citenamefont {Wan}\ \emph {et~al.}(2019)\citenamefont {Wan},
  \citenamefont {Kienzler}, \citenamefont {Erickson}, \citenamefont {Mayer},
  \citenamefont {Tan}, \citenamefont {Wu}, \citenamefont {Vasconcelos},
  \citenamefont {Glancy}, \citenamefont {Knill}, \citenamefont {Wineland} \emph
  {et~al.}}]{wan2019quantum}%
  \BibitemOpen
  \bibfield  {author} {\bibinfo {author} {\bibfnamefont {Y.}~\bibnamefont
  {Wan}}, \bibinfo {author} {\bibfnamefont {D.}~\bibnamefont {Kienzler}},
  \bibinfo {author} {\bibfnamefont {S.~D.}\ \bibnamefont {Erickson}}, \bibinfo
  {author} {\bibfnamefont {K.~H.}\ \bibnamefont {Mayer}}, \bibinfo {author}
  {\bibfnamefont {T.~R.}\ \bibnamefont {Tan}}, \bibinfo {author} {\bibfnamefont
  {J.~J.}\ \bibnamefont {Wu}}, \bibinfo {author} {\bibfnamefont {H.~M.}\
  \bibnamefont {Vasconcelos}}, \bibinfo {author} {\bibfnamefont
  {S.}~\bibnamefont {Glancy}}, \bibinfo {author} {\bibfnamefont
  {E.}~\bibnamefont {Knill}}, \bibinfo {author} {\bibfnamefont {D.~J.}\
  \bibnamefont {Wineland}}, \emph {et~al.},\ }\bibfield  {title} {\bibinfo
  {title} {Quantum gate teleportation between separated qubits in a trapped-ion
  processor},\ }\href@noop {} {\bibfield  {journal} {\bibinfo  {journal}
  {Science}\ }\textbf {\bibinfo {volume} {364}},\ \bibinfo {pages} {875}
  (\bibinfo {year} {2019})}\BibitemShut {NoStop}%
\bibitem [{\citenamefont {Myers}\ \emph {et~al.}(2007)\citenamefont {Myers},
  \citenamefont {Silva}, \citenamefont {Nemoto},\ and\ \citenamefont
  {Munro}}]{myers2007stabilizer}%
  \BibitemOpen
  \bibfield  {author} {\bibinfo {author} {\bibfnamefont {C.~R.}\ \bibnamefont
  {Myers}}, \bibinfo {author} {\bibfnamefont {M.}~\bibnamefont {Silva}},
  \bibinfo {author} {\bibfnamefont {K.}~\bibnamefont {Nemoto}},\ and\ \bibinfo
  {author} {\bibfnamefont {W.~J.}\ \bibnamefont {Munro}},\ }\bibfield  {title}
  {\bibinfo {title} {Stabilizer quantum error correction with quantum bus
  computation},\ }\href@noop {} {\bibfield  {journal} {\bibinfo  {journal}
  {Phys.~Rev.~A}\ }\textbf {\bibinfo {volume} {76}},\ \bibinfo {pages} {012303}
  (\bibinfo {year} {2007})}\BibitemShut {NoStop}%
\bibitem [{\citenamefont {McIntyre}\ and\ \citenamefont
  {Coish}(2024{\natexlab{a}})}]{mcintyre2024flying}%
  \BibitemOpen
  \bibfield  {author} {\bibinfo {author} {\bibfnamefont {Z.~M.}\ \bibnamefont
  {McIntyre}}\ and\ \bibinfo {author} {\bibfnamefont {W.~A.}\ \bibnamefont
  {Coish}},\ }\bibfield  {title} {\bibinfo {title} {Flying-cat parity checks
  for quantum error correction},\ }\href@noop {} {\bibfield  {journal}
  {\bibinfo  {journal} {Phys.~Rev.~Res.}\ }\textbf {\bibinfo {volume} {6}},\
  \bibinfo {pages} {023247} (\bibinfo {year} {2024}{\natexlab{a}})}\BibitemShut
  {NoStop}%
\bibitem [{\citenamefont {Eastin}\ and\ \citenamefont
  {Knill}(2009)}]{eastin2009restrictions}%
  \BibitemOpen
  \bibfield  {author} {\bibinfo {author} {\bibfnamefont {B.}~\bibnamefont
  {Eastin}}\ and\ \bibinfo {author} {\bibfnamefont {E.}~\bibnamefont {Knill}},\
  }\bibfield  {title} {\bibinfo {title} {Restrictions on transversal encoded
  quantum gate sets},\ }\href@noop {} {\bibfield  {journal} {\bibinfo
  {journal} {Phys.~Rev.~Lett.}\ }\textbf {\bibinfo {volume} {102}},\ \bibinfo
  {pages} {110502} (\bibinfo {year} {2009})}\BibitemShut {NoStop}%
\bibitem [{\citenamefont {Xu}\ \emph {et~al.}(2022)\citenamefont {Xu},
  \citenamefont {Seif}, \citenamefont {Yan}, \citenamefont {Mannucci},
  \citenamefont {Sane}, \citenamefont {Van~Meter}, \citenamefont {Cleland},\
  and\ \citenamefont {Jiang}}]{xu2022distributed}%
  \BibitemOpen
  \bibfield  {author} {\bibinfo {author} {\bibfnamefont {Q.}~\bibnamefont
  {Xu}}, \bibinfo {author} {\bibfnamefont {A.}~\bibnamefont {Seif}}, \bibinfo
  {author} {\bibfnamefont {H.}~\bibnamefont {Yan}}, \bibinfo {author}
  {\bibfnamefont {N.}~\bibnamefont {Mannucci}}, \bibinfo {author}
  {\bibfnamefont {B.~O.}\ \bibnamefont {Sane}}, \bibinfo {author}
  {\bibfnamefont {R.}~\bibnamefont {Van~Meter}}, \bibinfo {author}
  {\bibfnamefont {A.~N.}\ \bibnamefont {Cleland}},\ and\ \bibinfo {author}
  {\bibfnamefont {L.}~\bibnamefont {Jiang}},\ }\bibfield  {title} {\bibinfo
  {title} {Distributed quantum error correction for chip-level catastrophic
  errors},\ }\href@noop {} {\bibfield  {journal} {\bibinfo  {journal}
  {Phys.~Rev.~Lett.}\ }\textbf {\bibinfo {volume} {129}},\ \bibinfo {pages}
  {240502} (\bibinfo {year} {2022})}\BibitemShut {NoStop}%
\bibitem [{\citenamefont {MacKay}\ \emph {et~al.}(2004)\citenamefont {MacKay},
  \citenamefont {Mitchison},\ and\ \citenamefont
  {McFadden}}]{mackay2004sparse}%
  \BibitemOpen
  \bibfield  {author} {\bibinfo {author} {\bibfnamefont {D.~J.~C.}\
  \bibnamefont {MacKay}}, \bibinfo {author} {\bibfnamefont {G.}~\bibnamefont
  {Mitchison}},\ and\ \bibinfo {author} {\bibfnamefont {P.~L.}\ \bibnamefont
  {McFadden}},\ }\bibfield  {title} {\bibinfo {title} {Sparse-graph codes for
  quantum error correction},\ }\href@noop {} {\bibfield  {journal} {\bibinfo
  {journal} {IEEE Trans.~Inf.~Theory}\ }\textbf {\bibinfo {volume} {50}},\
  \bibinfo {pages} {2315} (\bibinfo {year} {2004})}\BibitemShut {NoStop}%
\bibitem [{\citenamefont {Kovalev}\ and\ \citenamefont
  {Pryadko}(2013)}]{kovalev2013quantum}%
  \BibitemOpen
  \bibfield  {author} {\bibinfo {author} {\bibfnamefont {A.~A.}\ \bibnamefont
  {Kovalev}}\ and\ \bibinfo {author} {\bibfnamefont {L.~P.}\ \bibnamefont
  {Pryadko}},\ }\bibfield  {title} {\bibinfo {title} {Quantum {K}ronecker
  sum-product low-density parity-check codes with finite rate},\ }\href@noop {}
  {\bibfield  {journal} {\bibinfo  {journal} {Phys.~Rev.~A}\ }\textbf {\bibinfo
  {volume} {88}},\ \bibinfo {pages} {012311} (\bibinfo {year}
  {2013})}\BibitemShut {NoStop}%
\bibitem [{\citenamefont {Panteleev}\ and\ \citenamefont
  {Kalachev}(2021)}]{panteleev2021degenerate}%
  \BibitemOpen
  \bibfield  {author} {\bibinfo {author} {\bibfnamefont {P.}~\bibnamefont
  {Panteleev}}\ and\ \bibinfo {author} {\bibfnamefont {G.}~\bibnamefont
  {Kalachev}},\ }\bibfield  {title} {\bibinfo {title} {Degenerate quantum
  {LDPC} codes with good finite length performance},\ }\href@noop {} {\bibfield
   {journal} {\bibinfo  {journal} {Quantum}\ }\textbf {\bibinfo {volume} {5}},\
  \bibinfo {pages} {585} (\bibinfo {year} {2021})}\BibitemShut {NoStop}%
\bibitem [{\citenamefont {Bravyi}\ \emph {et~al.}(2024)\citenamefont {Bravyi},
  \citenamefont {Cross}, \citenamefont {Gambetta}, \citenamefont {Maslov},
  \citenamefont {Rall},\ and\ \citenamefont {Yoder}}]{bravyi2024high}%
  \BibitemOpen
  \bibfield  {author} {\bibinfo {author} {\bibfnamefont {S.}~\bibnamefont
  {Bravyi}}, \bibinfo {author} {\bibfnamefont {A.~W.}\ \bibnamefont {Cross}},
  \bibinfo {author} {\bibfnamefont {J.~M.}\ \bibnamefont {Gambetta}}, \bibinfo
  {author} {\bibfnamefont {D.}~\bibnamefont {Maslov}}, \bibinfo {author}
  {\bibfnamefont {P.}~\bibnamefont {Rall}},\ and\ \bibinfo {author}
  {\bibfnamefont {T.~J.}\ \bibnamefont {Yoder}},\ }\bibfield  {title} {\bibinfo
  {title} {High-threshold and low-overhead fault-tolerant quantum memory},\
  }\href@noop {} {\bibfield  {journal} {\bibinfo  {journal} {Nature}\ }\textbf
  {\bibinfo {volume} {627}},\ \bibinfo {pages} {778} (\bibinfo {year}
  {2024})}\BibitemShut {NoStop}%
\bibitem [{\citenamefont {Duan}\ \emph {et~al.}(2005)\citenamefont {Duan},
  \citenamefont {Wang},\ and\ \citenamefont {Kimble}}]{duan2005robust}%
  \BibitemOpen
  \bibfield  {author} {\bibinfo {author} {\bibfnamefont {L.-M.}\ \bibnamefont
  {Duan}}, \bibinfo {author} {\bibfnamefont {B.}~\bibnamefont {Wang}},\ and\
  \bibinfo {author} {\bibfnamefont {H.}~\bibnamefont {Kimble}},\ }\bibfield
  {title} {\bibinfo {title} {Robust quantum gates on neutral atoms with
  cavity-assisted photon scattering},\ }\href@noop {} {\bibfield  {journal}
  {\bibinfo  {journal} {Phys.~Rev.~A}\ }\textbf {\bibinfo {volume} {72}},\
  \bibinfo {pages} {032333} (\bibinfo {year} {2005})}\BibitemShut {NoStop}%
\bibitem [{\citenamefont {Daiss}\ \emph {et~al.}(2021)\citenamefont {Daiss},
  \citenamefont {Langenfeld}, \citenamefont {Welte}, \citenamefont {Distante},
  \citenamefont {Thomas}, \citenamefont {Hartung}, \citenamefont {Morin},\ and\
  \citenamefont {Rempe}}]{daiss2021quantum}%
  \BibitemOpen
  \bibfield  {author} {\bibinfo {author} {\bibfnamefont {S.}~\bibnamefont
  {Daiss}}, \bibinfo {author} {\bibfnamefont {S.}~\bibnamefont {Langenfeld}},
  \bibinfo {author} {\bibfnamefont {S.}~\bibnamefont {Welte}}, \bibinfo
  {author} {\bibfnamefont {E.}~\bibnamefont {Distante}}, \bibinfo {author}
  {\bibfnamefont {P.}~\bibnamefont {Thomas}}, \bibinfo {author} {\bibfnamefont
  {L.}~\bibnamefont {Hartung}}, \bibinfo {author} {\bibfnamefont
  {O.}~\bibnamefont {Morin}},\ and\ \bibinfo {author} {\bibfnamefont
  {G.}~\bibnamefont {Rempe}},\ }\bibfield  {title} {\bibinfo {title} {A
  quantum-logic gate between distant quantum-network modules},\ }\href@noop {}
  {\bibfield  {journal} {\bibinfo  {journal} {Science}\ }\textbf {\bibinfo
  {volume} {371}},\ \bibinfo {pages} {614} (\bibinfo {year}
  {2021})}\BibitemShut {NoStop}%
\bibitem [{\citenamefont {Reiserer}\ \emph {et~al.}(2014)\citenamefont
  {Reiserer}, \citenamefont {Kalb}, \citenamefont {Rempe},\ and\ \citenamefont
  {Ritter}}]{reiserer2014quantum}%
  \BibitemOpen
  \bibfield  {author} {\bibinfo {author} {\bibfnamefont {A.}~\bibnamefont
  {Reiserer}}, \bibinfo {author} {\bibfnamefont {N.}~\bibnamefont {Kalb}},
  \bibinfo {author} {\bibfnamefont {G.}~\bibnamefont {Rempe}},\ and\ \bibinfo
  {author} {\bibfnamefont {S.}~\bibnamefont {Ritter}},\ }\bibfield  {title}
  {\bibinfo {title} {A quantum gate between a flying optical photon and a
  single trapped atom},\ }\href@noop {} {\bibfield  {journal} {\bibinfo
  {journal} {Nature}\ }\textbf {\bibinfo {volume} {508}},\ \bibinfo {pages}
  {237} (\bibinfo {year} {2014})}\BibitemShut {NoStop}%
\bibitem [{\citenamefont {Reiserer}\ and\ \citenamefont
  {Rempe}(2015)}]{reiserer2015cavity}%
  \BibitemOpen
  \bibfield  {author} {\bibinfo {author} {\bibfnamefont {A.}~\bibnamefont
  {Reiserer}}\ and\ \bibinfo {author} {\bibfnamefont {G.}~\bibnamefont
  {Rempe}},\ }\bibfield  {title} {\bibinfo {title} {Cavity-based quantum
  networks with single atoms and optical photons},\ }\href@noop {} {\bibfield
  {journal} {\bibinfo  {journal} {Rev.~Mod.~Phys.}\ }\textbf {\bibinfo {volume}
  {87}},\ \bibinfo {pages} {1379} (\bibinfo {year} {2015})}\BibitemShut
  {NoStop}%
\bibitem [{\citenamefont {Tomm}\ \emph {et~al.}(2024)\citenamefont {Tomm},
  \citenamefont {Antoniadis}, \citenamefont {Janovitch}, \citenamefont
  {Brunelli}, \citenamefont {Schott}, \citenamefont {Valentin}, \citenamefont
  {Wieck}, \citenamefont {Ludwig}, \citenamefont {Potts}, \citenamefont
  {Javadi} \emph {et~al.}}]{tomm2024realization}%
  \BibitemOpen
  \bibfield  {author} {\bibinfo {author} {\bibfnamefont {N.}~\bibnamefont
  {Tomm}}, \bibinfo {author} {\bibfnamefont {N.~O.}\ \bibnamefont
  {Antoniadis}}, \bibinfo {author} {\bibfnamefont {M.}~\bibnamefont
  {Janovitch}}, \bibinfo {author} {\bibfnamefont {M.}~\bibnamefont {Brunelli}},
  \bibinfo {author} {\bibfnamefont {R.}~\bibnamefont {Schott}}, \bibinfo
  {author} {\bibfnamefont {S.~R.}\ \bibnamefont {Valentin}}, \bibinfo {author}
  {\bibfnamefont {A.~D.}\ \bibnamefont {Wieck}}, \bibinfo {author}
  {\bibfnamefont {A.}~\bibnamefont {Ludwig}}, \bibinfo {author} {\bibfnamefont
  {P.~P.}\ \bibnamefont {Potts}}, \bibinfo {author} {\bibfnamefont
  {A.}~\bibnamefont {Javadi}}, \emph {et~al.},\ }\bibfield  {title} {\bibinfo
  {title} {Realization of a coherent and efficient one-dimensional atom},\
  }\href@noop {} {\bibfield  {journal} {\bibinfo  {journal} {Phys.~Rev.~Lett.}\
  }\textbf {\bibinfo {volume} {133}},\ \bibinfo {pages} {083602} (\bibinfo
  {year} {2024})}\BibitemShut {NoStop}%
\bibitem [{\citenamefont {Duan}\ and\ \citenamefont
  {Kimble}(2004)}]{duan2004scalable}%
  \BibitemOpen
  \bibfield  {author} {\bibinfo {author} {\bibfnamefont {L.-M.}\ \bibnamefont
  {Duan}}\ and\ \bibinfo {author} {\bibfnamefont {H.~J.}\ \bibnamefont
  {Kimble}},\ }\bibfield  {title} {\bibinfo {title} {Scalable photonic quantum
  computation through cavity-assisted interactions},\ }\href@noop {} {\bibfield
   {journal} {\bibinfo  {journal} {Phys.~Rev.~Lett.}\ }\textbf {\bibinfo
  {volume} {92}},\ \bibinfo {pages} {127902} (\bibinfo {year}
  {2004})}\BibitemShut {NoStop}%
\bibitem [{\citenamefont {Hacker}\ \emph {et~al.}(2016)\citenamefont {Hacker},
  \citenamefont {Welte}, \citenamefont {Rempe},\ and\ \citenamefont
  {Ritter}}]{hacker2016photon}%
  \BibitemOpen
  \bibfield  {author} {\bibinfo {author} {\bibfnamefont {B.}~\bibnamefont
  {Hacker}}, \bibinfo {author} {\bibfnamefont {S.}~\bibnamefont {Welte}},
  \bibinfo {author} {\bibfnamefont {G.}~\bibnamefont {Rempe}},\ and\ \bibinfo
  {author} {\bibfnamefont {S.}~\bibnamefont {Ritter}},\ }\bibfield  {title}
  {\bibinfo {title} {A photon--photon quantum gate based on a single atom in an
  optical resonator},\ }\href@noop {} {\bibfield  {journal} {\bibinfo
  {journal} {Nature}\ }\textbf {\bibinfo {volume} {536}},\ \bibinfo {pages}
  {193} (\bibinfo {year} {2016})}\BibitemShut {NoStop}%
\bibitem [{\citenamefont {Marquardt}(2007)}]{marquardt2007efficient}%
  \BibitemOpen
  \bibfield  {author} {\bibinfo {author} {\bibfnamefont {F.}~\bibnamefont
  {Marquardt}},\ }\bibfield  {title} {\bibinfo {title} {Efficient on-chip
  source of microwave photon pairs in superconducting circuit {QED}},\
  }\href@noop {} {\bibfield  {journal} {\bibinfo  {journal} {Phys.~Rev.~B}\
  }\textbf {\bibinfo {volume} {76}},\ \bibinfo {pages} {205416} (\bibinfo
  {year} {2007})}\BibitemShut {NoStop}%
\bibitem [{\citenamefont {Rigetti}\ and\ \citenamefont
  {Devoret}(2010)}]{rigetti2010fully}%
  \BibitemOpen
  \bibfield  {author} {\bibinfo {author} {\bibfnamefont {C.}~\bibnamefont
  {Rigetti}}\ and\ \bibinfo {author} {\bibfnamefont {M.}~\bibnamefont
  {Devoret}},\ }\bibfield  {title} {\bibinfo {title} {Fully microwave-tunable
  universal gates in superconducting qubits with linear couplings and fixed
  transition frequencies},\ }\href@noop {} {\bibfield  {journal} {\bibinfo
  {journal} {Phys.~Rev.~B}\ }\textbf {\bibinfo {volume} {81}},\ \bibinfo
  {pages} {134507} (\bibinfo {year} {2010})}\BibitemShut {NoStop}%
\bibitem [{\citenamefont {Song}\ \emph {et~al.}(2024)\citenamefont {Song},
  \citenamefont {Yang}, \citenamefont {Liu}, \citenamefont {Zhang},
  \citenamefont {Xue}, \citenamefont {Mi}, \citenamefont {Zhang}, \citenamefont
  {Yan}, \citenamefont {Jin},\ and\ \citenamefont {Yu}}]{song2024realization}%
  \BibitemOpen
  \bibfield  {author} {\bibinfo {author} {\bibfnamefont {J.}~\bibnamefont
  {Song}}, \bibinfo {author} {\bibfnamefont {S.}~\bibnamefont {Yang}}, \bibinfo
  {author} {\bibfnamefont {P.}~\bibnamefont {Liu}}, \bibinfo {author}
  {\bibfnamefont {H.-L.}\ \bibnamefont {Zhang}}, \bibinfo {author}
  {\bibfnamefont {G.-M.}\ \bibnamefont {Xue}}, \bibinfo {author} {\bibfnamefont
  {Z.-Y.}\ \bibnamefont {Mi}}, \bibinfo {author} {\bibfnamefont {W.-G.}\
  \bibnamefont {Zhang}}, \bibinfo {author} {\bibfnamefont {F.}~\bibnamefont
  {Yan}}, \bibinfo {author} {\bibfnamefont {Y.-R.}\ \bibnamefont {Jin}},\ and\
  \bibinfo {author} {\bibfnamefont {H.-F.}\ \bibnamefont {Yu}},\ }\bibfield
  {title} {\bibinfo {title} {Realization of high-fidelity perfect entangler
  between remote superconducting quantum processors},\ }\href@noop {}
  {\bibfield  {journal} {\bibinfo  {journal} {arXiv preprint arXiv:2407.20338}\
  } (\bibinfo {year} {2024})}\BibitemShut {NoStop}%
\bibitem [{\citenamefont {Cross}\ and\ \citenamefont
  {Gambetta}(2015)}]{cross2015optimized}%
  \BibitemOpen
  \bibfield  {author} {\bibinfo {author} {\bibfnamefont {A.~W.}\ \bibnamefont
  {Cross}}\ and\ \bibinfo {author} {\bibfnamefont {J.~M.}\ \bibnamefont
  {Gambetta}},\ }\bibfield  {title} {\bibinfo {title} {Optimized pulse shapes
  for a resonator-induced phase gate},\ }\href@noop {} {\bibfield  {journal}
  {\bibinfo  {journal} {Phys.~Rev.~A}\ }\textbf {\bibinfo {volume} {91}},\
  \bibinfo {pages} {032325} (\bibinfo {year} {2015})}\BibitemShut {NoStop}%
\bibitem [{\citenamefont {Deng}\ \emph {et~al.}(2025)\citenamefont {Deng},
  \citenamefont {Zheng}, \citenamefont {Liao}, \citenamefont {Zhou},
  \citenamefont {Ge}, \citenamefont {Zhao}, \citenamefont {Lan}, \citenamefont
  {Tan}, \citenamefont {Zhang}, \citenamefont {Li} \emph
  {et~al.}}]{deng2025long}%
  \BibitemOpen
  \bibfield  {author} {\bibinfo {author} {\bibfnamefont {X.}~\bibnamefont
  {Deng}}, \bibinfo {author} {\bibfnamefont {W.}~\bibnamefont {Zheng}},
  \bibinfo {author} {\bibfnamefont {X.}~\bibnamefont {Liao}}, \bibinfo {author}
  {\bibfnamefont {H.}~\bibnamefont {Zhou}}, \bibinfo {author} {\bibfnamefont
  {Y.}~\bibnamefont {Ge}}, \bibinfo {author} {\bibfnamefont {J.}~\bibnamefont
  {Zhao}}, \bibinfo {author} {\bibfnamefont {D.}~\bibnamefont {Lan}}, \bibinfo
  {author} {\bibfnamefont {X.}~\bibnamefont {Tan}}, \bibinfo {author}
  {\bibfnamefont {Y.}~\bibnamefont {Zhang}}, \bibinfo {author} {\bibfnamefont
  {S.}~\bibnamefont {Li}}, \emph {et~al.},\ }\bibfield  {title} {\bibinfo
  {title} {Long-range {ZZ} interaction via resonator-induced phase in
  superconducting qubits},\ }\href@noop {} {\bibfield  {journal} {\bibinfo
  {journal} {Phys.~Rev.~Lett.}\ }\textbf {\bibinfo {volume} {134}},\ \bibinfo
  {pages} {020801} (\bibinfo {year} {2025})}\BibitemShut {NoStop}%
\bibitem [{\citenamefont {Humphreys}\ \emph {et~al.}(2013)\citenamefont
  {Humphreys}, \citenamefont {Metcalf}, \citenamefont {Spring}, \citenamefont
  {Moore}, \citenamefont {Jin}, \citenamefont {Barbieri}, \citenamefont
  {Kolthammer},\ and\ \citenamefont {Walmsley}}]{humphreys2013linear}%
  \BibitemOpen
  \bibfield  {author} {\bibinfo {author} {\bibfnamefont {P.~C.}\ \bibnamefont
  {Humphreys}}, \bibinfo {author} {\bibfnamefont {B.~J.}\ \bibnamefont
  {Metcalf}}, \bibinfo {author} {\bibfnamefont {J.~B.}\ \bibnamefont {Spring}},
  \bibinfo {author} {\bibfnamefont {M.}~\bibnamefont {Moore}}, \bibinfo
  {author} {\bibfnamefont {X.-M.}\ \bibnamefont {Jin}}, \bibinfo {author}
  {\bibfnamefont {M.}~\bibnamefont {Barbieri}}, \bibinfo {author}
  {\bibfnamefont {W.~S.}\ \bibnamefont {Kolthammer}},\ and\ \bibinfo {author}
  {\bibfnamefont {I.~A.}\ \bibnamefont {Walmsley}},\ }\bibfield  {title}
  {\bibinfo {title} {Linear optical quantum computing in a single spatial
  mode},\ }\href@noop {} {\bibfield  {journal} {\bibinfo  {journal}
  {Phys.~Rev.~Lett.}\ }\textbf {\bibinfo {volume} {111}},\ \bibinfo {pages}
  {150501} (\bibinfo {year} {2013})}\BibitemShut {NoStop}%
\bibitem [{\citenamefont {Bouchard}\ \emph {et~al.}(2022)\citenamefont
  {Bouchard}, \citenamefont {England}, \citenamefont {Bustard}, \citenamefont
  {Heshami},\ and\ \citenamefont {Sussman}}]{bouchard2022quantum}%
  \BibitemOpen
  \bibfield  {author} {\bibinfo {author} {\bibfnamefont {F.}~\bibnamefont
  {Bouchard}}, \bibinfo {author} {\bibfnamefont {D.}~\bibnamefont {England}},
  \bibinfo {author} {\bibfnamefont {P.~J.}\ \bibnamefont {Bustard}}, \bibinfo
  {author} {\bibfnamefont {K.}~\bibnamefont {Heshami}},\ and\ \bibinfo {author}
  {\bibfnamefont {B.}~\bibnamefont {Sussman}},\ }\bibfield  {title} {\bibinfo
  {title} {Quantum communication with ultrafast time-bin qubits},\ }\href@noop
  {} {\bibfield  {journal} {\bibinfo  {journal} {PRX Quantum}\ }\textbf
  {\bibinfo {volume} {3}},\ \bibinfo {pages} {010332} (\bibinfo {year}
  {2022})}\BibitemShut {NoStop}%
\bibitem [{\citenamefont {Xie}\ \emph {et~al.}(2021)\citenamefont {Xie},
  \citenamefont {Liu}, \citenamefont {Mo}, \citenamefont {Li},\ and\
  \citenamefont {Li}}]{xie2021quantum}%
  \BibitemOpen
  \bibfield  {author} {\bibinfo {author} {\bibfnamefont {Z.}~\bibnamefont
  {Xie}}, \bibinfo {author} {\bibfnamefont {Y.}~\bibnamefont {Liu}}, \bibinfo
  {author} {\bibfnamefont {X.}~\bibnamefont {Mo}}, \bibinfo {author}
  {\bibfnamefont {T.}~\bibnamefont {Li}},\ and\ \bibinfo {author}
  {\bibfnamefont {Z.}~\bibnamefont {Li}},\ }\bibfield  {title} {\bibinfo
  {title} {Quantum entanglement creation for distant quantum memories via
  time-bin multiplexing},\ }\href@noop {} {\bibfield  {journal} {\bibinfo
  {journal} {Phys.~Rev.~A}\ }\textbf {\bibinfo {volume} {104}},\ \bibinfo
  {pages} {062409} (\bibinfo {year} {2021})}\BibitemShut {NoStop}%
\bibitem [{\citenamefont {Zheng}\ \emph {et~al.}(2022)\citenamefont {Zheng},
  \citenamefont {Sharma},\ and\ \citenamefont
  {Borregaard}}]{zheng2022entanglement}%
  \BibitemOpen
  \bibfield  {author} {\bibinfo {author} {\bibfnamefont {Y.}~\bibnamefont
  {Zheng}}, \bibinfo {author} {\bibfnamefont {H.}~\bibnamefont {Sharma}},\ and\
  \bibinfo {author} {\bibfnamefont {J.}~\bibnamefont {Borregaard}},\ }\bibfield
   {title} {\bibinfo {title} {Entanglement distribution with minimal memory
  requirements using time-bin photonic qudits},\ }\href@noop {} {\bibfield
  {journal} {\bibinfo  {journal} {PRX Quantum}\ }\textbf {\bibinfo {volume}
  {3}},\ \bibinfo {pages} {040319} (\bibinfo {year} {2022})}\BibitemShut
  {NoStop}%
\bibitem [{\citenamefont {Martinis}\ \emph {et~al.}(2005)\citenamefont
  {Martinis}, \citenamefont {Cooper}, \citenamefont {McDermott}, \citenamefont
  {Steffen}, \citenamefont {Ansmann}, \citenamefont {Osborn}, \citenamefont
  {Cicak}, \citenamefont {Oh}, \citenamefont {Pappas}, \citenamefont {Simmonds}
  \emph {et~al.}}]{martinis2005decoherence}%
  \BibitemOpen
  \bibfield  {author} {\bibinfo {author} {\bibfnamefont {J.~M.}\ \bibnamefont
  {Martinis}}, \bibinfo {author} {\bibfnamefont {K.~B.}\ \bibnamefont
  {Cooper}}, \bibinfo {author} {\bibfnamefont {R.}~\bibnamefont {McDermott}},
  \bibinfo {author} {\bibfnamefont {M.}~\bibnamefont {Steffen}}, \bibinfo
  {author} {\bibfnamefont {M.}~\bibnamefont {Ansmann}}, \bibinfo {author}
  {\bibfnamefont {K.~D.}\ \bibnamefont {Osborn}}, \bibinfo {author}
  {\bibfnamefont {K.}~\bibnamefont {Cicak}}, \bibinfo {author} {\bibfnamefont
  {S.}~\bibnamefont {Oh}}, \bibinfo {author} {\bibfnamefont {D.~P.}\
  \bibnamefont {Pappas}}, \bibinfo {author} {\bibfnamefont {R.~W.}\
  \bibnamefont {Simmonds}}, \emph {et~al.},\ }\bibfield  {title} {\bibinfo
  {title} {Decoherence in {J}osephson qubits from dielectric loss},\
  }\href@noop {} {\bibfield  {journal} {\bibinfo  {journal} {Phys.~Rev.~Lett.}\
  }\textbf {\bibinfo {volume} {95}},\ \bibinfo {pages} {210503} (\bibinfo
  {year} {2005})}\BibitemShut {NoStop}%
\bibitem [{\citenamefont {Wang}\ \emph {et~al.}(2015)\citenamefont {Wang},
  \citenamefont {Axline}, \citenamefont {Gao}, \citenamefont {Brecht},
  \citenamefont {Chu}, \citenamefont {Frunzio}, \citenamefont {Devoret},\ and\
  \citenamefont {Schoelkopf}}]{wang2015surface}%
  \BibitemOpen
  \bibfield  {author} {\bibinfo {author} {\bibfnamefont {C.}~\bibnamefont
  {Wang}}, \bibinfo {author} {\bibfnamefont {C.}~\bibnamefont {Axline}},
  \bibinfo {author} {\bibfnamefont {Y.~Y.}\ \bibnamefont {Gao}}, \bibinfo
  {author} {\bibfnamefont {T.}~\bibnamefont {Brecht}}, \bibinfo {author}
  {\bibfnamefont {Y.}~\bibnamefont {Chu}}, \bibinfo {author} {\bibfnamefont
  {L.}~\bibnamefont {Frunzio}}, \bibinfo {author} {\bibfnamefont
  {M.}~\bibnamefont {Devoret}},\ and\ \bibinfo {author} {\bibfnamefont {R.~J.}\
  \bibnamefont {Schoelkopf}},\ }\bibfield  {title} {\bibinfo {title} {Surface
  participation and dielectric loss in superconducting qubits},\ }\href@noop {}
  {\bibfield  {journal} {\bibinfo  {journal} {Appl.~Phys.~Lett.}\ }\textbf
  {\bibinfo {volume} {107}} (\bibinfo {year} {2015})}\BibitemShut {NoStop}%
\bibitem [{\citenamefont {De~Graaf}\ \emph {et~al.}(2018)\citenamefont
  {De~Graaf}, \citenamefont {Faoro}, \citenamefont {Burnett}, \citenamefont
  {Adamyan}, \citenamefont {Tzalenchuk}, \citenamefont {Kubatkin},
  \citenamefont {Lindstr{\"o}m},\ and\ \citenamefont
  {Danilov}}]{de2018suppression}%
  \BibitemOpen
  \bibfield  {author} {\bibinfo {author} {\bibfnamefont {S.}~\bibnamefont
  {De~Graaf}}, \bibinfo {author} {\bibfnamefont {L.}~\bibnamefont {Faoro}},
  \bibinfo {author} {\bibfnamefont {J.}~\bibnamefont {Burnett}}, \bibinfo
  {author} {\bibfnamefont {A.}~\bibnamefont {Adamyan}}, \bibinfo {author}
  {\bibfnamefont {A.~Y.}\ \bibnamefont {Tzalenchuk}}, \bibinfo {author}
  {\bibfnamefont {S.}~\bibnamefont {Kubatkin}}, \bibinfo {author}
  {\bibfnamefont {T.}~\bibnamefont {Lindstr{\"o}m}},\ and\ \bibinfo {author}
  {\bibfnamefont {A.}~\bibnamefont {Danilov}},\ }\bibfield  {title} {\bibinfo
  {title} {Suppression of low-frequency charge noise in superconducting
  resonators by surface spin desorption},\ }\href@noop {} {\bibfield  {journal}
  {\bibinfo  {journal} {Nat.~Commun.}\ }\textbf {\bibinfo {volume} {9}},\
  \bibinfo {pages} {1143} (\bibinfo {year} {2018})}\BibitemShut {NoStop}%
\bibitem [{\citenamefont {McRae}\ \emph {et~al.}(2020)\citenamefont {McRae},
  \citenamefont {Wang}, \citenamefont {Gao}, \citenamefont {Vissers},
  \citenamefont {Brecht}, \citenamefont {Dunsworth}, \citenamefont {Pappas},\
  and\ \citenamefont {Mutus}}]{mcrae2020materials}%
  \BibitemOpen
  \bibfield  {author} {\bibinfo {author} {\bibfnamefont {C.~R.~H.}\
  \bibnamefont {McRae}}, \bibinfo {author} {\bibfnamefont {H.}~\bibnamefont
  {Wang}}, \bibinfo {author} {\bibfnamefont {J.}~\bibnamefont {Gao}}, \bibinfo
  {author} {\bibfnamefont {M.~R.}\ \bibnamefont {Vissers}}, \bibinfo {author}
  {\bibfnamefont {T.}~\bibnamefont {Brecht}}, \bibinfo {author} {\bibfnamefont
  {A.}~\bibnamefont {Dunsworth}}, \bibinfo {author} {\bibfnamefont {D.~P.}\
  \bibnamefont {Pappas}},\ and\ \bibinfo {author} {\bibfnamefont
  {J.}~\bibnamefont {Mutus}},\ }\bibfield  {title} {\bibinfo {title} {Materials
  loss measurements using superconducting microwave resonators},\ }\href@noop
  {} {\bibfield  {journal} {\bibinfo  {journal} {Rev.~Sci.~Instrum.}\ }\textbf
  {\bibinfo {volume} {91}} (\bibinfo {year} {2020})}\BibitemShut {NoStop}%
\bibitem [{\citenamefont {Lisenfeld}\ \emph {et~al.}(2023)\citenamefont
  {Lisenfeld}, \citenamefont {Bilmes},\ and\ \citenamefont
  {Ustinov}}]{lisenfeld2023enhancing}%
  \BibitemOpen
  \bibfield  {author} {\bibinfo {author} {\bibfnamefont {J.}~\bibnamefont
  {Lisenfeld}}, \bibinfo {author} {\bibfnamefont {A.}~\bibnamefont {Bilmes}},\
  and\ \bibinfo {author} {\bibfnamefont {A.~V.}\ \bibnamefont {Ustinov}},\
  }\bibfield  {title} {\bibinfo {title} {Enhancing the coherence of
  superconducting quantum bits with electric fields},\ }\href@noop {}
  {\bibfield  {journal} {\bibinfo  {journal} {npj Quantum Inf.}\ }\textbf
  {\bibinfo {volume} {9}},\ \bibinfo {pages} {8} (\bibinfo {year}
  {2023})}\BibitemShut {NoStop}%
\bibitem [{\citenamefont {Crowley}\ \emph {et~al.}(2023)\citenamefont
  {Crowley}, \citenamefont {McLellan}, \citenamefont {Dutta}, \citenamefont
  {Shumiya}, \citenamefont {Place}, \citenamefont {Le}, \citenamefont {Gang},
  \citenamefont {Madhavan}, \citenamefont {Bland}, \citenamefont {Chang} \emph
  {et~al.}}]{crowley2023disentangling}%
  \BibitemOpen
  \bibfield  {author} {\bibinfo {author} {\bibfnamefont {K.~D.}\ \bibnamefont
  {Crowley}}, \bibinfo {author} {\bibfnamefont {R.~A.}\ \bibnamefont
  {McLellan}}, \bibinfo {author} {\bibfnamefont {A.}~\bibnamefont {Dutta}},
  \bibinfo {author} {\bibfnamefont {N.}~\bibnamefont {Shumiya}}, \bibinfo
  {author} {\bibfnamefont {A.~P.}\ \bibnamefont {Place}}, \bibinfo {author}
  {\bibfnamefont {X.~H.}\ \bibnamefont {Le}}, \bibinfo {author} {\bibfnamefont
  {Y.}~\bibnamefont {Gang}}, \bibinfo {author} {\bibfnamefont {T.}~\bibnamefont
  {Madhavan}}, \bibinfo {author} {\bibfnamefont {M.~P.}\ \bibnamefont {Bland}},
  \bibinfo {author} {\bibfnamefont {R.}~\bibnamefont {Chang}}, \emph {et~al.},\
  }\bibfield  {title} {\bibinfo {title} {Disentangling losses in tantalum
  superconducting circuits},\ }\href@noop {} {\bibfield  {journal} {\bibinfo
  {journal} {Phys.~Rev.~X}\ }\textbf {\bibinfo {volume} {13}},\ \bibinfo
  {pages} {041005} (\bibinfo {year} {2023})}\BibitemShut {NoStop}%
\bibitem [{\citenamefont {Chen}\ \emph {et~al.}(2024)\citenamefont {Chen},
  \citenamefont {Owens}, \citenamefont {Putterman}, \citenamefont
  {Sch{\"a}fer},\ and\ \citenamefont {Painter}}]{chen2024phonon}%
  \BibitemOpen
  \bibfield  {author} {\bibinfo {author} {\bibfnamefont {M.}~\bibnamefont
  {Chen}}, \bibinfo {author} {\bibfnamefont {J.~C.}\ \bibnamefont {Owens}},
  \bibinfo {author} {\bibfnamefont {H.}~\bibnamefont {Putterman}}, \bibinfo
  {author} {\bibfnamefont {M.}~\bibnamefont {Sch{\"a}fer}},\ and\ \bibinfo
  {author} {\bibfnamefont {O.}~\bibnamefont {Painter}},\ }\bibfield  {title}
  {\bibinfo {title} {Phonon engineering of atomic-scale defects in
  superconducting quantum circuits},\ }\href@noop {} {\bibfield  {journal}
  {\bibinfo  {journal} {Sci.~Adv.}\ }\textbf {\bibinfo {volume} {10}},\
  \bibinfo {pages} {eado6240} (\bibinfo {year} {2024})}\BibitemShut {NoStop}%
\bibitem [{\citenamefont {Liu}\ \emph {et~al.}(2024)\citenamefont {Liu},
  \citenamefont {Wang}, \citenamefont {Sheffer},\ and\ \citenamefont
  {Wang}}]{liu2024observation}%
  \BibitemOpen
  \bibfield  {author} {\bibinfo {author} {\bibfnamefont {B.-J.}\ \bibnamefont
  {Liu}}, \bibinfo {author} {\bibfnamefont {Y.-Y.}\ \bibnamefont {Wang}},
  \bibinfo {author} {\bibfnamefont {T.}~\bibnamefont {Sheffer}},\ and\ \bibinfo
  {author} {\bibfnamefont {C.}~\bibnamefont {Wang}},\ }\bibfield  {title}
  {\bibinfo {title} {Observation of discrete charge states of a coherent
  two-level system in a superconducting qubit},\ }\href@noop {} {\bibfield
  {journal} {\bibinfo  {journal} {Phys.~Rev.~Lett.}\ }\textbf {\bibinfo
  {volume} {133}},\ \bibinfo {pages} {160602} (\bibinfo {year}
  {2024})}\BibitemShut {NoStop}%
\bibitem [{\citenamefont {Zanuz}\ \emph {et~al.}(2024)\citenamefont {Zanuz},
  \citenamefont {Ficheux}, \citenamefont {Michaud}, \citenamefont {Orekhov},
  \citenamefont {Hanke}, \citenamefont {Flasby}, \citenamefont {Panah},
  \citenamefont {Norris}, \citenamefont {Kerschbaum}, \citenamefont {Remm}
  \emph {et~al.}}]{zanuz2024mitigating}%
  \BibitemOpen
  \bibfield  {author} {\bibinfo {author} {\bibfnamefont {D.~C.}\ \bibnamefont
  {Zanuz}}, \bibinfo {author} {\bibfnamefont {Q.}~\bibnamefont {Ficheux}},
  \bibinfo {author} {\bibfnamefont {L.}~\bibnamefont {Michaud}}, \bibinfo
  {author} {\bibfnamefont {A.}~\bibnamefont {Orekhov}}, \bibinfo {author}
  {\bibfnamefont {K.}~\bibnamefont {Hanke}}, \bibinfo {author} {\bibfnamefont
  {A.}~\bibnamefont {Flasby}}, \bibinfo {author} {\bibfnamefont {M.~B.}\
  \bibnamefont {Panah}}, \bibinfo {author} {\bibfnamefont {G.~J.}\ \bibnamefont
  {Norris}}, \bibinfo {author} {\bibfnamefont {M.}~\bibnamefont {Kerschbaum}},
  \bibinfo {author} {\bibfnamefont {A.}~\bibnamefont {Remm}}, \emph {et~al.},\
  }\bibfield  {title} {\bibinfo {title} {Mitigating losses of superconducting
  qubits strongly coupled to defect modes},\ }\href@noop {} {\bibfield
  {journal} {\bibinfo  {journal} {arXiv preprint arXiv:2407.18746}\ } (\bibinfo
  {year} {2024})}\BibitemShut {NoStop}%
\bibitem [{\citenamefont {Cirac}\ \emph {et~al.}(1997)\citenamefont {Cirac},
  \citenamefont {Zoller}, \citenamefont {Kimble},\ and\ \citenamefont
  {Mabuchi}}]{cirac1997quantum}%
  \BibitemOpen
  \bibfield  {author} {\bibinfo {author} {\bibfnamefont {J.~I.}\ \bibnamefont
  {Cirac}}, \bibinfo {author} {\bibfnamefont {P.}~\bibnamefont {Zoller}},
  \bibinfo {author} {\bibfnamefont {H.~J.}\ \bibnamefont {Kimble}},\ and\
  \bibinfo {author} {\bibfnamefont {H.}~\bibnamefont {Mabuchi}},\ }\bibfield
  {title} {\bibinfo {title} {Quantum state transfer and entanglement
  distribution among distant nodes in a quantum network},\ }\href@noop {}
  {\bibfield  {journal} {\bibinfo  {journal} {Phys.~Rev.~Lett.}\ }\textbf
  {\bibinfo {volume} {78}},\ \bibinfo {pages} {3221} (\bibinfo {year}
  {1997})}\BibitemShut {NoStop}%
\bibitem [{\citenamefont {Gerry}\ and\ \citenamefont
  {Eberly}(1990)}]{gerry1990dynamics}%
  \BibitemOpen
  \bibfield  {author} {\bibinfo {author} {\bibfnamefont {C.~C.}\ \bibnamefont
  {Gerry}}\ and\ \bibinfo {author} {\bibfnamefont {J.~H.}\ \bibnamefont
  {Eberly}},\ }\bibfield  {title} {\bibinfo {title} {Dynamics of a {R}aman
  coupled model interacting with two quantized cavity fields},\ }\href@noop {}
  {\bibfield  {journal} {\bibinfo  {journal} {Phys.~Rev.~A}\ }\textbf {\bibinfo
  {volume} {42}},\ \bibinfo {pages} {6805} (\bibinfo {year}
  {1990})}\BibitemShut {NoStop}%
\bibitem [{\citenamefont {Gorshkov}\ \emph {et~al.}(2007)\citenamefont
  {Gorshkov}, \citenamefont {Andr{\'e}}, \citenamefont {Lukin},\ and\
  \citenamefont {S{\o}rensen}}]{gorshkov2007photon}%
  \BibitemOpen
  \bibfield  {author} {\bibinfo {author} {\bibfnamefont {A.~V.}\ \bibnamefont
  {Gorshkov}}, \bibinfo {author} {\bibfnamefont {A.}~\bibnamefont {Andr{\'e}}},
  \bibinfo {author} {\bibfnamefont {M.~D.}\ \bibnamefont {Lukin}},\ and\
  \bibinfo {author} {\bibfnamefont {A.~S.}\ \bibnamefont {S{\o}rensen}},\
  }\bibfield  {title} {\bibinfo {title} {Photon storage in $\lambda$-type
  optically dense atomic media. {I.} {C}avity model},\ }\href@noop {}
  {\bibfield  {journal} {\bibinfo  {journal} {Phys.~Rev.~A}\ }\textbf {\bibinfo
  {volume} {76}},\ \bibinfo {pages} {033804} (\bibinfo {year}
  {2007})}\BibitemShut {NoStop}%
\bibitem [{\citenamefont {Morin}\ \emph {et~al.}(2019)\citenamefont {Morin},
  \citenamefont {K{\"o}rber}, \citenamefont {Langenfeld},\ and\ \citenamefont
  {Rempe}}]{morin2019deterministic}%
  \BibitemOpen
  \bibfield  {author} {\bibinfo {author} {\bibfnamefont {O.}~\bibnamefont
  {Morin}}, \bibinfo {author} {\bibfnamefont {M.}~\bibnamefont {K{\"o}rber}},
  \bibinfo {author} {\bibfnamefont {S.}~\bibnamefont {Langenfeld}},\ and\
  \bibinfo {author} {\bibfnamefont {G.}~\bibnamefont {Rempe}},\ }\bibfield
  {title} {\bibinfo {title} {Deterministic shaping and reshaping of
  single-photon temporal wave functions},\ }\href@noop {} {\bibfield  {journal}
  {\bibinfo  {journal} {Phys.~Rev.~Lett.}\ }\textbf {\bibinfo {volume} {123}},\
  \bibinfo {pages} {133602} (\bibinfo {year} {2019})}\BibitemShut {NoStop}%
\bibitem [{\citenamefont {Ilves}\ \emph {et~al.}(2020)\citenamefont {Ilves},
  \citenamefont {Kono}, \citenamefont {Sunada}, \citenamefont {Yamazaki},
  \citenamefont {Kim}, \citenamefont {Koshino},\ and\ \citenamefont
  {Nakamura}}]{ilves2020demand}%
  \BibitemOpen
  \bibfield  {author} {\bibinfo {author} {\bibfnamefont {J.}~\bibnamefont
  {Ilves}}, \bibinfo {author} {\bibfnamefont {S.}~\bibnamefont {Kono}},
  \bibinfo {author} {\bibfnamefont {Y.}~\bibnamefont {Sunada}}, \bibinfo
  {author} {\bibfnamefont {S.}~\bibnamefont {Yamazaki}}, \bibinfo {author}
  {\bibfnamefont {M.}~\bibnamefont {Kim}}, \bibinfo {author} {\bibfnamefont
  {K.}~\bibnamefont {Koshino}},\ and\ \bibinfo {author} {\bibfnamefont
  {Y.}~\bibnamefont {Nakamura}},\ }\bibfield  {title} {\bibinfo {title}
  {On-demand generation and characterization of a microwave time-bin qubit},\
  }\href@noop {} {\bibfield  {journal} {\bibinfo  {journal} {npj Quantum Inf.}\
  }\textbf {\bibinfo {volume} {6}},\ \bibinfo {pages} {34} (\bibinfo {year}
  {2020})}\BibitemShut {NoStop}%
\bibitem [{\citenamefont {Pechal}\ \emph {et~al.}(2014)\citenamefont {Pechal},
  \citenamefont {Huthmacher}, \citenamefont {Eichler}, \citenamefont
  {Zeytino{\u{g}}lu}, \citenamefont {Abdumalikov~Jr}, \citenamefont {Berger},
  \citenamefont {Wallraff},\ and\ \citenamefont
  {Filipp}}]{pechal2014microwave}%
  \BibitemOpen
  \bibfield  {author} {\bibinfo {author} {\bibfnamefont {M.}~\bibnamefont
  {Pechal}}, \bibinfo {author} {\bibfnamefont {L.}~\bibnamefont {Huthmacher}},
  \bibinfo {author} {\bibfnamefont {C.}~\bibnamefont {Eichler}}, \bibinfo
  {author} {\bibfnamefont {S.}~\bibnamefont {Zeytino{\u{g}}lu}}, \bibinfo
  {author} {\bibfnamefont {A.~A.}\ \bibnamefont {Abdumalikov~Jr}}, \bibinfo
  {author} {\bibfnamefont {S.}~\bibnamefont {Berger}}, \bibinfo {author}
  {\bibfnamefont {A.}~\bibnamefont {Wallraff}},\ and\ \bibinfo {author}
  {\bibfnamefont {S.}~\bibnamefont {Filipp}},\ }\bibfield  {title} {\bibinfo
  {title} {Microwave-controlled generation of shaped single photons in circuit
  quantum electrodynamics},\ }\href@noop {} {\bibfield  {journal} {\bibinfo
  {journal} {Phys.~Rev.~X}\ }\textbf {\bibinfo {volume} {4}},\ \bibinfo {pages}
  {041010} (\bibinfo {year} {2014})}\BibitemShut {NoStop}%
\bibitem [{\citenamefont {Zeytino{\u{g}}lu}\ \emph {et~al.}(2015)\citenamefont
  {Zeytino{\u{g}}lu}, \citenamefont {Pechal}, \citenamefont {Berger},
  \citenamefont {Abdumalikov~Jr}, \citenamefont {Wallraff},\ and\ \citenamefont
  {Filipp}}]{zeytinouglu2015microwave}%
  \BibitemOpen
  \bibfield  {author} {\bibinfo {author} {\bibfnamefont {S.}~\bibnamefont
  {Zeytino{\u{g}}lu}}, \bibinfo {author} {\bibfnamefont {M.}~\bibnamefont
  {Pechal}}, \bibinfo {author} {\bibfnamefont {S.}~\bibnamefont {Berger}},
  \bibinfo {author} {\bibfnamefont {A.~A.}\ \bibnamefont {Abdumalikov~Jr}},
  \bibinfo {author} {\bibfnamefont {A.}~\bibnamefont {Wallraff}},\ and\
  \bibinfo {author} {\bibfnamefont {S.}~\bibnamefont {Filipp}},\ }\bibfield
  {title} {\bibinfo {title} {Microwave-induced amplitude-and phase-tunable
  qubit-resonator coupling in circuit quantum electrodynamics},\ }\href@noop {}
  {\bibfield  {journal} {\bibinfo  {journal} {Phys.~Rev.~A}\ }\textbf {\bibinfo
  {volume} {91}},\ \bibinfo {pages} {043846} (\bibinfo {year}
  {2015})}\BibitemShut {NoStop}%
\bibitem [{\citenamefont {Gardiner}\ and\ \citenamefont
  {Collett}(1985)}]{gardiner1985input}%
  \BibitemOpen
  \bibfield  {author} {\bibinfo {author} {\bibfnamefont {C.~W.}\ \bibnamefont
  {Gardiner}}\ and\ \bibinfo {author} {\bibfnamefont {M.~J.}\ \bibnamefont
  {Collett}},\ }\bibfield  {title} {\bibinfo {title} {Input and output in
  damped quantum systems: {Q}uantum stochastic differential equations and the
  master equation},\ }\href@noop {} {\bibfield  {journal} {\bibinfo  {journal}
  {Phys.~Rev.~A}\ }\textbf {\bibinfo {volume} {31}},\ \bibinfo {pages} {3761}
  (\bibinfo {year} {1985})}\BibitemShut {NoStop}%
\bibitem [{\citenamefont {Ranjan}\ \emph {et~al.}(2020)\citenamefont {Ranjan},
  \citenamefont {Probst}, \citenamefont {Albanese}, \citenamefont {Doll},
  \citenamefont {Jacquot}, \citenamefont {Flurin}, \citenamefont {Heeres},
  \citenamefont {Vion}, \citenamefont {Esteve}, \citenamefont {Morton} \emph
  {et~al.}}]{ranjan2020pulsed}%
  \BibitemOpen
  \bibfield  {author} {\bibinfo {author} {\bibfnamefont {V.}~\bibnamefont
  {Ranjan}}, \bibinfo {author} {\bibfnamefont {S.}~\bibnamefont {Probst}},
  \bibinfo {author} {\bibfnamefont {B.}~\bibnamefont {Albanese}}, \bibinfo
  {author} {\bibfnamefont {A.}~\bibnamefont {Doll}}, \bibinfo {author}
  {\bibfnamefont {O.}~\bibnamefont {Jacquot}}, \bibinfo {author} {\bibfnamefont
  {E.}~\bibnamefont {Flurin}}, \bibinfo {author} {\bibfnamefont
  {R.}~\bibnamefont {Heeres}}, \bibinfo {author} {\bibfnamefont
  {D.}~\bibnamefont {Vion}}, \bibinfo {author} {\bibfnamefont {D.}~\bibnamefont
  {Esteve}}, \bibinfo {author} {\bibfnamefont {J.~J.~L.}\ \bibnamefont
  {Morton}}, \emph {et~al.},\ }\bibfield  {title} {\bibinfo {title} {Pulsed
  electron spin resonance spectroscopy in the {P}urcell regime},\ }\href@noop
  {} {\bibfield  {journal} {\bibinfo  {journal} {J.~Magn.~Reson.~}\ }\textbf
  {\bibinfo {volume} {310}},\ \bibinfo {pages} {106662} (\bibinfo {year}
  {2020})}\BibitemShut {NoStop}%
\bibitem [{\citenamefont {McIntyre}\ and\ \citenamefont
  {Coish}(2022)}]{mcintyre2022non}%
  \BibitemOpen
  \bibfield  {author} {\bibinfo {author} {\bibfnamefont {Z.}~\bibnamefont
  {McIntyre}}\ and\ \bibinfo {author} {\bibfnamefont {W.~A.}\ \bibnamefont
  {Coish}},\ }\bibfield  {title} {\bibinfo {title} {Non-{M}arkovian transient
  spectroscopy in cavity {QED}},\ }\href@noop {} {\bibfield  {journal}
  {\bibinfo  {journal} {Phys.~Rev.~Res.}\ }\textbf {\bibinfo {volume} {4}},\
  \bibinfo {pages} {L042039} (\bibinfo {year} {2022})}\BibitemShut {NoStop}%
\bibitem [{\citenamefont {Peterer}\ \emph {et~al.}(2015)\citenamefont
  {Peterer}, \citenamefont {Bader}, \citenamefont {Jin}, \citenamefont {Yan},
  \citenamefont {Kamal}, \citenamefont {Gudmundsen}, \citenamefont {Leek},
  \citenamefont {Orlando}, \citenamefont {Oliver},\ and\ \citenamefont
  {Gustavsson}}]{peterer2015coherence}%
  \BibitemOpen
  \bibfield  {author} {\bibinfo {author} {\bibfnamefont {M.~J.}\ \bibnamefont
  {Peterer}}, \bibinfo {author} {\bibfnamefont {S.~J.}\ \bibnamefont {Bader}},
  \bibinfo {author} {\bibfnamefont {X.}~\bibnamefont {Jin}}, \bibinfo {author}
  {\bibfnamefont {F.}~\bibnamefont {Yan}}, \bibinfo {author} {\bibfnamefont
  {A.}~\bibnamefont {Kamal}}, \bibinfo {author} {\bibfnamefont {T.~J.}\
  \bibnamefont {Gudmundsen}}, \bibinfo {author} {\bibfnamefont {P.~J.}\
  \bibnamefont {Leek}}, \bibinfo {author} {\bibfnamefont {T.~P.}\ \bibnamefont
  {Orlando}}, \bibinfo {author} {\bibfnamefont {W.~D.}\ \bibnamefont
  {Oliver}},\ and\ \bibinfo {author} {\bibfnamefont {S.}~\bibnamefont
  {Gustavsson}},\ }\bibfield  {title} {\bibinfo {title} {Coherence and decay of
  higher energy levels of a superconducting transmon qubit},\ }\href@noop {}
  {\bibfield  {journal} {\bibinfo  {journal} {Phys.~Rev.~Lett.}\ }\textbf
  {\bibinfo {volume} {114}},\ \bibinfo {pages} {010501} (\bibinfo {year}
  {2015})}\BibitemShut {NoStop}%
\bibitem [{\citenamefont {Besse}\ \emph {et~al.}(2018)\citenamefont {Besse},
  \citenamefont {Gasparinetti}, \citenamefont {Collodo}, \citenamefont
  {Walter}, \citenamefont {Kurpiers}, \citenamefont {Pechal}, \citenamefont
  {Eichler},\ and\ \citenamefont {Wallraff}}]{besse2018single}%
  \BibitemOpen
  \bibfield  {author} {\bibinfo {author} {\bibfnamefont {J.-C.}\ \bibnamefont
  {Besse}}, \bibinfo {author} {\bibfnamefont {S.}~\bibnamefont {Gasparinetti}},
  \bibinfo {author} {\bibfnamefont {M.~C.}\ \bibnamefont {Collodo}}, \bibinfo
  {author} {\bibfnamefont {T.}~\bibnamefont {Walter}}, \bibinfo {author}
  {\bibfnamefont {P.}~\bibnamefont {Kurpiers}}, \bibinfo {author}
  {\bibfnamefont {M.}~\bibnamefont {Pechal}}, \bibinfo {author} {\bibfnamefont
  {C.}~\bibnamefont {Eichler}},\ and\ \bibinfo {author} {\bibfnamefont
  {A.}~\bibnamefont {Wallraff}},\ }\bibfield  {title} {\bibinfo {title}
  {Single-shot quantum nondemolition detection of individual itinerant
  microwave photons},\ }\href@noop {} {\bibfield  {journal} {\bibinfo
  {journal} {Phys.~Rev.~X}\ }\textbf {\bibinfo {volume} {8}},\ \bibinfo {pages}
  {021003} (\bibinfo {year} {2018})}\BibitemShut {NoStop}%
\bibitem [{\citenamefont {Hacker}\ \emph {et~al.}(2019)\citenamefont {Hacker},
  \citenamefont {Welte}, \citenamefont {Daiss}, \citenamefont {Shaukat},
  \citenamefont {Ritter}, \citenamefont {Li},\ and\ \citenamefont
  {Rempe}}]{hacker2019deterministic}%
  \BibitemOpen
  \bibfield  {author} {\bibinfo {author} {\bibfnamefont {B.}~\bibnamefont
  {Hacker}}, \bibinfo {author} {\bibfnamefont {S.}~\bibnamefont {Welte}},
  \bibinfo {author} {\bibfnamefont {S.}~\bibnamefont {Daiss}}, \bibinfo
  {author} {\bibfnamefont {A.}~\bibnamefont {Shaukat}}, \bibinfo {author}
  {\bibfnamefont {S.}~\bibnamefont {Ritter}}, \bibinfo {author} {\bibfnamefont
  {L.}~\bibnamefont {Li}},\ and\ \bibinfo {author} {\bibfnamefont
  {G.}~\bibnamefont {Rempe}},\ }\bibfield  {title} {\bibinfo {title}
  {Deterministic creation of entangled atom--light {S}chr{\"o}dinger-cat
  states},\ }\href@noop {} {\bibfield  {journal} {\bibinfo  {journal}
  {Nat.~Photonics}\ }\textbf {\bibinfo {volume} {13}},\ \bibinfo {pages} {110}
  (\bibinfo {year} {2019})}\BibitemShut {NoStop}%
\bibitem [{\citenamefont {Besse}\ \emph {et~al.}(2020)\citenamefont {Besse},
  \citenamefont {Gasparinetti}, \citenamefont {Collodo}, \citenamefont
  {Walter}, \citenamefont {Remm}, \citenamefont {Krause}, \citenamefont
  {Eichler},\ and\ \citenamefont {Wallraff}}]{besse2020parity}%
  \BibitemOpen
  \bibfield  {author} {\bibinfo {author} {\bibfnamefont {J.-C.}\ \bibnamefont
  {Besse}}, \bibinfo {author} {\bibfnamefont {S.}~\bibnamefont {Gasparinetti}},
  \bibinfo {author} {\bibfnamefont {M.~C.}\ \bibnamefont {Collodo}}, \bibinfo
  {author} {\bibfnamefont {T.}~\bibnamefont {Walter}}, \bibinfo {author}
  {\bibfnamefont {A.}~\bibnamefont {Remm}}, \bibinfo {author} {\bibfnamefont
  {J.}~\bibnamefont {Krause}}, \bibinfo {author} {\bibfnamefont
  {C.}~\bibnamefont {Eichler}},\ and\ \bibinfo {author} {\bibfnamefont
  {A.}~\bibnamefont {Wallraff}},\ }\bibfield  {title} {\bibinfo {title} {Parity
  detection of propagating microwave fields},\ }\href@noop {} {\bibfield
  {journal} {\bibinfo  {journal} {Phys.~Rev.~X}\ }\textbf {\bibinfo {volume}
  {10}},\ \bibinfo {pages} {011046} (\bibinfo {year} {2020})}\BibitemShut
  {NoStop}%
\bibitem [{\citenamefont {Kono}\ \emph {et~al.}(2018)\citenamefont {Kono},
  \citenamefont {Koshino}, \citenamefont {Tabuchi}, \citenamefont {Noguchi},\
  and\ \citenamefont {Nakamura}}]{kono2018quantum}%
  \BibitemOpen
  \bibfield  {author} {\bibinfo {author} {\bibfnamefont {S.}~\bibnamefont
  {Kono}}, \bibinfo {author} {\bibfnamefont {K.}~\bibnamefont {Koshino}},
  \bibinfo {author} {\bibfnamefont {Y.}~\bibnamefont {Tabuchi}}, \bibinfo
  {author} {\bibfnamefont {A.}~\bibnamefont {Noguchi}},\ and\ \bibinfo {author}
  {\bibfnamefont {Y.}~\bibnamefont {Nakamura}},\ }\bibfield  {title} {\bibinfo
  {title} {Quantum non-demolition detection of an itinerant microwave photon},\
  }\href@noop {} {\bibfield  {journal} {\bibinfo  {journal} {Nat.~Phys.}\
  }\textbf {\bibinfo {volume} {14}},\ \bibinfo {pages} {546} (\bibinfo {year}
  {2018})}\BibitemShut {NoStop}%
\bibitem [{\citenamefont {Wang}\ \emph {et~al.}(2022)\citenamefont {Wang},
  \citenamefont {Bao}, \citenamefont {Wu}, \citenamefont {Li}, \citenamefont
  {Cai}, \citenamefont {Wang}, \citenamefont {Ma}, \citenamefont {Cai},
  \citenamefont {Han}, \citenamefont {Wang} \emph {et~al.}}]{wang2022flying}%
  \BibitemOpen
  \bibfield  {author} {\bibinfo {author} {\bibfnamefont {Z.}~\bibnamefont
  {Wang}}, \bibinfo {author} {\bibfnamefont {Z.}~\bibnamefont {Bao}}, \bibinfo
  {author} {\bibfnamefont {Y.}~\bibnamefont {Wu}}, \bibinfo {author}
  {\bibfnamefont {Y.}~\bibnamefont {Li}}, \bibinfo {author} {\bibfnamefont
  {W.}~\bibnamefont {Cai}}, \bibinfo {author} {\bibfnamefont {W.}~\bibnamefont
  {Wang}}, \bibinfo {author} {\bibfnamefont {Y.}~\bibnamefont {Ma}}, \bibinfo
  {author} {\bibfnamefont {T.}~\bibnamefont {Cai}}, \bibinfo {author}
  {\bibfnamefont {X.}~\bibnamefont {Han}}, \bibinfo {author} {\bibfnamefont
  {J.}~\bibnamefont {Wang}}, \emph {et~al.},\ }\bibfield  {title} {\bibinfo
  {title} {A flying {S}chr{\"o}dinger’s cat in multipartite entangled
  states},\ }\href@noop {} {\bibfield  {journal} {\bibinfo  {journal}
  {Sci.~Adv.}\ }\textbf {\bibinfo {volume} {8}},\ \bibinfo {pages} {eabn1778}
  (\bibinfo {year} {2022})}\BibitemShut {NoStop}%
\bibitem [{\citenamefont {Pechal}\ \emph {et~al.}(2016)\citenamefont {Pechal},
  \citenamefont {Besse}, \citenamefont {Mondal}, \citenamefont {Oppliger},
  \citenamefont {Gasparinetti},\ and\ \citenamefont
  {Wallraff}}]{pechal2016superconducting}%
  \BibitemOpen
  \bibfield  {author} {\bibinfo {author} {\bibfnamefont {M.}~\bibnamefont
  {Pechal}}, \bibinfo {author} {\bibfnamefont {J.-C.}\ \bibnamefont {Besse}},
  \bibinfo {author} {\bibfnamefont {M.}~\bibnamefont {Mondal}}, \bibinfo
  {author} {\bibfnamefont {M.}~\bibnamefont {Oppliger}}, \bibinfo {author}
  {\bibfnamefont {S.}~\bibnamefont {Gasparinetti}},\ and\ \bibinfo {author}
  {\bibfnamefont {A.}~\bibnamefont {Wallraff}},\ }\bibfield  {title} {\bibinfo
  {title} {Superconducting switch for fast on-chip routing of quantum microwave
  fields},\ }\href@noop {} {\bibfield  {journal} {\bibinfo  {journal}
  {Phys.~Rev.~Appl.}\ }\textbf {\bibinfo {volume} {6}},\ \bibinfo {pages}
  {024009} (\bibinfo {year} {2016})}\BibitemShut {NoStop}%
\bibitem [{\citenamefont {Li}\ \emph {et~al.}(2024)\citenamefont {Li},
  \citenamefont {Bao}, \citenamefont {Wang}, \citenamefont {Wu}, \citenamefont
  {Wang}, \citenamefont {Yang}, \citenamefont {Xiong}, \citenamefont {Song},
  \citenamefont {Zhang},\ and\ \citenamefont {Duan}}]{li2024quantum}%
  \BibitemOpen
  \bibfield  {author} {\bibinfo {author} {\bibfnamefont {Y.}~\bibnamefont
  {Li}}, \bibinfo {author} {\bibfnamefont {Z.}~\bibnamefont {Bao}}, \bibinfo
  {author} {\bibfnamefont {Z.}~\bibnamefont {Wang}}, \bibinfo {author}
  {\bibfnamefont {Y.}~\bibnamefont {Wu}}, \bibinfo {author} {\bibfnamefont
  {J.}~\bibnamefont {Wang}}, \bibinfo {author} {\bibfnamefont {J.}~\bibnamefont
  {Yang}}, \bibinfo {author} {\bibfnamefont {H.}~\bibnamefont {Xiong}},
  \bibinfo {author} {\bibfnamefont {Y.}~\bibnamefont {Song}}, \bibinfo {author}
  {\bibfnamefont {H.}~\bibnamefont {Zhang}},\ and\ \bibinfo {author}
  {\bibfnamefont {L.}~\bibnamefont {Duan}},\ }\bibfield  {title} {\bibinfo
  {title} {Quantum switch for itinerant microwave single photons with
  superconducting quantum circuits},\ }\href@noop {} {\bibfield  {journal}
  {\bibinfo  {journal} {Phys.~Rev.~Appl.}\ }\textbf {\bibinfo {volume} {21}},\
  \bibinfo {pages} {044030} (\bibinfo {year} {2024})}\BibitemShut {NoStop}%
\bibitem [{\citenamefont {Noh}\ \emph {et~al.}(2023)\citenamefont {Noh},
  \citenamefont {Xiao}, \citenamefont {Jin}, \citenamefont {Cicak},
  \citenamefont {Doucet}, \citenamefont {Aumentado}, \citenamefont {Govia},
  \citenamefont {Ranzani}, \citenamefont {Kamal},\ and\ \citenamefont
  {Simmonds}}]{noh2023strong}%
  \BibitemOpen
  \bibfield  {author} {\bibinfo {author} {\bibfnamefont {T.}~\bibnamefont
  {Noh}}, \bibinfo {author} {\bibfnamefont {Z.}~\bibnamefont {Xiao}}, \bibinfo
  {author} {\bibfnamefont {X.~Y.}\ \bibnamefont {Jin}}, \bibinfo {author}
  {\bibfnamefont {K.}~\bibnamefont {Cicak}}, \bibinfo {author} {\bibfnamefont
  {E.}~\bibnamefont {Doucet}}, \bibinfo {author} {\bibfnamefont
  {J.}~\bibnamefont {Aumentado}}, \bibinfo {author} {\bibfnamefont {L.~C.~G.}\
  \bibnamefont {Govia}}, \bibinfo {author} {\bibfnamefont {L.}~\bibnamefont
  {Ranzani}}, \bibinfo {author} {\bibfnamefont {A.}~\bibnamefont {Kamal}},\
  and\ \bibinfo {author} {\bibfnamefont {R.~W.}\ \bibnamefont {Simmonds}},\
  }\bibfield  {title} {\bibinfo {title} {Strong parametric dispersive shifts in
  a statically decoupled two-qubit cavity {QED} system},\ }\href@noop {}
  {\bibfield  {journal} {\bibinfo  {journal} {Nat.~Phys.}\ }\textbf {\bibinfo
  {volume} {19}},\ \bibinfo {pages} {1445} (\bibinfo {year}
  {2023})}\BibitemShut {NoStop}%
\bibitem [{\citenamefont {Gard}\ \emph {et~al.}(2024)\citenamefont {Gard},
  \citenamefont {Parrott}, \citenamefont {Jacobs}, \citenamefont {Aumentado},\
  and\ \citenamefont {Simmonds}}]{gard2024fast}%
  \BibitemOpen
  \bibfield  {author} {\bibinfo {author} {\bibfnamefont {B.~T.}\ \bibnamefont
  {Gard}}, \bibinfo {author} {\bibfnamefont {Z.}~\bibnamefont {Parrott}},
  \bibinfo {author} {\bibfnamefont {K.}~\bibnamefont {Jacobs}}, \bibinfo
  {author} {\bibfnamefont {J.}~\bibnamefont {Aumentado}},\ and\ \bibinfo
  {author} {\bibfnamefont {R.~W.}\ \bibnamefont {Simmonds}},\ }\bibfield
  {title} {\bibinfo {title} {Fast high-fidelity quantum nondemolition readout
  of a superconducting qubit with tunable transverse couplings},\ }\href@noop
  {} {\bibfield  {journal} {\bibinfo  {journal} {Phys.~Rev.~Appl.}\ }\textbf
  {\bibinfo {volume} {21}},\ \bibinfo {pages} {024008} (\bibinfo {year}
  {2024})}\BibitemShut {NoStop}%
\bibitem [{\citenamefont {Jerger}\ \emph {et~al.}(2016)\citenamefont {Jerger},
  \citenamefont {Macha}, \citenamefont {Hamann}, \citenamefont {Reshitnyk},
  \citenamefont {Juliusson},\ and\ \citenamefont
  {Fedorov}}]{jerger2016realization}%
  \BibitemOpen
  \bibfield  {author} {\bibinfo {author} {\bibfnamefont {M.}~\bibnamefont
  {Jerger}}, \bibinfo {author} {\bibfnamefont {P.}~\bibnamefont {Macha}},
  \bibinfo {author} {\bibfnamefont {A.~R.}\ \bibnamefont {Hamann}}, \bibinfo
  {author} {\bibfnamefont {Y.}~\bibnamefont {Reshitnyk}}, \bibinfo {author}
  {\bibfnamefont {K.}~\bibnamefont {Juliusson}},\ and\ \bibinfo {author}
  {\bibfnamefont {A.}~\bibnamefont {Fedorov}},\ }\bibfield  {title} {\bibinfo
  {title} {Realization of a binary-outcome projection measurement of a
  three-level superconducting quantum system},\ }\href@noop {} {\bibfield
  {journal} {\bibinfo  {journal} {Phys.~Rev.~Appl.}\ }\textbf {\bibinfo
  {volume} {6}},\ \bibinfo {pages} {014014} (\bibinfo {year}
  {2016})}\BibitemShut {NoStop}%
\bibitem [{\citenamefont {Soudagar}\ \emph {et~al.}(2007)\citenamefont
  {Soudagar}, \citenamefont {Bussi{\`e}res}, \citenamefont {Berl{\'\i}n},
  \citenamefont {Lacroix}, \citenamefont {Fernandez},\ and\ \citenamefont
  {Godbout}}]{soudagar2007cluster}%
  \BibitemOpen
  \bibfield  {author} {\bibinfo {author} {\bibfnamefont {Y.}~\bibnamefont
  {Soudagar}}, \bibinfo {author} {\bibfnamefont {F.}~\bibnamefont
  {Bussi{\`e}res}}, \bibinfo {author} {\bibfnamefont {G.}~\bibnamefont
  {Berl{\'\i}n}}, \bibinfo {author} {\bibfnamefont {S.}~\bibnamefont
  {Lacroix}}, \bibinfo {author} {\bibfnamefont {J.~M.}\ \bibnamefont
  {Fernandez}},\ and\ \bibinfo {author} {\bibfnamefont {N.}~\bibnamefont
  {Godbout}},\ }\bibfield  {title} {\bibinfo {title} {Cluster-state quantum
  computing in optical fibers},\ }\href@noop {} {\bibfield  {journal} {\bibinfo
   {journal} {JOSA B}\ }\textbf {\bibinfo {volume} {24}},\ \bibinfo {pages}
  {226} (\bibinfo {year} {2007})}\BibitemShut {NoStop}%
\bibitem [{\citenamefont {Lo}\ \emph {et~al.}(2020)\citenamefont {Lo},
  \citenamefont {Ikuta}, \citenamefont {Matsuda}, \citenamefont {Honjo},
  \citenamefont {Munro},\ and\ \citenamefont {Takesue}}]{lo2020quantum}%
  \BibitemOpen
  \bibfield  {author} {\bibinfo {author} {\bibfnamefont {H.-P.}\ \bibnamefont
  {Lo}}, \bibinfo {author} {\bibfnamefont {T.}~\bibnamefont {Ikuta}}, \bibinfo
  {author} {\bibfnamefont {N.}~\bibnamefont {Matsuda}}, \bibinfo {author}
  {\bibfnamefont {T.}~\bibnamefont {Honjo}}, \bibinfo {author} {\bibfnamefont
  {W.~J.}\ \bibnamefont {Munro}},\ and\ \bibinfo {author} {\bibfnamefont
  {H.}~\bibnamefont {Takesue}},\ }\bibfield  {title} {\bibinfo {title} {Quantum
  process tomography of a controlled-phase gate for time-bin qubits},\
  }\href@noop {} {\bibfield  {journal} {\bibinfo  {journal} {Phys.~Rev.~Appl.}\
  }\textbf {\bibinfo {volume} {13}},\ \bibinfo {pages} {034013} (\bibinfo
  {year} {2020})}\BibitemShut {NoStop}%
\bibitem [{\citenamefont {Bussi{\`e}res}\ \emph {et~al.}(2010)\citenamefont
  {Bussi{\`e}res}, \citenamefont {Slater}, \citenamefont {Jin}, \citenamefont
  {Godbout},\ and\ \citenamefont {Tittel}}]{bussieres2010testing}%
  \BibitemOpen
  \bibfield  {author} {\bibinfo {author} {\bibfnamefont {F.}~\bibnamefont
  {Bussi{\`e}res}}, \bibinfo {author} {\bibfnamefont {J.~A.}\ \bibnamefont
  {Slater}}, \bibinfo {author} {\bibfnamefont {J.}~\bibnamefont {Jin}},
  \bibinfo {author} {\bibfnamefont {N.}~\bibnamefont {Godbout}},\ and\ \bibinfo
  {author} {\bibfnamefont {W.}~\bibnamefont {Tittel}},\ }\bibfield  {title}
  {\bibinfo {title} {Testing nonlocality over 12.4 km of underground fiber with
  universal time-bin qubit analyzers},\ }\href@noop {} {\bibfield  {journal}
  {\bibinfo  {journal} {Phys.~Rev.~A}\ }\textbf {\bibinfo {volume} {81}},\
  \bibinfo {pages} {052106} (\bibinfo {year} {2010})}\BibitemShut {NoStop}%
\bibitem [{\citenamefont {Tissot}\ and\ \citenamefont
  {Burkard}(2024)}]{tissot2024efficient}%
  \BibitemOpen
  \bibfield  {author} {\bibinfo {author} {\bibfnamefont {B.}~\bibnamefont
  {Tissot}}\ and\ \bibinfo {author} {\bibfnamefont {G.}~\bibnamefont
  {Burkard}},\ }\bibfield  {title} {\bibinfo {title} {Efficient high-fidelity
  flying qubit shaping},\ }\href@noop {} {\bibfield  {journal} {\bibinfo
  {journal} {Phys.~Rev.~Res.}\ }\textbf {\bibinfo {volume} {6}},\ \bibinfo
  {pages} {013150} (\bibinfo {year} {2024})}\BibitemShut {NoStop}%
\bibitem [{\citenamefont {Combes}\ \emph {et~al.}(2017)\citenamefont {Combes},
  \citenamefont {Kerckhoff},\ and\ \citenamefont {Sarovar}}]{combes2017slh}%
  \BibitemOpen
  \bibfield  {author} {\bibinfo {author} {\bibfnamefont {J.}~\bibnamefont
  {Combes}}, \bibinfo {author} {\bibfnamefont {J.}~\bibnamefont {Kerckhoff}},\
  and\ \bibinfo {author} {\bibfnamefont {M.}~\bibnamefont {Sarovar}},\
  }\bibfield  {title} {\bibinfo {title} {The {SLH} framework for modeling
  quantum input-output networks},\ }\href@noop {} {\bibfield  {journal}
  {\bibinfo  {journal} {Adv.~Phys.:~X}\ }\textbf {\bibinfo {volume} {2}},\
  \bibinfo {pages} {784} (\bibinfo {year} {2017})}\BibitemShut {NoStop}%
\bibitem [{\citenamefont {McIntyre}\ and\ \citenamefont
  {Coish}(2024{\natexlab{b}})}]{mcintyre2024photonic}%
  \BibitemOpen
  \bibfield  {author} {\bibinfo {author} {\bibfnamefont {Z.~M.}\ \bibnamefont
  {McIntyre}}\ and\ \bibinfo {author} {\bibfnamefont {W.~A.}\ \bibnamefont
  {Coish}},\ }\bibfield  {title} {\bibinfo {title} {Photonic which-path
  entangler based on longitudinal cavity-qubit coupling},\ }\href@noop {}
  {\bibfield  {journal} {\bibinfo  {journal} {Phys.~Rev.~Lett.}\ }\textbf
  {\bibinfo {volume} {132}},\ \bibinfo {pages} {093603} (\bibinfo {year}
  {2024}{\natexlab{b}})}\BibitemShut {NoStop}%
\end{thebibliography}
\end{document}